\RequirePackage{fix-cm}
\documentclass{svjour3}                     
\smartqed  
\usepackage{graphicx}
\graphicspath{{./figures/}}
\usepackage[width=17cm]{geometry}
\usepackage{amsmath,amssymb,physics}
\usepackage[skip=0.333\baselineskip]{caption}
\usepackage{subcaption,float}
\usepackage{xcolor}
\usepackage{hyperref}
\usepackage{blindtext}
\usepackage{enumitem}
\usepackage{xcolor}
\usepackage{nicefrac,xfrac}

\usepackage{booktabs}
\usepackage{multirow}

\DeclareSymbolFont{yhlargesymbols}{OMX}{yhex}{m}{n} 
\DeclareMathAccent{\widerhat}{\mathord}{yhlargesymbols}{"62}
\journalname{Journal of Elasticity}
\begin{document}

\title{Deformation of a planar ferromagnetic elastic ribbon
}


\author{G. R. Krishna Chand Avatar         \and
	Vivekanand Dabade 
}


\institute{Vivekanand Dabade \at
	Department of Aerospace Engineering \\
	Indian Institute of Science\\
	Bengaluru, India \\
	\email{dabade@iisc.ac.in}           
}

\date{Received: date / Accepted: date}

\maketitle

\begin{abstract}

In this paper we explore the influence of magnetisation on the deformation of planar ferromagnetic elastic ribbons. We begin the investigation by deriving the leading-order magnetic energy associated with a curved planar ferromagnetic elastic ribbon. The sum of the magnetic  and the elastic energy is the total energy of the ribbon. We derive the equilibrium equations by taking the first variation of the total energy. We then systematically determine and analyse solutions to these equilibrium equations under various canonical boundary conditions. We also determine the stability of the equilibrium solutions. Comparing our findings with the well-studied Euler's elastica provides insights into the magnetic effects on the deformation behaviour of elastic ribbons. Our analysis contributes to a deeper understanding of the interplay between magnetisation and the mechanical response of planar ferromagnetic structures, and offers valuable insights for both theoretical and practical applications.
	
	\keywords{Magnetoelastic slender structures \and Soft ferromagnets \and Hard ferromagnets \and Ribbons  \and Elastica}
	\subclass{74B20
		 \and 74F15
		  \and 74G60}
\end{abstract}

\section{Introduction}\label{sec:intro}

Ferromagnetic ribbons and slender structures exhibit complex coupling between magnetism and deformation or elasticity and they offer the potential to achieve substantial deformations using remote external magnetic fields. Novel deformations, unachievable in purely elastic slender structures, find interesting applications in advanced actuators and sensors fabricated from ferromagnetic slender structures \cite{samourgkanidis2020characterization,Ramachandran2016,lum2016shape}. For instance, the concept of a stowed preloaded structure, which can transition into novel configurations upon the application of a small external magnetic field, holds promise for diverse applications in deployable structures. In this paper, we employ concepts drawn from the theory of micromagnetics  and Euler's elastica to construct a model aimed at analysing the planar deformation of ferromagnetic ribbons.

Motivated by the aforementioned applications and prospects of ferromagnetic slender structures, our study focuses on analysing the interplay between magnetism and deformation in ferromagnetic ribbons. We adopt a variational approach for this problem and begin our analysis by deriving the total energy of the ferromagnetic ribbon undergoing planar deformation. The total energy of a ferromagnetic slender ribbon is given by the sum of the elastic energy, magnetic energy of the ribbon and energy due to mechanical loading device. The elastic energy is equal to the bending energy, as we shall assume that the ribbon is inextensible. The rationale for selecting this particular model for the elastic energy of the ferromagnetic planar ribbon is justified in section \ref{sec:mechanical-energy}. The magnetic energy is formulated based on the theory of micromagnetics.

 Micromagnetics is a continuum theory for magnetism that has proven to be a robust model capable of explaining and predicting the diverse array of magnetic domain structures observed in ferromagnetic materials \cite{desimone2006recent,brown1963micromagnetics}.  The magnetisation vector $\vb*{m}(\vb*{x})$ is the primary variable in the micromagnetics functional. The magnetisation vector  \(\vb*{m}(\vb*{x})\) is a unit-normed vector supported on a ferromagnetic body. The micromagnetics functional is expressed as the sum of exchange, magnetocrystalline anisotropy, magnetostatic (abbreviated as demag.), and Zeeman energy, see \cite{desimone2006recent} and \cite{dabade2019micromagnetics}. The exchange energy is characterized by a material constant known as the exchange constant, denoted as $A$. Similarly, the anisotropy energy is characterized by a material constant known as the anisotropy constant, denoted as $K_{a}$. The demag. and Zeeman energy are characterized by a material constant known as the magnetostatic energy density constant, denoted as $K_{d}$. Thus, the constants $A$, $K_{a}$, and $K_{d}$ describe the ferromagnet according to the micromagnetics functional. We refer the reader to Section (\ref{sec:magnetic-energy}) for a concise discussion on the micromagnetics functional.

Ferromagnets are commonly categorised into soft and hard ferromagnets, depending on the relative magnitudes of $(K_{a})$ and $(K_{d})$. In soft ferromagnets $K_{d} \gg K_{a}$, while in hard ferromagnets $K_{a} \gg K_{d}$. In this study, we investigate both soft and hard ferromagnetic ribbons. In soft ferromagnetic ribbons, the demag. energy influences the deformation of the ribbon.  The demag. energy, strongly depends on the shape of the body and is computed by solving Maxwell’s equations of magnetostatics. Due to the non local nature of the demag. energy, calculating it is computationally expensive and challenging for general three dimensional bodies, as it requires the calculation of the magnetic field in the entire space surrounding the body. However, for slender structures, the leading order demag. energy is local in nature and assumes simpler forms, offering an opportunity to explore these problems (semi-) analytically. In hard ferromagnetic ribbons, the Zeeman energy influences the deformation of the ribbon.  In such materials, magnetization remains fixed along specific preferred directions known as the easy axis of the material. The Zeeman energy seeks to orient the body such that these preferred directions are aligned with the external magnetic field. Hence, our analysis focuses on the deformation of a slender body resulting from the interplay of demag. energy and Zeeman energy along with elastic energy. 


The demagnetization (and Zeeman) energy scales very differently compared to the elastic energy in bulk versus slender ferromagnets. This disparity in scaling leads to significantly larger displacements in slender ferromagnetic structures compared to bulk ferromagnets under an identical external magnetic field. In bulk ferromagnetic materials, the elastic energy density scales as the Young's modulus $E~(\sim 10^{11} J/m^3)$, whereas the demag. energy scales as $K_{d}~(\sim10^5 J/m^3)$. Here, $K_{d}$ and $E$ represent the magnetostatic energy constant and Young's modulus of the material, respectively, with both being material parameters. Bulk ferromagnetic solids (Iron, Nickel, Galfenol etc.) are known to deform when magnetised. This phenomenon is known as Joule magnetostriction. However, even under strong external magnetic fields, magnetostriction strains are small, typically in the order of $10^{-5}$. Our work does not consider Joule magnetostriction, we are primarily interested in understanding the deformation in slender ferromagnetic structures due to the effects of demag. energy and Zeeman energy. The elastic energy and magnetic energy are comparable in slender ferromagnetic structures, even though the elastic energy density is typically much greater than the magnetic energy density in bulk ferromagnetic materials. Hence, we can obtain large displacements in ferromagnetic slender structures while the strains in bulk ferromagnetic solids due to Joule magnetostriction are very small.

\begin{figure}[h!]
	\centering
	\includegraphics[width=0.6\linewidth]{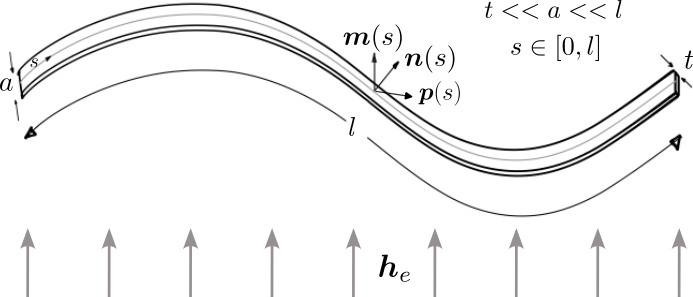}
	\caption{Geometry of the ferromagnetic ribbon illustrating the representation of magnetisation vector \(\vb*{m}(s)\), easy axis \(\vb*{p}(s)\), and normal vector \(\vb*{n}(s)\) = \(\vb*{d}_2(s)\).}
	\label{fig:planar_ribbon}
\end{figure}

In slender ferromagnetic structures, the interplay between elastic and magnetic energy becomes significant. The dissimilar scaling of these energies with respect to the aspect ratio allows their magnitudes to become comparable. The leading-order demag. energy of a slender ferromagnetic ribbon scales as $\mathcal{O}(K_{d}atl)$, and the elastic energy scales as $\mathcal{O}\left(\frac{EI}{2l}\right)$. Here, thickness, width, and length of the ferromagnetic ribbon are denoted as \(t\), \(a\), and \(l\), respectively, and $I=\frac{at^3}{12}$, is the area moment of inertia, see Fig \ref{fig:planar_ribbon}. Hence, ferromagnetic slender structures offer an opportunity to tune the aspect ratio, allowing the interaction between demag. energy and elastic energy. A balance of these energies implies that our analysis is valid for ribbons with an aspect ratio of $\mathcal{O}\left(\sqrt{\frac{E}{24K_d}}\right)$.

In this study, we consider a ferromagnetic planar ribbon composed of a ferromagnetic material such as Iron or Nickel. Our focus lies in investigating the influence of magnetization on ribbon deformation as the applied load varies quasi-statically. In particular, we shall study the two following cases:
\begin{itemize}
	\item {\it{Case 1. (Soft ferromagnet)}} : Magnetisation vector $\vb*{m(x)}$ is spatially constant and does not change with deformation. This case characterises soft ferromagnetic materials such as Permalloy, with an external magnetic field large enough to saturate the magnetisation uniformly throughout the  deformed planar ribbon.
	
	\item {\it{Case 2. (Hard ferromagnet)}} :  Magnetisation vector $\vb*{m(x)}$ makes a constant angle with respect to the tangent of the deformed planar curve. Here, the tangent vector, and hence the magnetisation vector vary spatially as the slender body undergoes deformation. This case characterises hard ferromagnetic materials such as Neodymium or Samarium-Cobalt.
\end{itemize}

To recapitulate, the total energy of a one-dimensional ferromagnetic ribbon is the summation of its elastic energy, exchange energy, anisotropy energy, demag. energy, Zeeman energy, and mechanical loading device energy. Calculating the elastic energy, anisotropy energy, Zeeman energy, and mechanical loading device energy are straightforward. However, determining the demag. and exchange energies for a curved ferromagnetic ribbon requires computations to be performed within the curvilinear local material frame. We obtain the leading order demag. energy for a curved ferromagnetic ribbon by employing a local material frame aligned with the tangent line of the deformed curve. Working in the deformed configuration alleviates the need to define a pull-back for the magnetisation vector $\vb*{m(x)}$ and solving the Maxwell's equations of magnetostatics in the reference configuration. Various pullbacks for magnetisation are defined in the literature, posing challenges in the rational selection of the appropriate one, \cite{James2002,Keip2019}. The leading order demag. energy is  local in nature and hence amenable for analysis. We also derive the exchange energy for our one dimensional ferromagnetic ribbon. Our exchange energy matches with findings reported in the field of curvilinear micromagnetics \cite{streubel2016magnetism,sheka2022fundamentals} to leading order. With the effective energy functional of the ribbon derived, we proceed to the next step of determining the energy-minimizing deformations as the external load is increased in a quasi-static manner.

We determine the deformed configuration by solving the equilibrium equations obtained by taking the first variation of the total energy. The equilibrium equations show that the magnetisation produces a body couple along the length of the curve that is dependent on the local orientation of the curve, in addition to the conventional terms originating from elastic energy, see Eqn. \ref{eqn:soft-magnetic-elastica-equilibrium-equations}. 
The equilibrium equations are solved numerically for various canonical boundary conditions, using \textsc{Auto}-07p, a standard software used in continuation and bifurcation problems \cite{Doedel1991a,Doedel1991b}. Further, we also perform a stability analysis of our computed solutions. The stability analysis involves casting the second variation of the total energy as a Sturm-Liouville boundary value problem. The eigenvalues of this boundary value problem, as the load is varied continuously, are used to determine the stability of the deformed configurations. 

We obtain the equilibrium path and the stability of the first few relevant modes, under quasi-static load control simulation. Our results provide us with a stable deformed state of the ferromagnetic ribbon, beginning from zero load and, in certain instances, a moderate tensile load, and extending to an arbitrarily large compressive load. The main findings from our analysis are as follows:

\noindent Case 1: Soft ferromagnetic ribbon:
\begin{description}[font=$\bullet$~\normalfont\scshape\color{red!50!black}]
	\item Critical buckling load is determined to be tensile under all canonical boundary conditions. Refer to Figs. \ref{fig:vertical-field-fixed-free-kdbar-100}, \ref{fig:vertical-field-fixed-fixed-kdbar-100} and \ref{fig:vertical-field-pinned-pinned-kdbar-100}.
	\item Fixed-fixed boundary conditions reveal the emergence of novel stable curves on the mode-2 branch. In Fig. \ref{fig:vertical-field-fixed-fixed-kdbar-100} (a), we highlight the segment of the mode-2 branch that corresponds to novel stable configurations, observed in a ferromagnetic ribbon but absent in Euler's elastica. A novel stable mode shape featuring two self-intersection points is shown in red in Fig. \ref{fig:vertical-field-fixed-fixed-kdbar-100} (a), and additional illustrations can be found in Fig. \ref{fig:vertical-field-fixed-fixed-kdbar-100} (b).
\end{description}

\noindent Case 2: Hard ferromagnetic ribbon:
\begin{description}[font=$\bullet$~\normalfont\scshape\color{red!50!black}]
	\item Introduction of magnetization alters the compressive buckling load compared to the elastica. See Figs. \ref{fig:hard-tangent-fixed-free-kdbarhe-100}, \ref{fig:hard-tangent-fixed-fixed-kdbarhe-100} and \ref{fig:hard-tangent-pinned-pinned-kdbarhe-100}.
	\item Comparing the deformed configuration of the hard ferromagnetic ribbon to Euler's elastica, no discernible difference in shape is observed, despite the presence of a change in the vertical reaction force at the supports. The expressions for the vertical reaction forces at the supports are provided in Equations \ref{eqn:reaction-transverse-he-hard-magnet-1} and \ref{eqn:reaction-transverse-he-hard-magnet-2}.\\
\end{description}

Before proceeding into the main body of the paper, we compare and contextualize our work with the existing literature. Several researchers have studied the deformation of ferromagnetic ribbons under an external magnetic field, primarily focusing on cantilever configurations, that is with fixed-free boundary conditions. Some of the foundational research related to deformation of planar ferromagnetic ribbons traces its origins back to the 1960s, where Moon et al. \cite{Moon1968,moon1984magneto} investigated the onset of buckling instabilities in a cantilevered beam-plate under magnetic loading \cite{Moon1968,moon1984magneto}. The distributed magnetic torque is dependent on $\theta$. Here, $\theta$ represents the angle formed by the local tangent with respect to the horizontal, see Figure \ref{fig:elastica-problem-setup}. Our model recovers this limit for $\theta \ll 1$. Furthermore, the scaling for critical magnetic load, $B_c \sim \left(\frac{l}{t}\right)^2$ matches with our prediction (See Section \ref{sec:model-comparison} for detailed comparison). The buckling of slender ferromagnetic structures has also garnered attention in the recent past. Singh et al. \cite{Singh2013} studied nonlinear elastic deformations in a vertically cantileverd Euler-Bernoulli beam with a permanent magnet attached to the free end. Their work explored supercritical and subcritical bifurcation behaviors under combined magnetic field and hard magnet influence. Gerbal et al. \cite{Gerbal2015} investigated magnetoelastic buckling transition in ferromagnetic cantilevered rods. Similar to \cite{Gerbal2015}, Wang et al. \cite{Wang2020} developed a mathematical model for hard magnetic cantilevered beam and determined critical load and the mode deformation under external magnetic fields varying in intensity and direction. We have recovered the equilibrium equation corresponding to the  hard magnetisation distribution of a magnetically loaded cantilevered beam which matches with that derived in \cite{Gerbal2015,Wang2020}. Singh et al. \cite{Singh2018} studied the occurrence of magnetically-induced instability in cantilever beams using small deflection assumptions. The main focus of these investigations has been to determine the critical external magnetic field at which the cantilever buckles, while either maintaining a fixed external load or having no load applied. In contrast, our work has centered on maintaining a constant external magnetic field while quasi-statically varying the applied load.

\subsection{Organisation of the paper}
The paper is organised as follows: In Section \ref{sec:one-dimensional-model-for-ferromagnetic-ribbon}, we present the mathematical model for the ferromagnetic planar ribbon. Beginning with the geometry and kinematics, we then derive the total energy of a ferromagnetic planar ribbons.  In Section \ref{sec:equilibrium-equations}, we  obtain the total energy functionals for soft and hard ferromagnetic planar ribbons, and the corresponding equilibrium equations for various canonical boundary conditions. In Section \ref{sec:continuation-method}, we describe the numerical continuation method employed for solving the equilibrium equations. Section \ref{sec:bifurcation-analysis}, focuses on the stability analysis of the solutions to the equilibrium equations. Section \ref{sec:results} details the numerical solutions for various canonical loading scenarios. We comprehensively analyse the deformation behaviour and bifurcation patterns for various canonical loading scenarios. The end of Section \ref{sec:results} is devoted to a discussion in the limit as $\bar{K_{d}}$ approaches infinity. Here, $\bar{K_{d}} = \frac{K_d atl^2}{EI}$ represents the ratio of the magnetostatic energy density to the elastic energy density of the ferromagnetic ribbon. We close our paper with the conclusions and remarks in Section \ref{sec:conclusion}.

\section{One-dimensional model for a ferromagnetic ribbon}\label{sec:one-dimensional-model-for-ferromagnetic-ribbon}

We begin with the kinematic description of ribbons.
Ribbons are characterized by three disparate length scales  -- length ($l$), width ($a$), thickness ($t$) -- such that $t \ll a \ll l$. Geometrically, ribbon is described in terms of a three-coordinate set -- centerline arc length ($s$), lateral ($\tilde{a}$), transverse ($\tilde{t}$) coordinates. 
In the reference configuration, the ribbon is represented in standard basis of $(\vb*{e}_1, \vb*{e}_2, \vb*{e}_3)$ as $\vb*{x}_0(s,\tilde{a},\tilde{t}) = s\vb*{e}_3 + \tilde{a}\vb*{e}_1 + \tilde{t}\vb*{e}_2$. We assume that the centreline of the ribbon lies in the $\vb*{e}_2-\vb*{e}_3$ plane without undergoing any twist in the deformed configuration. The centreline representation of the deformed planar ribbon is given by
\begin{align}
	\vb*{x}(s,\tilde{a},\tilde{t}) &= \vb*{r}(s) + \tilde{a} \vb*{d}_1(s) + \tilde{t} \vb*{d}_2(s).
\end{align}

Here, $\vb*{r}(s)$ denotes the position vector of a material point on the centreline, $(\vb*{d}_1(s), \vb*{d}_2(s), \vb*{d}_3(s))$ is the orthonormal material frame basis, and $s \in [0,l],~ \tilde{a} \in \left[-\frac{a}{2},\frac{a}{2}\right],~ \tilde{t} \in \left[-\frac{t}{2},\frac{t}{2}\right]$.   The ribbon is assumed to be inextensible, uniform and obeys Kirchhoff's hypothesis, that is, normal sections to the centreline remain normal after deformation \cite{Bigoni2015}. The condition of inextensibility implies $\dv{\vb*{r}}{s}(s) = \vb*{d}_3(s)$. 
Also, since the ribbon does not twist, we have $\vb*{d}_1(s) =\vb*{e}_1$. The material basis vectors evolves with arc length coordinate ($s$) as
\begin{equation}
\begin{split}
	\vb*{d}_1'(s) &= 0, \\
	\vb*{d}_2'(s) &= \kappa^{(1)} \vb*{d}_3(s) = \kappa \vb*{d}_3(s), \\
	\vb*{d}_3'(s) &= -\kappa^{(1)} \vb*{d}_2(s) = -\kappa \vb*{d}_2(s). 
\end{split}
\end{equation}
Here, $\kappa^{(1)}=\kappa(s)$ represents the material curvature associated with $\vb*{d}_1(s)$.

\begin{figure}[h!]
	\centering
	\includegraphics[width=0.8\linewidth]{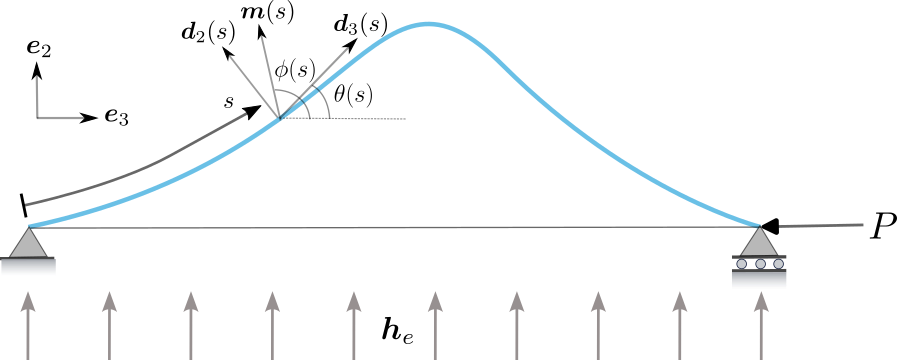}
	\caption{Schematic setup of the ferromagnetic ribbon.}
	\label{fig:elastica-problem-setup}
\end{figure}

The schematic of the  centerline in the deformed configuration is shown in  Fig. \ref{fig:elastica-problem-setup} and the coordinates of the centerline in $(\vb*{e}_1, \vb*{e}_2, \vb*{e}_3)$ basis is given as follows: 
\begin{equation}
		x(s) = 0, \qquad y(s) = \int_{0}^{s} \sin\theta(t)~ dt,\qquad
		z(s) = \int_{0}^{s} \cos\theta(t)~ dt,     \label{eqn:arclength-parametrisation}
\end{equation}
where $s$ denotes arc length coordinate, $\theta(s)$ is the angle between $\vb*{d}_3(s)$ and $\vb*{e}_3$ basis vector at $s$. The tangent vector field $\vb*{d}_3(s)$ (or $\vb*{t}(s)$) is $\vb*{d}_3(s) = (0,\sin\theta (s),\cos\theta (s))$ and so the normal vector field \cite{Pressley2010} is $\vb*{n}(s) = \vb*{d}_2(s) = \dv{\vb*{t}}{s}/\abs{\dv{\vb*{t}}{s}} = (0,\cos\theta (s),-\sin\theta (s))$. Note that the curvature $\kappa(s)=\theta'(s)$.

The magnetic variables involved in the description of the ferromagnetic ribbon are the magnetisation vector $\vb*{m}(s)$ and the externally applied magnetic field $\vb*{h}_e$. For both soft and hard ferromagnetic ribbons, we justify that the magnetization vector is solely determined by the centerline arc-length coordinate, denoted as $s$. The angle formed by the magnetization vector $\vb*{m}(s)$ and the reference direction $\vb*{e}_3$ is represented by $\phi(s)$, as depicted in Figure \ref{fig:elastica-problem-setup}. In this paper, we consider the external applied magnetic field in the transverse direction, represented by $\vb*{h}_e=(0,h_{e},0)$, where $h_{e}$ denotes the magnitude of the external magnetic field. We now proceed to formulate the mechanical and the magnetic energy of the ferromagnetic ribbon. 

\subsection{Mechanical energy}\label{sec:mechanical-energy}
The free energy functional of ferromagnetic ribbon is equal to the sum of its magnetic energy, elastic energy and the loading device energy. Given our assumption that the ribbon is made of an inextensible ferromagnetic material, the primary source of the elastic energy is the bending energy. We denote the sum of the bending energy and the loading device energy as $\mathcal{E}_{\text{elastica}}$ which is given as follows:
\begin{align}
	\mathcal{E}_{\text{elastica}} &=  \underbrace{\frac{1}{2} \int_{0}^{l} EI \kappa^2(s) ds}_{\text{Elastic (or bending) energy}} + ~ \underbrace{P \left(1 - \int_{0}^{l} \cos\theta(s) ds\right)}_{\text{Loading energy}}, \label{eqn:E-elastica}
\end{align}
subject to the integral constraint
\begin{equation}
	\begin{split}
		y(l) = \int_{0}^{l} \sin\theta(s)~ ds = 0.
	\end{split}
	\label{eqn:constraints}
\end{equation}
Here, $\kappa(s)=\dv{\theta (s)}{s}$ is the bending curvature, $E$ is the Young's modulus, $I = \frac{at^3}{12}$ is the area moment of inertia ($EI$ is the bending stiffness of the cross section), and $P$ is the horizontal load.

We use Kirchhoff rod theory to describe the deformation of our ferromagnetic ribbon. The Kirchhoff rod theory serves as an appropriate framework for describing the deformation of ribbons, when the dimensionless parameter $\bar{\kappa} \sim \frac{\kappa a^2}{t} \ll 1$, as discussed in \cite{Audoly2021}. For a ferromagnetic ribbon that is considered in our work, we can estimate $\bar{\kappa}$ as follows:
$\bar{\kappa} = \frac{\kappa a^2}{t} \ll \frac{\kappa_{c} a^2}{t} = \left(\frac{a}{t}\right)^2\sqrt{\frac{K_d}{E}}$. Here, $\kappa_{c}$ is the maximum curvature of the ferromagnetic ribbon and is estimated as $\kappa_{c}\sim \mathcal{O}\bigg( \frac{1}{t}\sqrt{\frac{K_{d}}{E}}\bigg)$, see section \ref{subsec:kd-infinity} for the derivation of $\kappa_{c}$. For a typical ferromagnetic material, $K_d \sim \mathcal{O}(10^5  \nicefrac{J}{m^3})$, and $E \sim \mathcal{O}(10^{11}  \nicefrac{J}{m^3})$ implying that $\frac{K_d}{E} \sim \mathcal{O}(10^{-6})$ and thus $\bar{\kappa}<<1$ for $\frac{a}{t} \sim \mathcal{O}(10^1)$. Therefore, the Kirchhoff rod model serves well for a typical ferromagnetic ribbon of aspect ratio 10 or smaller.


\subsection{Magnetic energy}\label{sec:magnetic-energy}

We now proceed to write down the magnetic energy of the ribbon. We employ the theoretical framework of micromagnetics to formulate the magnetic energy associated with a ferromagnetic ribbon. Recall that the magnetisation vector ($\vb*{m}(\vb*{x})$), is the primary variable in the micromagnetics functional.  The magnetisation vector $\vb*{m}(\vb*{x})$ is supported on $\Omega$ and $|\vb*{m}(\vb*{x})|=1, \vb*{x} \in \Omega$. In our analysis, $\Omega$ represents the deformed planar ferromagnetic ribbon. We now examine the various terms of the micromagnetic functional when applied to our ferromagnetic ribbon. For a detailed exposition on the micromagnetic functional, we refer the reader to \cite{dabade2019micromagnetics}.\\
The micromagnetic energy functional comprises of the following components
\begin{align}
   \mathcal{E}_{\text{magnetic}} &= \underbrace{A\int_{\Omega} \abs{\nabla{\vb*{m}(\vb*{x})}}^{2}d\vb*{x}}_{\mathcal{E}_{ex}} + \underbrace{K_a \int_{\Omega} \phi(\vb*{m}) d\vb*{x}}_{\mathcal{E}_{\text{anisotropy}}} + \underbrace{K_d\int_{\mathbb{R}^3} \abs{\vb*{h}_m}^2~d\vb*{x}}_{\mathcal{E}_{\text{demag.}}} - \underbrace{2K_d \int_{\Omega} \vb*{h}_e \cdot \vb*{m} d\vb*{x}}_{\mathcal{E}_{\text{Zeeman}}}, \label{eqn:micromagnetics-functional}
\end{align} 
where $\mathcal{E}_{ex}$ denotes the exchange energy, $\mathcal{E}_{\text{anisotropy}}$ the magnetocrystalline anisotropy energy, $\mathcal{E}_{\text{demag.}}$ is the demag. energy, and $\mathcal{E}_{\text{Zeeman}}$ is the Zeeman energy. We now briefly discuss each energy component in the magnetic energy functional.

\noindent The Zeeman energy in minimised when magnetisation vector $\vb*{m}$ is parallel and aligned along the external magnetic field $\vb*{h}_e$, and is expressed as follows for the ribbon: 
\begin{equation}
	\mathcal{E}_{\text{Zeeman}} = -2K_d at \int_{0}^{l} \vb*{h}_e \cdot \vb*{m}(s) ds. \label{eqn:zeeman-energy-functional}
\end{equation}
\vspace{2mm}

\noindent The magnetocrystalline anisotropy energy governs the favoured orientations for the magnetisation vector within the ferromagnetic sample, and assumes the following form:
\begin{equation}
	\mathcal{E}_{\text{anisotropy}} = K_a \int_{\Omega} \left(1 - (\vb*{m}(s)\cdot\vb*{p}(s))^2\right) d\tilde{t}d\tilde{a}ds = K_a at\int_{0}^{l} \left(1 - (\vb*{m}(s)\cdot\vb*{p}(s))^2\right)ds, \label{eqn:anisotropy-energy-functional}
\end{equation}
where the material parameter $K_{a}$ is known as the anisotropy constant, and $\vb*{p}(s)$ is the easy axis. We have the following two scenarios:
\begin{itemize}
 \item {\it{Case 1. (soft ferromagnet)}}: Here, $K_{a} \ll K_{d}$ and the magnetocrystalline anisotropy energy is very small, and the magnetisation has no preferred direction of orientation and aligns parallel with the external magnetic field $\vb*{h}_e$ for a moderate external field. Since, the external magnetic field $\vb*{h}_e$ is applied along the $\vb*{e}_2$ direction, we will assume that $\vb*{m}(\vb*{x})  = \vb*{m}(s) = \vb*{e}_2$ for a soft ferromagnetic ribbon.

 \item {\it{Case 2. (hard ferromagnet)}}: Here, $K_{a} \gg K_{d}$ and the magnetocrystalline anisotropy energy is very large and and the magnetisation vector makes a fixed angle with tangent along the length of the ribbon, i.e., $\vb*{m}(\vb*{x})  = \vb*{m}(s) = \vb*{p}(s) \in \{\vb*{t}(s), \vb*{n}(s)\}$. 
\end{itemize}
\vspace{4mm}

\noindent The exchange energy penalises the gradient of the magnetisation and is expressed as follows:
\begin{align}
	\mathcal{E}_{ex}=  A\int_{\Omega} \abs{\nabla{\vb*{m}(\vb*{x})}}^{2} d\tilde{t}d\tilde{a}ds = A\int_{s=0}^{l}\int_{\tilde{a}=-a/2}^{a/2}\int_{\tilde{t}=-t/2}^{t/2} \abs{\nabla{\vb*{m}(\vb*{x})}}^{2} d\tilde{t}d\tilde{a}ds, \label{eqn:exchange-energy-functional}
\end{align}
where $A$ is a material parameter known as the exchange constant. 
We consider the representation of magnetisation distribution in material frame basis, that is, $\vb*{m}(s)\rvert_{\{\vb*{d}_i\}} = \big(\vb*{m}(s)\cdot\vb*{d}_1(s), \vb*{m}(s)\cdot\vb*{d}_2(s), \vb*{m}(s)\cdot\vb*{d}_3(s)\big) = \left(m_{d_1}(s),m_{d_2}(s), m_{d_3}(s)\right)$. Incorporating this into Eqn. \ref{eqn:exchange-energy-functional}, the exchange energy expression reduces to 
\begin{align}
	\mathcal{E}_{ex}=&  Aa\int_{s=0}^{l}\int_{\tilde{t}=-t/2}^{t/2} \frac{1}{(1+\kappa \tilde{t})^2} \left[\left(\pdv{m_{d_1}}{s}\right)^2 + \left(\pdv{m_{d_2}}{s}-\kappa m_{d_3}\right)^2 + \left(\pdv{m_{d_3}}{s} + \kappa m_{d_2}\right)^2 \right] d\tilde{t}ds, \label{eqn:exchange-energy-functional-reduced} \\
	=& \,4 Aat\int_{s=0}^{l} \frac{1}{(4-\kappa^{2} t^2)} \left[\left(\pdv{m_{d_1}}{s}\right)^2 + \left(\pdv{m_{d_2}}{s}-\kappa m_{d_3}\right)^2 + \left(\pdv{m_{d_3}}{s} + \kappa m_{d_2}\right)^2 \right] ds. \label{eqn:exchange-energy-functional-reduced-1.5}
\end{align}

Given the non-negativity of the exchange energy, the integrand of Eqn. \ref{eqn:exchange-energy-functional-reduced-1.5} implies that $\kappa (s) < \nicefrac{2}{t}$. By retaining the leading-order terms in thickness, the exchange energy can be expressed as follows:
\begin{align}
	\mathcal{E}_{ex} &=  A a t\int_{0}^{l} \left[\left(\pdv{m_{d_1}}{s}\right)^2 + \left(\pdv{m_{d_2}}{s}-\kappa m_{d_3}\right)^2 + \left(\pdv{m_{d_3}}{s} + \kappa m_{d_2}\right)^2 \right] ds + \mathcal{O}(t^2) \notag \\
	&= A a t\int_{0}^{l} \biggr[\underbrace{\left(\pdv{m_{d_1}}{s}\right)^2 + \left(\pdv{m_{d_2}}{s}\right)^2 + \left(\pdv{m_{d_3}}{s}\right)^2}_{\text{isotropic}} + \underbrace{2\kappa \left(m_{d_2}\pdv{m_{d_3}}{s} - m_{d_3}\pdv{m_{d_2}}{s}\right)}_{\text{chiral}} + \underbrace{\kappa^2 \left(m_{d_2}^2 + m_{d_3}^2 \right)}_{\text{anisotropic}} \biggr] ds + \mathcal{O}(t^2). \label{eqn:exchange-energy-functional-reduced-1}
\end{align}
We recover the isotropic, chiral and anisotropic components of the exchange energy, which are extensively documented in the literature on curvilinear micromagnetics \cite{sheka2022fundamentals,streubel2016magnetism}. The detailed derivation of exchange energy is given in Appendix \ref{app:exchange}. 

\vspace{3mm}

\noindent We now write down the demag. energy of the planar ribbon. The demag. energy associated with planar ribbon is computed by solving Maxwell's equations of magnetostatics, namely,
\begin{equation}
	\begin{split}
		\curl{\vb*{h}_m(\vb*{x})} &= \vb*{0}, \\
		\div(\vb*{h}_m(\vb*{x}) + \vb*{m}(\vb*{x})) &= 0.
	\end{split} 
\end{equation}
Here, $\vb*{m}(\vb*{x})$ is the magnetisation vector defined on the ferromagnetic body and is equal to zero outside the body. The field induced due to the magnetisation $\vb*{m}(\vb*{x})$ is denoted by $\vb*{h}_m(\vb*{x})$. The induced magnetic field $\vb*{h}_m(\vb*{x})$ can be found by solving Maxwell's equations of magnetostatics.

The demag. energy is evaluated by computing the square of the $L^2$-norm of $\vb*{h}_m(\vb*{x})$ on all of $\mathbb{R}^3$, it can be conveniently calculated in terms of Fourier transform of $\vb*{m}(\vb*{x})$, as follows: 
\begin{align}
	\mathcal{E}_{\text{demag.}} = K_d\int_{\mathbb{R}^3} \abs{\vb*{h}_m}^2~d\vb*{x} &= K_d\int_{\mathbb{R}^3} \frac{\abs{\widehat{\div{\vb*{m}}}(\vb*{\xi})}^2}{\abs{\vb*{\xi}}^2} d\vb*{\xi},
\end{align}
where $K_d = \frac{M_s^2}{2\mu_0}$ is the magnetostatic energy constant, $M_s$ is the saturation magnetization of the material, and $\mu_0$ is the permeability of the free space. We carry out the above integration in the material frame $(\vb*{d}_1(s), \vb*{d}_2(s), \vb*{d}_3(s))$. In the material frame, $\vb*{x} \mapsto \vb*{r}(s) + \tilde{a} \vb*{d}_1(s) + \tilde{t}\vb*{d}_2(s)$, and $s \in [0,l],~ \tilde{a} \in \left[-\frac{a}{2},\frac{a}{2}\right],~\tilde{t} \in \left[-\frac{t}{2},\frac{t}{2}\right]$, and $\vb*{r}'(s) = \vb*{d}_3(s)$. The Fourier transform of $\vb*{m}$ in the material frame as follows:
\begin{equation}
	\begin{split}
	 \widehat{\div{\vb*{m}}} (\vb*{\xi})&= \int (\div{\vb*{m}}) \exp(-i2\pi\vb*{x}\cdot\vb*{\xi}) d\vb*{x} \\
	 &= \int \frac{1}{J} \left(\partial_{\tilde{a}} m_{d_1} + \partial_{\tilde{t}} m_{d_2} + \partial_s m_{d_3}        + \kappa [\tilde{t} \partial_{\tilde{a}} m_{d_1} + \tilde{t}\partial_{\tilde{t}} m_{d_2} + m_{d_2}] \right) \exp(-i2\pi(s\xi_{d_3} + \tilde{a} \xi_{d_1} + \tilde{t}\xi_{d_2}))J ds d\tilde{a} d\tilde{t}.	
	\end{split}
\end{equation} 
Here, $J (=1+\tilde{t}\kappa(s))$ is the Jacobian involved in computing the divergence of $\vb*{m}$ and in the change of variables from $(x_1,x_2,x_3)$ to $(s,\tilde{a},\tilde{t})$. Furthermore, $m_{d_i} = \vb*{m}\cdot\vb*{d}_i(s),~ \xi_{d_i} = \vb*{\xi}\cdot \vb*{d}_i(s)$. Since the ribbon under study is narrow, we shall assume $\vb*{m}(\vb*{x}) = \vb*{m} (s)$ and in the above computation, the leading order term ($\mathcal{O}(t)$) of magnetostatic or demag. energy simplifies as follows:
\begin{equation}
		\mathcal{E}_{\text{demag.}} = K_dat \int_{\mathbb{R}} \abs{\widehat{m_{d_3}}(\xi_{d_2})}^2 d\xi_{d_3} = K_dat \int_{0}^l (m_{d_2})^2 ds  = K_d at \int_{0}^{l} (\vb*{m}\cdot\vb*{d}_2)^2 ds. \label{eqn:demag-energy-functional}
\end{equation}

Details of the above calculation and other computations involved in deriving the leading order demag. energy can be found in Appendix \ref{app:magnetostatic}.


\subsection{Total energy of ferromagnetic elastic ribbon}
The total energy is the sum of $\mathcal{E}_{\text{magnetic}}$ (Eqns. \ref{eqn:exchange-energy-functional}, \ref{eqn:anisotropy-energy-functional}, \ref{eqn:demag-energy-functional}, and \ref{eqn:zeeman-energy-functional}) and $\mathcal{E}_{\text{elastica}}$ (Eqn. \ref{eqn:E-elastica}) and is given as follows:  
\begin{multline}
	\mathcal{E}(\theta,\vb*{m}) = \frac{EI}{2} \int_{0}^{l} \left(\dv{\theta(s)}{s}\right)^2 ds + P \left(l - \int_{0}^{l} \cos\theta(s) ds\right) + A\int_{s=0}^{l}\int_{\tilde{a}=-a/2}^{a/2}\int_{\tilde{t}=-t/2}^{t/2} \abs{\nabla{\vb*{m}(s)}}^{2} d\tilde{t}d\tilde{a}ds \\ + K_a at\int_{0}^{l} \left(1 - (\vb*{m}(s)\cdot\vb*{p}(s))^2\right)ds +  K_d at \int_{0}^{l} (\vb*{m}(s)\cdot\vb*{n}(s))^2 ds - 2K_d at \int_{0}^l \vb*{h}_e \cdot \vb*{m}(s) ds.
\end{multline}
Incorporating the integral constraint (Eqn. \ref{eqn:constraints}) with the help of a Lagrange multiplier $R$, the augmented energy functional for magnetoelastic ribbon, here, ferromagnetic ribbon, is expressed as 
\begin{multline}
	\mathcal{E}(\theta,\vb*{m}) = \frac{EI}{2} \int_{0}^{l} \left(\dv{\theta(s)}{s}\right)^2 ds  +  A\int_{s=0}^{l}\int_{\tilde{a}=-a/2}^{a/2}\int_{\tilde{t}=-t/2}^{t/2} \abs{\nabla{\vb*{m}(s)}}^{2} d\tilde{t}d\tilde{a}ds + K_a at\int_{0}^{l} \left(1 - (\vb*{m}(s)\cdot\vb*{p}(s))^2\right)ds  \\ +  K_d at \int_{0}^{l} (\vb*{m}(s)\cdot\vb*{n}(s))^2 ds - 2K_d at \int_{0}^{l} \vb*{h}_e \cdot \vb*{m}(s) ds + P \left(l - \int_{0}^{l} \cos\theta(s) ds\right) - R \int_{0}^{l} \sin\theta(s) ds. 
	\label{eqn:elastica-energy-functional-augmented-1}
\end{multline}
Non-dimensionalising the augmented energy functional gives:
\begin{multline}
	\bar{\mathcal{E}}(\theta,\vb*{m}) = \frac{\mathcal{E}(\theta,\vb*{m})}{EI/l} = \frac{1}{2} \int_{0}^{1} \left(\theta'(\bar{s})\right)^2 d\bar{s}  +  \bar{A}\int_{0}^{1}\int_{\bar{a} = -1/2}^{1/2} \int_{\bar{\tilde{t}} = -1/2}^{1/2} \abs{\grad{\vb*{m}}(\bar{s})}^{2} d\bar{t}d\bar{a}d\bar{s} + \bar{K}_a \int_{0}^{1} \left(1 - (\vb*{m}(\bar{s})\cdot\vb*{p}(\bar{s}))^2\right)d\bar{s} \\ + \bar{K}_d \int_{0}^{1} (\vb*{m}(\bar{s})\cdot\vb*{n}(\bar{s}))^2 d\bar{s}  - 2\bar{K}_d \int_{0}^{1} \vb*{h}_e \cdot \vb*{m}(\bar{s}) d\bar{s} + \bar{P} \left(1 - \int_{0}^{1} \cos\theta(\bar{s}) d\bar{s}\right) - \bar{R} \int_{0}^{1} \sin\theta(\bar{s}) d\bar{s}. \label{eqn:elastica-energy-functional-non-dimensionalized}
\end{multline}
where, $\bar{A}$, $\bar{K}_a$, $\bar{K}_d$, $\bar{P}$ and $\bar{Q}$ are non-dimensional parameters defined as follows
\begin{equation*}
	 \bar{A} = \frac{A at/l}{EI/l} = \frac{12A}{Et^2}, \quad \bar{K}_a = \frac{K_a atl^2}{EI}, \quad \bar{K}_d  = \frac{K_d a t l^2}{EI} = \frac{12K_d}{E}\left(\frac{l}{t}\right)^2, \quad \bar{P} = \frac{P l^2}{EI} \text{, and} \quad \bar{R} = \frac{R l^2}{EI},
\end{equation*}
while $\bar{a} = \frac{\tilde{a}}{a}$, and  $\bar{t} = \frac{\tilde{t}}{t}$. 
Note that $\bar{K}_d$ and $\bar{K}_a$ are the two important non-dimensional numbers in our analysis; $\bar{K}_d$ represents the ratio of the demag. energy and the elastic energy, and $\bar{K}_a$ denotes the ratio of the anisotropy energy and the elastic energy of the ferromagnetic ribbon. They depend on the material properties and the aspect ratio ($\frac{t}{l}$) of the ferromagnetic ribbon. 

\section{Equilibrium equations} \label{sec:equilibrium-equations}

We will now derive our energy functional for the two following scenarios: 1. soft ferromagnetic ribbon and 2. hard ferromagnetic ribbon.

%

\subsection{Case 1. Soft ferromagnetic ribbon: $\left(K_a at \ll K_d at \sim \frac{EI}{l}\right)$ or $(\bar{K}_a\rightarrow 0$ and $\bar{K}_d \sim \mathcal{O}(1)$)} 

\begin{figure}[h!]
	\centering
	\includegraphics[width=0.6\linewidth]{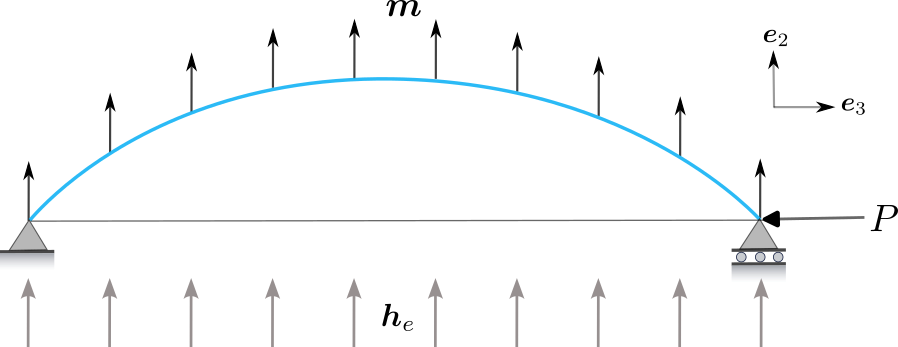}
	\caption{Soft ferromagnetic ribbon under the influence of a transverse external magnetic field.}
	\label{fig:elastica-soft-magnetic-vertical-field}
\end{figure}

A soft ferromagnet is a ferromagnet for which the anisotropy energy can be neglected, \cite{Dabade2018}. The magnetisation vector has no preferred direction of alignment within the ferromagnetic solid. Thus, the magnetisation vector rotates and aligns along the externally applied magnetic field for sufficiently large fields (Fig. \ref{fig:elastica-soft-magnetic-vertical-field}), that is, $\vb*{m}(\bar{s}) \parallel \vb*{h}_e$, and $\vb*{m}(\bar{s}) = \vb*{m}$, a constant vector. For the case when external magnetic field is oriented along $\vb*{e}_2$-axis, that is, $\vb*{h}_e = h_e\vb*{e}_2$, we have $\vb*{m}(\bar{s}) = \vb*{e}_2$. Here, $\it{h_e}$ denotes the strength of the externally applied magnetic field. 
Hence, the total energy of a soft ferromagnetic ribbon is given by
\begin{multline}
	\bar{\mathcal{E}}(\theta, \phi) = \frac{1}{2} \int_{0}^{1} \left(\theta'(\bar{s})\right)^2 d\bar{s}  +  \bar{A}\int_{0}^{1}\int_{\bar{a} = -1/2}^{1/2} \int_{\bar{t} = -1/2}^{1/2} \underbrace{\abs{\grad{\vb*{m}}}^{2}}_{=0}d\bar{t}d\bar{a}d\bar{s}  + \bar{K}_d \int_{0}^{1} \cos^2\theta(\bar{s}) d\bar{s}  \\ + \bar{P} \int_{0}^{1} \cos\theta(\bar{s}) d\bar{s} - \bar{R}\int_{0}^{1} \sin\theta(\bar{s}) d\bar{s} - \bar{P} - 2 \bar{K}_d h_e. \label{eqn:soft-magnetic-energy-functional}
\end{multline}
For a constant magnetisation $\vb*{m}$, the exchange energy is zero, and the Zeeman energy is a constant. Consequently, the deformation in this case is driven by the interplay among demag. energy, bending energy, and loading device energy. 

We derive the Euler-Lagrange equations, i.e., equilibrium equations, and generic boundary conditions, for a soft ferromagnetic ribbon as follows:
\begin{equation}
	\begin{aligned}
		\theta''(\bar{s}) + \bar{K}_d\sin 2\theta(\bar{s}) + \bar{P}\sin\theta(\bar{s}) + \bar{R}\cos\theta(\bar{s}) &= 0, \\
		\theta'(\bar{s}) \eta(\bar{s})\rvert_{\bar{s} = 0}  &= 0,  \\
		\theta'(\bar{s}) \eta(\bar{s})\rvert_{\bar{s} = 1}  &= 0.
	\end{aligned} \label{eqn:soft-magnetic-elastica-equilibrium-equations}
\end{equation}
where $\eta (\bar{s})$ is the admissible perturbation added to $\theta(s)$, which is subject to the constraint:
\begin{equation}
	\begin{split}
	\bar{y}(1) = \int_{0}^{1} \sin\theta(\bar{s})~ d\bar{s} = 0.
	\end{split}
	\label{eqn:constraints-non-dimensionalized}
\end{equation}

\noindent When $\bar{K}_d = 0$, we obtain the equilibrium equation for the classical Euler's elastica as follows:
\begin{equation}
	\begin{aligned}
		\theta''(\bar{s}) + \bar{P}\sin\theta(\bar{s}) + \bar{R}_e\cos\theta(\bar{s}) &= 0,
	\end{aligned} \label{eqn:pure-elastica-equilibrium-equations}
\end{equation}
where $\bar{R}_e$ denotes $\bar{R}$-value for the Euler's elastica. \\
We consider three canonical boundary conditions for our analysis, namely
\begin{itemize}
	\item fixed-free: $\theta(\bar{s} = 0) = 0, \theta'(\bar{s}=1) = 0$,
	\item fixed-fixed: $\theta(\bar{s} = 0) = 0, \theta(\bar{s}=1) = 0$,
	\item pinned-pinned: $\theta'(\bar{s} = 0) = 0, \theta'(\bar{s}=1) = 0$.
\end{itemize}
Note that the constraint (Eqn. \ref{eqn:constraints-non-dimensionalized}) does not apply for fixed-free condition.

\subsection{Case 2. Hard ferromagnetic ribbon: $\left(K_d at \ll K_a at \sim \frac{EI}{l} \sim K_d at h_{e} \right)$ or ($\bar{K}_d\rightarrow 0,~\bar{K}_dh_e \sim \mathcal{O}(1)$ \text{ and} $\bar{K}_a \sim \mathcal{O}(1)$)}

A hard ferromagnet is a ferromagnet with very large magnetic anisotropy energy ($K_{a} \gg 1$).  The magnetic anisotropy energy is minimised when the magnetization vector aligns along a preferred direction within the ferromagnetic solid. Thus, in a hard ferromagnetic ribbon, magnetization vector ($\vb*{m}(\bar{s})$) makes a constant angle with the tangent ($\vb*{d}_3(\bar{s})$) along the entire length of the curve. $\vb*{m}(\bar{s})$ varies spatially along the length of the curve to maintain a constant angle with the tangent of the curve as the ribbon deforms. In this case, the magnetization direction reorients as a result of induced deformation due to the presence of $\vb*{h}_e$. 

\begin{figure}[hbt!]
	\begin{subfigure}{.49\linewidth}
		\includegraphics[width=\linewidth]{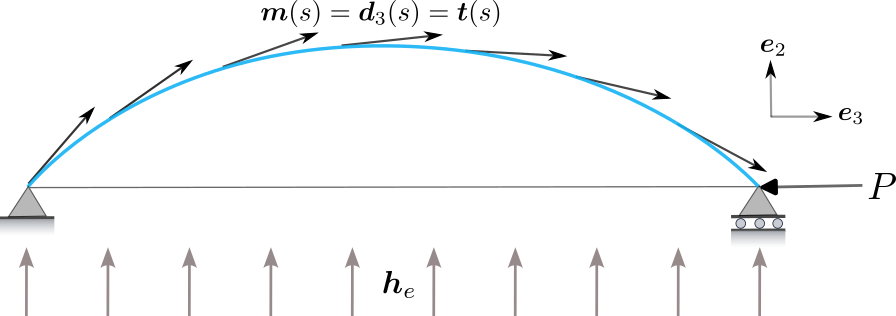}
		\caption{}
		\label{fig:elastica-hard-magnetic-tangential}
	\end{subfigure}\hfill 
	\begin{subfigure}{.49\linewidth}
		\includegraphics[width=\linewidth]{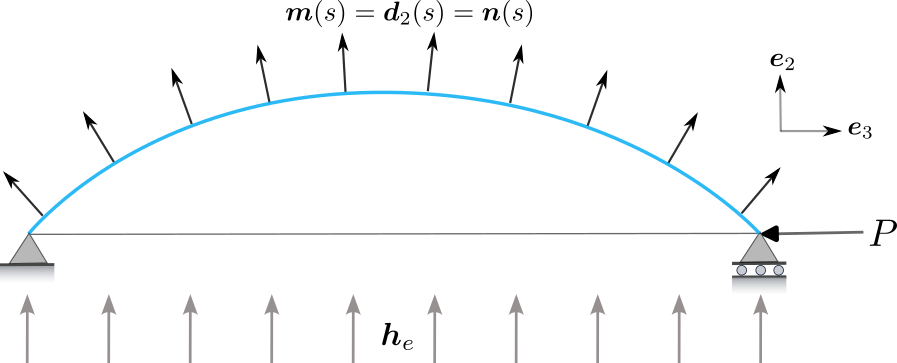}
		\caption{}
		\label{fig:elastica-hard-magnetic-normal}
	\end{subfigure}
	\caption{Magnetically hard ferromagnetic ribbon under the influence of an external magnetic field: (a) Magnetization along the tangential direction, (b) Magnetization along the normal direction.}
	\label{fig:elastica-hard-magnetic}
\end{figure}

\noindent As the magnetisation maintains a constant angle with respect to the tangent to the centerline, the components of $\vb*{m}(\bar{s})$ in material frame basis remain constant, that is, $\pdv{m_{d_i}}{\bar{s}} = 0,~i=1,2,3$. Incorporating these in Eqn. \ref{eqn:exchange-energy-functional-reduced-1}, we obtain the form of exchange energy as:
\begin{equation}
		\mathcal{E}_{ex}=  Aat\int_{0}^{l}\kappa^2 \left[\left(m_{d_2}\right)^2 + \left(m_{d_3}\right)^2 \right] ds. 
\end{equation}
We will consider that $\vb*{m}(\bar{s})$ is either aligned along the normal ($\vb*{m}(\bar{s}) \perp \vb*{d}_2(\bar{s})$) or that it is aligned along the tangent ($\vb*{m}(\bar{s}) \parallel\vb*{d}_3(\bar{s})$) of the curve ${\vb*{r}}(\bar{s})$, see Figs. \ref{fig:elastica-hard-magnetic-tangential} and \ref{fig:elastica-hard-magnetic-normal}.  
Accordingly, the components of $\vb*{m}(\bar{s})$ are $(m_{d_1}, m_{d_2}, m{d_3})$ = $(0,1,0)$ or $(0,0,1)$, for normal or tangential magnetisation distributions respectively. For either case, exchange energy expression reduces to  
\begin{equation}
	\mathcal{E}_{ex} = Aat\int_{0}^{l} \kappa^2 ds.
\end{equation}
Non-dimensionalising it with respect to $EI/l$, in the same manner as Eqn. \ref{eqn:elastica-energy-functional-non-dimensionalized}, results in
\begin{equation}
	\bar{\mathcal{E}}_{ex}(\theta) = \frac{\mathcal{E}_{ex}(\theta)}{EI/l} =  \bar{A} \int_{0}^{1} \left(\theta'(\bar{s})\right)^2 d\bar{s}. \label{eqn:hard-magnetic-exchange-energy-non-dimensionalised}
\end{equation}
Also, the magnetisation is aligned along the easy axis, such that $\vb*{m}(s) = \vb*{p}(s) \in \{\vb*{t}(s), \vb*{n}(s)\}$. 
Due to the unity nature of $\vb*{m}(\bar{s})$, that is, $\abs{\vb*{m}(s)}=1$, we adopt the following general expression for $\vb*{m}(\bar{s})$ in the Cartesian representation:  
\begin{equation}
	\vb*{m}(\bar{s}) = (0, \sin\phi(\bar{s}),\cos\phi(\bar{s})), \label{eqn:magnetisation-distribution-expression}
\end{equation}
where $\phi(\bar{s})$ is the angle between $\vb*{m}(\bar{s})$ and $\vb*{e}_3$ at $\bar{s}$. The transverse external magnetic field can be represented as 
$\vb*{h}_e =  h_e\vb*{e}_2$ or $\vb*{h}_e = -h_e\vb*{e}_2$ depending on whether the field is applied along positive or negative $\vb*{e}_2$-direction. 
 Upon substituting the expressions of $\vb*{m}(\bar{s})$ (Eqn. \ref{eqn:magnetisation-distribution-expression}) and the exchange energy (Eqn. \ref{eqn:hard-magnetic-exchange-energy-non-dimensionalised}) in Eqn. \ref{eqn:elastica-energy-functional-non-dimensionalized}, we obtain
\begin{multline}
	\bar{\mathcal{E}}(\theta, \psi) = \left(\frac{1}{2} + \bar{A}\right) \int_{0}^{l} \left(\theta'(\bar{s})\right)^2 d\bar{s} + \bar{K}_a \int_{0}^{1} \underbrace{\left(1 - (\vb*{m}(\bar{s})\cdot\vb*{p}(\bar{s}))^2\right)}_{= 0}d\bar{s} +  \bar{K}_d \int_{0}^{1} \underbrace{(\vb*{m}(\bar{s})\cdot\vb*{n}(\bar{s}))^2}_{\text{constant}} d\bar{s} \\  +  \bar{P} \int_{0}^{1} \cos\theta(\bar{s}) d\bar{s} \mp 2\bar{K}_d h_e \int_{0}^{1} \cos\phi(\bar{s}) d\bar{s} - \bar{Q}\int_{0}^{1} \sin\theta(\bar{s}) d\bar{s} - \bar{P} . \label{eqn:hard-magnetic-energy-functional}
\end{multline}
where $\bar{Q}$ is a Lagrange multiplier introduced to distinguish from $\bar{R}$ or $\bar{R}_e$. Recall $\bar{R}_e$ is the vertical reaction for the Euler's elastica (see Eqn. \ref{eqn:pure-elastica-equilibrium-equations}).

Note that, in this case $(\vb*{m}(\bar{s}) \cdot \vb*{n}(\bar{s}))^2$ is constant and hence the demag. energy does not participate in energy minimization. Large $K_a$ ensures $\vb*{m}(\bar{s})$ is parallel and aligned along $\vb*{p}(\bar{s})$. The contribution of exchange energy is negligible since $\bar{A} \ll 1$. Therefore, the deformation in this case is driven by the interplay among bending energy, loading device energy and the Zeeman energy. 

Following similar steps as in soft magnetization scenario, we arrive at the following Euler-Lagrange equations for hard magnetization case as 
\begin{equation}
	\begin{aligned}
		\theta''(\bar{s}) + \bar{P}\sin\theta(\bar{s}) \pm 2\bar{K}_d h_e\sin(\phi(\bar{s}) - \psi) + \bar{Q}\cos\theta(\bar{s}) &= 0, \\
		\theta'(\bar{s}) \eta(\bar{s})\rvert_{\bar{s} = 0}  &= 0, \\
		\theta'(\bar{s}) \eta(\bar{s})\rvert_{\bar{s} = 1}  &= 0.
	\end{aligned} \label{eqn:hard-magnetic-elastica-equilibrium-equations}
\end{equation}
We consider tangential and normal uniform magnetization distributions such that $\phi(\bar{s}) = \theta(\bar{s}), \theta(\bar{s})+\frac{\pi}{2}$ respectively, see Fig. \ref{fig:elastica-hard-magnetic}. Also, we expose the hard ferromagnetic ribbon to transverse ($\vb*{h}_e = h_e \vb*{e}_2,~ \vb*{h}_e = -h_e \vb*{e}_2$) external magnetic field. The corresponding equilibrium equations are enumerated as
\begin{equation}
	\begin{split}
		\vb*{h}_e = \pm h_e\vb*{e}_2: \vb*{m}(\bar{s}) = \vb*{t}(\bar{s}):  & \quad \theta''(\bar{s})  + \bar{P} \sin \theta(\bar{s}) + (\bar{Q} \pm 2\bar{K}_d h_e) \cos \theta (\bar{s}) = 0,   \\
		\vb*{m}(\bar{s}) = \vb*{n}(\bar{s}):  & \quad \theta''(\bar{s})  + (\bar{P}  \mp 2\bar{K}_d h_e) \sin \theta(\bar{s}) + \bar{Q} \cos \theta (\bar{s}) = 0. 
	\end{split}
	\label{eqn:hard-magnetic-elastica-cases}
\end{equation}

\subsection{Comparison of our model against Moon et al. \cite{Moon1968}}\label{sec:model-comparison}
\begin{figure}[h!]
	\centering
	\includegraphics[width=0.4\linewidth]{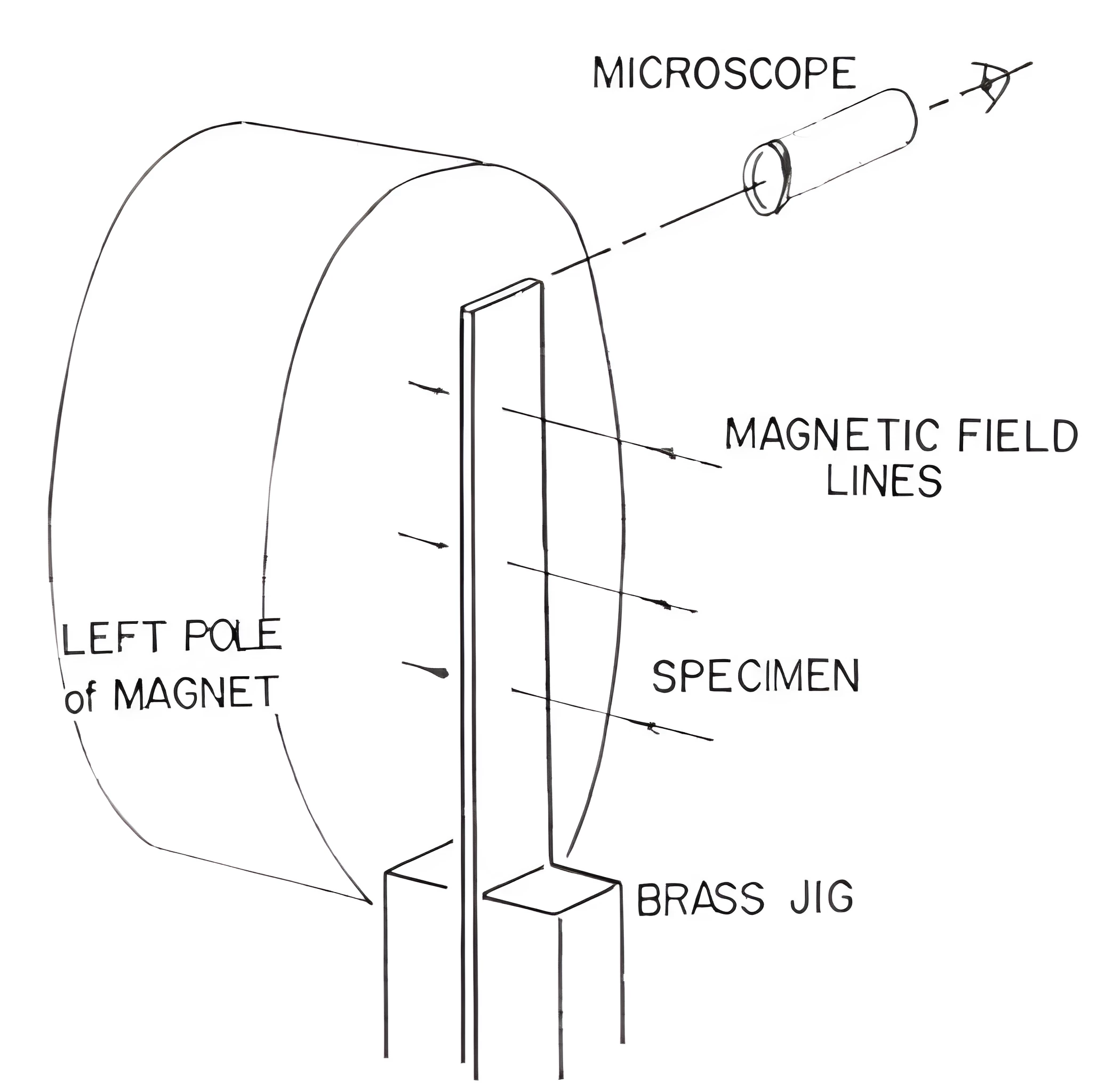}
	\caption{Cantilevered beam immersed in magnetic field; adapted from \cite{Moon1968}.}
	\label{fig:moon-and-pao-cantilevered-plate}
\end{figure}

The work done by Moon et al. \cite{Moon1968} comprised a cantilevered Euler beam with hard magnetization (see Fig. \ref{fig:moon-and-pao-cantilevered-plate}). 
We consider the equilibrium equation for hard magnetic ribbon with tangential magnetization (Eqn. \ref{eqn:hard-magnetic-elastica-cases}$_1$) subjected to pure magnetic loading under fixed-free boundary conditions, $\bar{Q} = \bar{P} = 0$, as
\begin{equation}
	\theta''(\bar{s})  + 2\bar{K}_d h_e\sin \theta(\bar{s}) = 0 \label{eqn:comparison-1}
\end{equation}
The above equation matches with the hard cantilevered Euler beam considered in Wang et al. \cite{Wang2020}. We linearise Eqn. \ref{eqn:comparison-1} about the undeformed configuration $\theta(\bar{s}) = 0$. From the non-dimensionalised arclength parametrisation (Eqn. \ref{eqn:arclength-parametrisation}), we have
\begin{equation}
	\dv{\bar{y}}{\bar{s}} = \sin\theta(\bar{s}), \qquad \dv{\bar{z}}{\bar{s}} = \cos\theta(\bar{s}). \label{eqn:comparison-2}
\end{equation}
 The small angle approximation would result in 
 \begin{equation}
 	 \cos\theta(\bar{z}) \approx 1,~\bar{z} \approx \bar{s},~\sin\theta(\bar{z}) \approx \theta(\bar{z}). \label{eqn:comparison-3}
 \end{equation}
Upon substituting Eqns. \ref{eqn:comparison-3} in Eqn. \ref{eqn:comparison-1}, and differentiating it with respect to $\bar{z}$, we have
\begin{equation}
	\dv[3]{\theta}{\bar{z}}  + 2\bar{K}_d h_e \dv{\theta}{\bar{z}} = 0 \label{eqn:comparison-4}
\end{equation}
From Eqns. \ref{eqn:comparison-2}, \ref{eqn:comparison-3} and \ref{eqn:comparison-4}, we obtain the linearisation of Eqn. \ref{eqn:comparison-1} as
\begin{equation}
	\dv[4]{\bar{y}}{\bar{z}}(\bar{z})  + 2\bar{K}_d h_e \dv[2]{\bar{y}}{\bar{z}}(\bar{z}) = 0
\end{equation}
which is identical to the equation presented in Moon et al. \cite{Moon1968}. Furthermore, the critical magnetic load varies as
\begin{equation}
   (K_dh_e)_c \sim  \left(\frac{t}{l}\right)^2.
\end{equation}
We can also recover all the critical loads as well as the deformation modeshapes presented in Table 1 of Moon et al. \cite{Moon1968}.

\section{Path continuation methodology} \label{sec:continuation-method}
We determine numerical solutions to the equilibrium equations as the loading parameter $\bar{P}$ is quasi-statically varied for various loading scenarios. We use a continuation method called pseudo arc length-based technique  as discussed in \cite{Peletan2014} to determine the deformed configurations. The technique constrains the incremental deformation measure and increment in load parameters ($P,R$) via a hyperspheric constraint. 
\subsection{Discretization of equilibrium equations}
We discretize Eqn. \ref{eqn:soft-magnetic-elastica-equilibrium-equations} for fixed-fixed boundary conditions. We discretize the domain into $N+1$ nodal points as $\bar{s}_i = ih; i=0,1,\dots,N+1$ where $h = \frac{1}{N+1}$. We utilize second-order accurate central difference scheme to approximate the second-order derivative in  Eqn. \ref{eqn:soft-magnetic-elastica-equilibrium-equations}. The discrete system of equations alongwith the boundary conditions is 
\begin{equation}
	\begin{aligned}
		\frac{\theta_{i-1} - 2\theta_i + \theta_{i+1}}{h^2} + \bar{P} \sin\theta_i + \bar{K}_d\sin 2\theta_i + \bar{R} \cos\theta_i &= 0; \qquad i=1,2,\dots, N, \\
		 \theta_0 = 0, \theta_{N+1} = 0, &
	\end{aligned}
    \label{eqn:elastica-governing-equations-discretized}
\end{equation}
where $\theta_i$ denotes the numerical counterpart of $\theta(\bar{s} = \bar{s}_i)$.
In matrix-vector form, the discretized system can be written for all three cases (for fixed-free case where $\bar{R}=0$) as 
\begin{align}
	\vb*{K\theta} + \bar{P}\sin\vb*{\theta} + \bar{K}_d \sin 2\vb*{\theta} + \bar{R} \cos\vb*{\theta} = \vb*{0} \text{ (see Appendix \ref{app:equilibrium} for $\vb*{K}$ and $\vb*{\theta}$}).
\end{align}
Incorporating the discretized form of integral constraint (Eqn. \ref{eqn:constraints-non-dimensionalized}) results in the combined nonlinear system of equations (for fixed-fixed and pinned-pinned conditions)
\begin{equation}
	\begin{split}
	  \begin{pmatrix}
		\vb*{K\theta} + \bar{P}\sin\vb*{\theta} + \bar{K}_d \sin 2\vb*{\theta} + \bar{R} \cos\vb*{\theta} \\
		\sin\theta_1 + \dots +  \sin\theta_N
	\end{pmatrix} &=
	\begin{pmatrix}
		\vb*{0} \\
		0
	\end{pmatrix},  \\
     \text{ or, } \qquad \qquad \vb*{f}(\vb*{\theta}, \bar{R}, \bar{P}) = \vb*{0}. &
    \end{split}
    \label{eqn:nonlinear-discrete-system-for-magnetoelastica}
\end{equation}

\subsection{Pseudo arc length-based continuation method}
We proceed to devise a strategy to solve the nonlinear system of equations (Eqn. \ref{eqn:nonlinear-discrete-system-for-magnetoelastica}) using an arc length-based continuation method. Let us describe $k$-th configuration of the planar-deforming ferromagnetic ribbon as $(\vb*{\theta}_{k}, \bar{R}_{k}, \bar{P}_{k})$ where $k$ denotes the configuration index. The algorithm works in two steps: prediction step and correction step.

\paragraph{Prediction step}: We expand the equilibrium equation (Eqn. \ref{eqn:nonlinear-discrete-system-for-magnetoelastica}) at $(k+1)$-th index in terms of Taylor series expansion about $k$ as
\begin{align}
	\underbrace{\vb*{f}(\vb*{\theta}_{k+1}, \bar{R}_{k+1}, \bar{P}_{k+1})}_{=\vb*{0}} = \underbrace{\vb*{f}(\vb*{\theta}_{k}, \bar{R}_{k}, \bar{P}_{k})}_{=\vb*{0}} + \pdv{\vb*{f}}{\vb*{\theta}}\bigg\rvert_k \Delta \vb*{\theta}_{k} + \pdv{\vb*{f}}{\bar{R}}\bigg\rvert_k \Delta \bar{R}_k + \pdv{\vb*{f}}{\bar{P}}\bigg\rvert_k \Delta \bar{P}_k + \underbrace{\mathcal{O}(\Delta \vb*{\theta}_{k}^2, \Delta \bar{R}_k^2, \Delta \bar{P}_k^2)}_{\text{ignore}}, \\
	\implies \pdv{\vb*{f}}{\vb*{\theta}}\bigg\rvert_k \Delta \vb*{\theta}_{k} + \pdv{\vb*{f}}{\bar{R}}\bigg\rvert_k \Delta \bar{R}_k + \pdv{\vb*{f}}{\bar{P}}\bigg\rvert_k \Delta \bar{P}_k &= \vb*{0}. \label{eqn:continuation-method-elastica-1}
\end{align}
We define a tangent vector as
\begin{align}
	\vb*{e}_k &= \begin{pmatrix}
		\Delta \vb*{\theta}_{k} \\ \Delta \bar{R}_k \\ \Delta \bar{P}_k
	\end{pmatrix}, \label{eqn:continuation-method-elastica-2}
\end{align}
whose norm squared is
\begin{align}
	\norm{\vb*{e}_k}^2 = (\Delta \vb*{\theta}_k)^T(\Delta \vb*{\theta}_k) + (\Delta\bar{R}_k)^2 + (\Delta \bar{P}_k)^2.
\end{align}
We make the following substitution 
\begin{equation}
	\Delta \bar{P}_k = a_k,~ \Delta \bar{R}_k = a_k \Delta r_k,~ \Delta \vb*{\theta}_k = a_k \Delta \vb*{\phi}_k. 
\end{equation}
Setting the norm of $\vb*{e}_k$ to unity, we obtain the hyperspheric constraint which yields an expression for $a_k$ 
\begin{align}
	a_k^2(\Delta \vb*{\phi}_k)^T(\Delta \vb*{\phi}_k) + a_k^2 (\Delta r_k)^2 + a_k^2 &= 1, \\
	\implies a_k &= \pm \frac{1}{\sqrt{(\Delta \vb*{\phi}_k)^T(\Delta \vb*{\phi}_k) + (\Delta r_k)^2 + 1}}.  \label{eqn:continuation-method-elastica-3}
\end{align}
The sign of $a_k$ is chosen positive if $\vb*{e}_{k-1}^T\vb*{e}_k > 0$ or else $a_k$ is taken to be negative. Substituting the expressions for $\Delta \vb*{\theta}_k, \Delta \bar{R}_k$ and $\Delta \bar{P}_k$ in Eqn. \ref{eqn:continuation-method-elastica-1}, we have
\begin{align}
	\pdv{\vb*{f}}{\vb*{\theta}}\bigg\rvert_k a_k \Delta \vb*{\phi}_k  + \pdv{\vb*{f}}{\bar{R}}\bigg\rvert_k a_k \Delta r_k + \pdv{\vb*{f}}{\bar{P}}\bigg\rvert_k a_k &= \vb*{0}, \\
	\implies \begin{bmatrix}
		\pdv{\vb*{f}}{\vb*{\theta}}\bigg\rvert_k & \pdv{\vb*{f}}{\bar{R}}\bigg\rvert_k 
	\end{bmatrix} \begin{pmatrix}
		\Delta \vb*{\phi}_k \\ \Delta r_k 
	\end{pmatrix} =
	- \pdv{\vb*{f}}{\bar{P}}\bigg\rvert_k.  \label{eqn:continuation-method-elastica-4}
\end{align}
We use Eqn. \ref{eqn:continuation-method-elastica-4} to solve for $\Delta \vb*{\phi}_k, \Delta r_k $ and then compute $a_k$ (Eqn. \ref{eqn:continuation-method-elastica-3}) and $\vb*{e}_k$ (Eqn. \ref{eqn:continuation-method-elastica-2}). We now construct the initial guess for $(k+1)$-configuration as
\begin{align}
	\begin{pmatrix}
		\vb*{\theta}_{k+1}^{(0)} \\ \bar{R}_{k+1}^{(0)} \\ \bar{P}_{k+1}^{(0)} 
	\end{pmatrix} &=
	\begin{pmatrix}
		\vb*{\theta}_{k}\\ \bar{R}_{k} \\ \bar{P}_{k} 
	\end{pmatrix}  + \Delta s \cdot \vb*{e}_k, \label{eqn:continuation-method-elastica-5}
\end{align}
where $\Delta s$ is the radius of the hypersphere, and also the {arc length step size} .  $\Delta s$ is small enough to capture the bifurcation along the equilibrium path.

\begin{figure}[h!]
	\centering
	\includegraphics[width=0.5\linewidth]{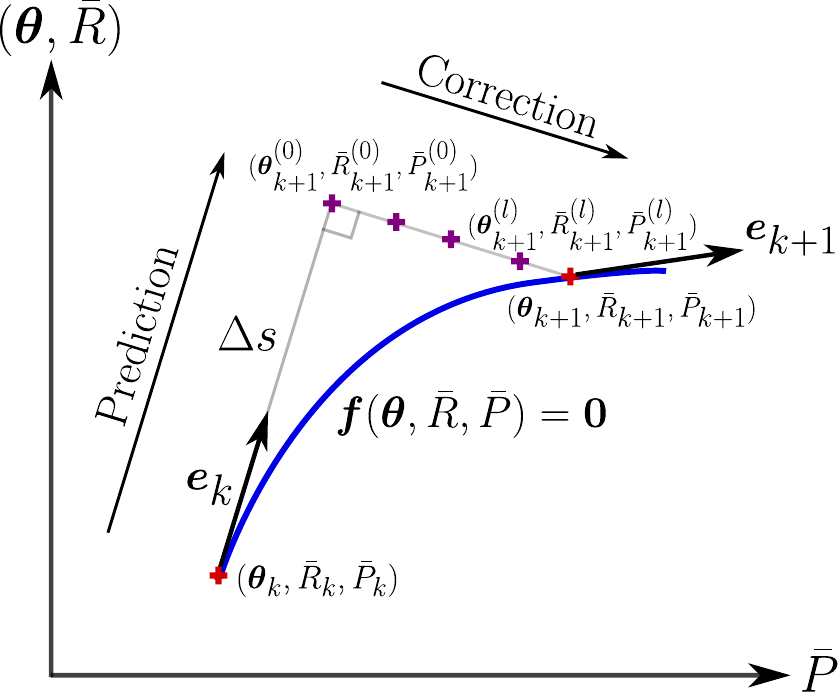}
	\caption{Schematic of pseudo arc length-based continuation procedure showing a curve in blue on which each point represents an equilibrium configuration.}
	\label{fig:continuation-method-schematic}
\end{figure}

\paragraph{Correction step}: We subject the predicted initial guess (Eqn. \ref{eqn:continuation-method-elastica-5}) of $(k+1)$-configuration to a sequence of corrective iterations using the widely-used Newton-Raphson technique. 
Denoting $(k+1)$-configuration by $(\vb*{\theta}_{k+1}, \bar{R}_{k+1}, \bar{P}_{k+1})$, we obtain the following Taylor series expansion about $l$-iteration level 
\begin{align}
	\underbrace{\vb*{f}(\vb*{\theta}_{k+1}, \bar{R}_{k+1}, \bar{P}_{k+1})}_{=\vb*{0}} \approx  \vb*{f}(\vb*{\theta}_{k+1}^{(l)}, \bar{R}_{k+1}^{(l)}, \bar{P}_{k+1}^{(l)}) + \pdv{\vb*{f}}{\vb*{\theta}}\bigg\rvert_{k+1}^{(l)} \Delta \vb*{\theta}_{k+1}^{(l+1)} + \pdv{\vb*{f}}{\bar{R}}\bigg\rvert_{k+1}^{(l)} \Delta \bar{R}_{k+1}^{(l+1)} + \pdv{\vb*{f}}{\bar{P}}\bigg\rvert_{k+1}^{(l)} \Delta \bar{P}_{k+1}^{(l+1)}, \\
	\implies  \pdv{\vb*{f}}{\vb*{\theta}}\bigg\rvert_{k+1}^{(l)} \Delta \vb*{\theta}_{k+1}^{(l+1)} + \pdv{\vb*{f}}{\bar{R}}\bigg\rvert_{k+1}^{(l)} \Delta \bar{R}_{k+1}^{(l+1)} + \pdv{\vb*{f}}{\bar{P}}\bigg\rvert_{k+1}^{(l)} \Delta \bar{P}_{k+1}^{(l+1)} = - \vb*{f}(\vb*{\theta}_{k+1}^{(l)}, \bar{R}_{k+1}^{(l)}, \bar{P}_{k+1}^{(l)}). \label{eqn:continuation-method-elastica-6}
\end{align}
We enforce the orthogonality constraint on tangent vector at every iteration level $l$, which is
\begin{align}
	\vb*{e}_k^T\vb*{e}_{k+1}^{(l+1)} &= 0, \\
	\implies  (\Delta \vb*{\theta}_k)^T(\Delta \vb*{\theta}_{k+1}^{(l+1)}) + (\Delta\bar{R}_k)(\Delta\bar{R}_{k+1}^{(l+1)}) + (\Delta \bar{P}_k)(\Delta\bar{P}_{k+1}^{(l+1)}) &= 0. \label{eqn:continuation-method-elastica-7}
\end{align}
Combining Eqns. \ref{eqn:continuation-method-elastica-6} and \ref{eqn:continuation-method-elastica-7} to construct the following iterative scheme 
\begin{align}
	\begin{bmatrix}
		\pdv{\vb*{f}}{\vb*{\theta}}\bigg\rvert_{k+1}^{(l)} & \pdv{\vb*{f}}{\bar{R}}\bigg\rvert_{k+1}^{(l)} & \pdv{\vb*{f}}{\bar{P}}\bigg\rvert_{k+1}^{(l)} \\ 
		(\Delta \vb*{\theta}_k)^T &  \Delta\bar{R}_k & \Delta\bar{P}_k 
	\end{bmatrix} \begin{pmatrix}
		\Delta \vb*{\theta}_{k+1}^{(l+1)} \\ \Delta\bar{R}_{k+1}^{(l+1)} \\ \Delta \bar{P}_{k+1}^{(l+1)}
	\end{pmatrix} =
	\begin{pmatrix}
		- \vb*{f}(\vb*{\theta}_{k+1}^{(l)}, \bar{R}_{k+1}^{(l)}, \bar{P}_{k+1}^{(l)}) \\
		0
	\end{pmatrix}. \label{eqn:continuation-method-elastica-8}
\end{align}
We solve the matrix equation (Eqn. \ref{eqn:continuation-method-elastica-8}) and then carry out the update procedure as
\begin{align}
	\begin{pmatrix}
		\vb*{\theta}_{k+1}^{(l+1)} \\ \bar{R}_{k+1}^{(l+1)} \\  \bar{P}_{k+1}^{(l+1)}
	\end{pmatrix} = \begin{pmatrix}
		\vb*{\theta}_{k+1}^{(l)} \\ \bar{R}_{k+1}^{(l)} \\ \bar{P}_{k+1}^{(l)}
	\end{pmatrix}  + 
	\begin{pmatrix}
		\Delta \vb*{\theta}_{k+1}^{(l+1)} \\ \Delta\bar{R}_{k+1}^{(l+1)} \\ \Delta \bar{P}_{k+1}^{(l+1)}
	\end{pmatrix}.
\end{align} 
The correction steps ($l=1,2,\dots$) are performed till the increments meet a prescribed tolerance. We then increment the configuration index ($k \leftarrow k+1$) and repeat the prediction and correction steps until $\bar{P}_{\text{max}}$ is reached. The adopted continuation procedure is illustrated in Fig. \ref{fig:continuation-method-schematic}.

\section{Bifurcation analysis of equilibrium configuration} \label{sec:bifurcation-analysis}
We now determine whether a given deformed equilibrium configuration is stable with respect to infinitesimal perturbations. We expect stable perturbations to be physically realizable. This requires evaluation of the second variational derivative of the energy functional at the critical points. The critical points are the solutions to the Euler-Lagrange equations. \\

We proceed to determine the second variational derivative of the functional, $\delta^2\mathcal{\overline{E}} (\theta)$ for the soft ferromagnetic ribbon when $\vb*{h}_e$ is applied along $\vb*{e}_2$-axis such that $\vb*{m}(\bar{s}) = \vb*{e}_2$. Introducing first order perturbation in the assumed extremum $\theta$ as $\hat{\theta}(\bar{s}) = \theta (\bar{s}) + \epsilon \eta (\bar{s})$ where $\eta (\bar{s})$ is a kinematically admissible planar variation and $\epsilon$ is a small parameter, and substituting it into Eqn. \ref{eqn:elastica-energy-functional-non-dimensionalized}
\begin{multline}
	\mathcal{\overline{E}}  (\theta + \epsilon \eta) =  \frac{1}{2} \int_{0}^{1} (\theta' + \epsilon \eta')^2 d\bar{s} + \bar{K}_d \int_{0}^{1} \cos^2 (\theta + \epsilon \eta) at d\bar{s} + \bar{P} \int_{0}^{1} \cos (\theta + \epsilon \eta) d\bar{s} \\ - \bar{R} \int_{0}^{1} \sin (\theta + \epsilon \eta ) d\bar{s} + \text{constant}.
\end{multline}
Simplifying the above expansion and taking into account the boundary conditions, we have
\begin{align}
 \delta^2\mathcal{\overline{E}} (\theta) = \dv[2]{\mathcal{\overline{E}}  (\theta + \epsilon \eta)}{\epsilon}\bigg\rvert_{\epsilon=0} &= - \int_{0}^{1} [\eta'' + 2\bar{K}_d\cos 2\theta \eta + \bar{P} \cos \theta \eta - \bar{R}\sin \theta \eta] \eta d\bar{s}, \label{eqn:second-variational-derivative}
\end{align}
for all kinematically admissible functions $\eta(\bar{s})$. The stability criterion states that
\begin{align}
	\delta^2\mathcal{\overline{E}}  (\theta) \begin{cases}
		> 0 & \text{ stable}, \\
		< 0 & \text{ unstable}.
	\end{cases}
\end{align}
Further, we also introduce the first variation of the integral constraint $\int_{0}^{1} \sin\theta(\bar{s})d\bar{s} = 0$ given as:
\begin{align}
	\int_{0}^{1} \cos \theta(\bar{s}) \eta(\bar{s}) d\bar{s} = 0. 
	\label{eqn:constraints-first-variation}
\end{align}

\subsection{Construction of Sturm-Liouville problem}
Following \cite{Levyakov2009,Bigoni2015}, we construct an equivalent Sturm-Liouville problem for the second variation $\delta^2\mathcal{\overline{E}} (\theta)$ as follows:
\begin{align}
	\phi_n''(\bar{s}) + \lambda_n L(\bar{s}) \phi_n (\bar{s}) = C_{R_n} \cos \theta (\bar{s}), \label{eqn:sturm-liouville-problem}
\end{align}
where $\lambda_n$ are the eigenvalues, $\phi_n$ the corresponding eigenmodes of Eqn. \ref{eqn:sturm-liouville-problem} and $L(\bar{s}) = (2\bar{K}_d\cos 2\theta(\bar{s}) + \bar{P} \cos\theta(\bar{s}) - \bar{R} \sin\theta(\bar{s}))$ denotes the weight function. $C_{R_n}$ is a Lagrange parameter introduced to enforce the isoperimetric constraint (Eqn. \ref{eqn:constraints-first-variation}). The conditions on $\phi_n(\bar{s})$ are
\begin{itemize}
	\item fixed-fixed case: $\phi_n(0) = \phi_n(1) = 0$ and $\int_{0}^{1} \cos \theta(\bar{s}) \phi_n(\bar{s}) d\bar{s} = 0$,
	\item pinned-pinned case: $\phi_n'(0) = \phi_n'(1) = 0$ and $\int_{0}^{1} \cos \theta(\bar{s}) \phi_n(\bar{s}) d\bar{s} = 0$,  
	\item fixed-free case:  $\phi_n(0) = \phi_n'(1) = 0$.
\end{itemize}
We obtain the following conditions for two arbitrary eigenmodes $\phi_m(\bar{s})$ and $\phi_n(\bar{s})$ of the Sturm-Liouville problem (Eqn. \ref{eqn:sturm-liouville-problem})
\begin{align}
  \lambda_n  \int_{0}^{1} L(\bar{s}) \phi_n^2(\bar{s})d\bar{s} &= \int_{0}^{1} \phi_n'^2d\bar{s}, \label{eqn:sturm-liouville-condition-1}
\end{align}
and the orthogonality condition
\begin{align}
	\int_{0}^{1} L(\bar{s})\phi_n(\bar{s})\phi_m(\bar{s})d\bar{s} &= 0. \label{eqn:sturm-liouville-condition-2}
\end{align}
The details of the above calculation can be found in Appendix \ref{app:bifurcation}.
\paragraph{Spectral decomposition}: Let us use $\phi_n(\bar{s})$ alongwith the weight function $L(\bar{s})$ to construct a Fourier series representation (converging in the mean) to the square-integrable function $\eta(\bar{s})$,
\begin{align}
	\eta (\bar{s}) &= \sum_{n=1}^\infty c_n \phi_n (\bar{s}), \qquad c_n \text{ are Fourier coefficients}.
\end{align}
Substituting the above representation in Eqn. \ref{eqn:second-variational-derivative} and invoking the conditions (Eqns. \ref{eqn:sturm-liouville-condition-1} and \ref{eqn:sturm-liouville-condition-2}), we obtain the stability criterion:
\begin{align}
  \delta^2\mathcal{\overline{E}}  (\theta) = \sum_{n=1}^{\infty} c_n^2 \bigg(1 - \frac{1}{\lambda_n}\bigg)\int_{0}^{1} (\phi_n'(\bar{s}))^2 d\bar{s} \quad  \begin{cases}
		> 0 \text{ if } \lambda_n \notin [0,1]  & \text{ stable}, \\
		< 0 \text{ if } \lambda_n \in [0,1] & \text{ unstable}.
	\end{cases}
\end{align}
The detailed bifurcation analysis technique is presented in Appendix \ref{app:bifurcation}.

\subsubsection{Numerical bifurcation analysis}
We describe the numerical procedure for the fixed-fixed case, which can be easily adapted to pinned-pinned and fixed-free cases. Given, Eqn. \ref{eqn:sturm-liouville-problem} as
\begin{align}
	\phi_m''(\bar{s}) + \lambda_m L_m (\bar{s}) \phi_m (\bar{s}) &= C_{R_m} N(\bar{s}), \label{eqn:sturm-liouville-problem-1}
\end{align}
where, the coefficient functions are
\begin{equation}
	L_m(\bar{s}) = 2\bar{K}_d \cos 2\theta(\bar{s}) + \bar{P} \cos \theta (\bar{s}) - \bar{R} \sin \theta(\bar{s}), \qquad N(\bar{s}) = \cos \theta(\bar{s}).
\end{equation}
Eqn. \ref{eqn:sturm-liouville-problem-1} is also subjected to the boundary conditions $\phi_m (0) = 0, \phi_m (1) = 0$ and the additional constraint
\begin{align}
	\int_{0}^{1} \phi_m(\bar{s}) N(\bar{s}) d\bar{s} &= 0. \label{eqn:sturm-liouville-problem-constraint}
\end{align}
\paragraph{Numerical procedure to compute eigenvalues $\lambda_m$}: Partition the interval $\bar{s} \in [0,1]$ segments of equal length $h = \frac{1}{n+1}$ such that the starting points of the segments can be denoted by $\bar{s}_{i-1} = (i-1)h; i=1,2,\dots,n+1$. For $i$-th segment, the functions $L(\bar{s})$ and $N(\bar{s})$ are approximated by averaging their corresponding values at nodal indices $i$ and $i+1$ resulting in $L_i$ and $N_i$ respectively. Substituting these averaged quantities, Eqn. \ref{eqn:sturm-liouville-problem-1} becomes
\begin{align}
	\phi_m''(\bar{s}) + \lambda_m L_i \phi_m (\bar{s}) = C_{R_m} N_i, \label{eqn:sturm-liouville-problem-2}
\end{align}
which is an ordinary differential equation (ODE) with constant coefficients. The solution to this ODE is 
\begin{align}
	\phi_m (\bar{s}) &= A_{1i} F_{1i} (\bar{s} - \bar{s}_{i-1}) + A_{2i} F_{2i} (\bar{s} - \bar{s}_{i-1}) + C_{R_m} \frac{N_i}{\lambda_m L_i}, 
\end{align}
where $A_{1i}$ and $A_{2i}$ are constants, and the functions $F_{1i}$ and $F_{2i}$ are defined as 
\begin{enumerate}
	\item for $\lambda_m L_i > 0$ as
	$F_{1i}(\bar{s} - \bar{s}_{i-1}) = \cos a_i (\bar{s} - \bar{s}_{i-1}), \quad F_{2i}(\bar{s}-\bar{s}_{i-1}) = \sin a_i(\bar{s} - \bar{s}_{i-1})$,
	
	\item for $\lambda_m L_i < 0$ as
	$F_{1i}(\bar{s} - \bar{s}_{i-1}) = \cosh a_i (\bar{s} - \bar{s}_{i-1}), \quad F_{2i}(\bar{s} - \bar{s}_{i-1}) = \sin a_i(\bar{s} - \bar{s}_{i-1})$
	with $a_i = \sqrt{\abs{\lambda L_i}}$.
\end{enumerate}
The constants $A_{1i}$ and $A_{2i}$ can be obtained from the matching conditions 
\begin{equation}
	\phi_m(\bar{s}_{i-1}) = \phi_{m,i-1}, \qquad \phi_m'(\bar{s}_{i-1}) = \phi_{m,i-1}'
\end{equation}
as
\begin{equation}
	A_{1i} = \phi_{m,i-1} - C_{R_m} \frac{N_i}{\lambda_m L_i}, \qquad A_{2i} = \frac{\phi_{m,i-1}'}{a_i}.
\end{equation}
The quantity $\phi_m(\bar{s}_i) = \phi_{m,i}$ at the right end of the segment is computed alongwith its derivative as
\begin{equation}
	\begin{split}
		 \phi_{m,i} &= \phi_{m,i-1}F_{1i}(h) + \frac{\phi_{m,i-1}'}{a_i} F_{2i}(h) + C_{R_m} N_i \frac{[1-F_{1i}(h)]}{\lambda_m L_i}, \\
		 \phi'_{m,i} &= \phi_{m,i-1}F_{1i}'(h) + \frac{\phi_{m,i-1}'}{a_i} F_{2i}'(h) - C_{R_m} N_i \frac{F_{1i}'(h)}{\lambda_m L_i}.
	\end{split}
   \label{eqn:sturm-liouville-problem-recurrence-relation}
\end{equation}
The general solution of Eqn. \ref{eqn:sturm-liouville-problem-2} can be constructed by using Eqns. \ref{eqn:sturm-liouville-problem-recurrence-relation}, and the continuity requirement of $\phi_m$ and $\phi_m'$ at the extremities of every integration segment. Since Eqn. \ref{eqn:sturm-liouville-problem-1} is linear, its general solution can be written as a combination of three particular solutions
\begin{align}
	\phi_m(\bar{s}) &= c_1 \psi_1(\bar{s}) + c_2 \psi_2(\bar{s}) + C_{R_m}\psi_3(\bar{s}), \label{eqn:sturm-liouville-problem-general-solution}
\end{align}
where $c_1$ and $c_2$ are constants.
We use the following initial data
\begin{equation}
	\begin{split}
		&\psi_1(0) = 1, \quad \psi_1'(0) = 0, \quad C_{R_m} = 0, \\
		&\psi_2(0) = 0, \quad \psi_2'(0) = 1, \quad C_{R_m} = 0, \\
		&\psi_3(0) = 0, \quad \psi_3'(0) = 0, \quad C_{R_m} = 1.
	\end{split}
\end{equation}
The functions $\psi_i(\bar{s}), i=1,2,3$ can be constructed separately using the recurrence relations Eqn. \ref{eqn:sturm-liouville-problem-recurrence-relation}. 
Using boundary conditions, we have
\begin{equation}
	\begin{split}
	  	\phi_m(0) = 0 &\implies c_1 = 0, \\
	  \phi_m(1) = 0  &\implies c_2 \psi_2(1) + C_{R_m}\psi_3(1) = 0.
	\end{split}
\end{equation}
Substituting Eqn. \ref{eqn:sturm-liouville-problem-general-solution} in Eqn. \ref{eqn:sturm-liouville-problem-constraint}, the constraint can be rewritten as
\begin{align}
    c_2\int_{0}^{1} \psi_2(\bar{s})N(\bar{s}) d\bar{s}  + C_{R_m}\int_{0}^{1} \psi_3(\bar{s})N(\bar{s}) d\bar{s}  &= 0.
\end{align}
We construct the following homogeneous system of equations
\begin{align}
	\begin{bmatrix}
		\psi_2(1) & \psi_3(1)  \\
		\int_{0}^{1} \psi_2(\bar{s})N(\bar{s}) d\bar{s} & \int_{0}^{1} \psi_3(\bar{s})N(\bar{s}) d\bar{s}
	\end{bmatrix}\begin{pmatrix}
		c_2 \\ C_{R_m}
	\end{pmatrix} = 
	\begin{pmatrix}
		0 \\ 0
	\end{pmatrix}.
\end{align}
For non-trivial solution to the above equation, the determinant,
\begin{align}
	\Delta =& \psi_2 (1) \int_{0}^{1} \psi_3(\bar{s}) N(\bar{s}) d\bar{s} - \psi_3(1)\int_{0}^{1} \psi_2(\bar{s})N(\bar{s})d\bar{s}
\end{align}
must be zero. The integrals are evaluated numerically. \\
We vary the value of $\lambda_m$ from 0 to 1 and monitor the changes in $\Delta$.
\begin{itemize}
	\item If at least one value of $\lambda_m \in [0,1]$ exists resulting in $\Delta = 0$, then the non-trivial solution of Eqn. \ref{eqn:sturm-liouville-problem-general-solution} exists which also satisfies the boundary conditions. The corresponding equilibrium configuration is unstable.
	\item It no $\lambda_m \in [0,1]$ results in $\Delta = 0$, it implies that there exists a trivial solution of Eqn. \ref{eqn:sturm-liouville-problem-general-solution}. The corresponding equilibrium configuration is therefore stable.
\end{itemize}

\section{Results and Discussion}\label{sec:results}

\begin{figure}[h!]
	\centering
	\includegraphics[width=0.35\linewidth]{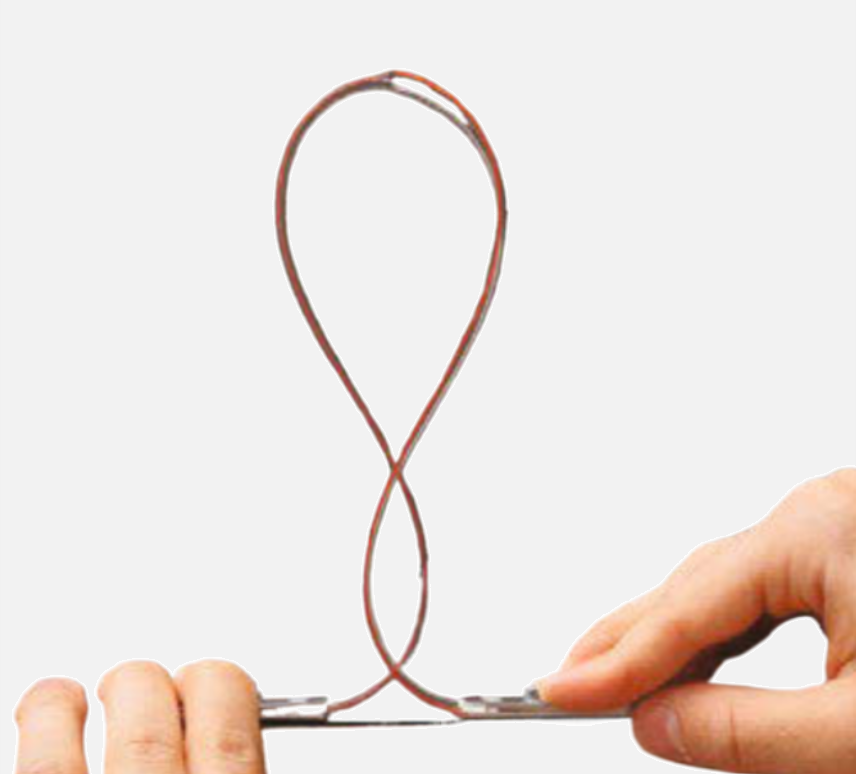}
	\caption{Experiment conducted by Bigoni et al. \cite{Bigoni2015} showing a self-intersecting Mode-1 deformation of an elastic ribbon subjected to fixed-fixed boundary conditions. The dimensions of the ribbon in the experiment are: $t = 1.5$ mm, $a =25$ mm and  $l =490$ mm.}
	\label{fig:self-intersection-bigoni}
\end{figure}

In this section, we begin with the validation of our numerical framework by comparing our results with the well-studied Euler's elastica from existing literature \cite{Bigoni2015}. We consider three canonical boundary conditions, namely, fixed-fixed, pinned-pinned, and fixed-free. Our results are presented as load-displacement curves, also known as equilibrium curves, which have been determined using the continuation algorithm described in Section \ref{sec:continuation-method}.  We use the well-known mathematical analysis tool, \textsc{Auto}-07p, to validate numerical results. We then explore solutions for both soft and hard ferromagnetic ribbons under different boundary conditions. We have summarised all the cases considered in Table \ref{tab:cases}. The width of the ferromagnetic ribbon is assumed to be very small, allowing it to deform into self-intersecting loops while remaining planar.  Self-intersecting loops have been observed in experiments involving elastic ribbons with meticulous longitudinal cuts, as depicted in Fig. \ref{fig:self-intersection-bigoni}.


\begin{table}[h!]
	\centering
	\resizebox{0.8\textwidth}{!}{%
		\begin{tabular}{@{}l|c|c|cc@{}}
			\toprule
			Sections & \ref{subsec:purely-elastic} & \ref{subsec:soft-magnetic} & \multicolumn{2}{c}{\ref{subsec:hard-magnetic}} \\ \midrule
			\multirow{3}{*}{Description} & \multirow{3}{*}{Euler's elastica} & Case 1 Soft ferromagnet & \multicolumn{2}{c}{Case 2: Hard ferromagnet} \\ \cmidrule(l){3-5} 
			&  & \multirow{2}{*}{Transverse $\vb*{h}_e$} & \multicolumn{2}{c}{Transverse $\vb*{h}_e$}  \\ \cmidrule(l){4-5} 
			&  &  & \multicolumn{1}{c|}{$\vb*{m} = \vb*{t}$} & \multicolumn{1}{c}{$\vb*{m} = \vb*{n}$}  \\ \midrule
			\multirow{3}{*}{Figures} & \multicolumn{1}{l|}{Fixed-Free (\ref{fig:purely-elastic-fixed-free})} & (\ref{fig:vertical-field-fixed-free-kdbar-100}) &  \multicolumn{1}{l|}{(\ref{fig:purely-elastic-fixed-free}) but different $\bar{R}$} &  \multicolumn{1}{c}{(\ref{fig:hard-tangent-fixed-free-kdbarhe-100})}  \\
			& \multicolumn{1}{l|}{Fixed-Fixed (\ref{fig:purely-elastic-fixed-fixed})} & (\ref{fig:vertical-field-fixed-fixed-kdbar-100}) & \multicolumn{1}{l|}{(\ref{fig:purely-elastic-fixed-fixed})  $\qquad$ ,, } &  \multicolumn{1}{c}{(\ref{fig:hard-tangent-fixed-fixed-kdbarhe-100})}   \\
			& \multicolumn{1}{l|}{Pinned-Pinned (\ref{fig:purely-elastic-pinned-pinned}) } & (\ref{fig:vertical-field-pinned-pinned-kdbar-100}) & \multicolumn{1}{l|}{(\ref{fig:purely-elastic-pinned-pinned})  $\qquad$ ,, } & \multicolumn{1}{c}{(\ref{fig:hard-tangent-pinned-pinned-kdbarhe-100})}   \\ \bottomrule
		\end{tabular}%
	}
  \caption{Various considered cases in the study.}
   \label{tab:cases}
\end{table}

\subsection{Euler's elastica} \label{subsec:purely-elastic}
\begin{figure}[h!]
	\begin{subfigure}{.49\linewidth}
		\includegraphics[width=\linewidth]{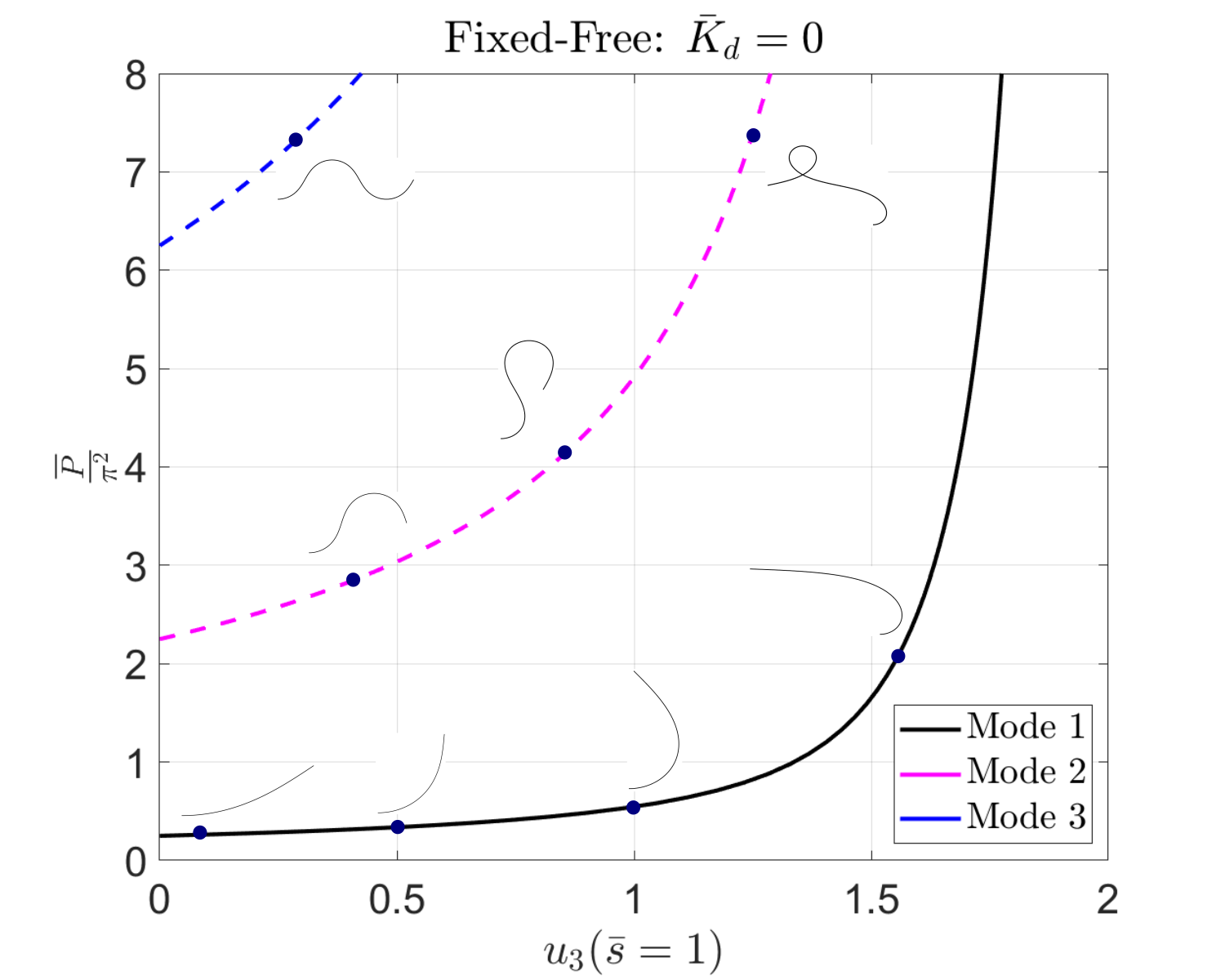}
		\caption{}
		\label{fig:purely-elastic-fixed-free-stability}
	\end{subfigure}\hfill 
	\begin{subfigure}{.49\linewidth}
		\includegraphics[width=\linewidth]{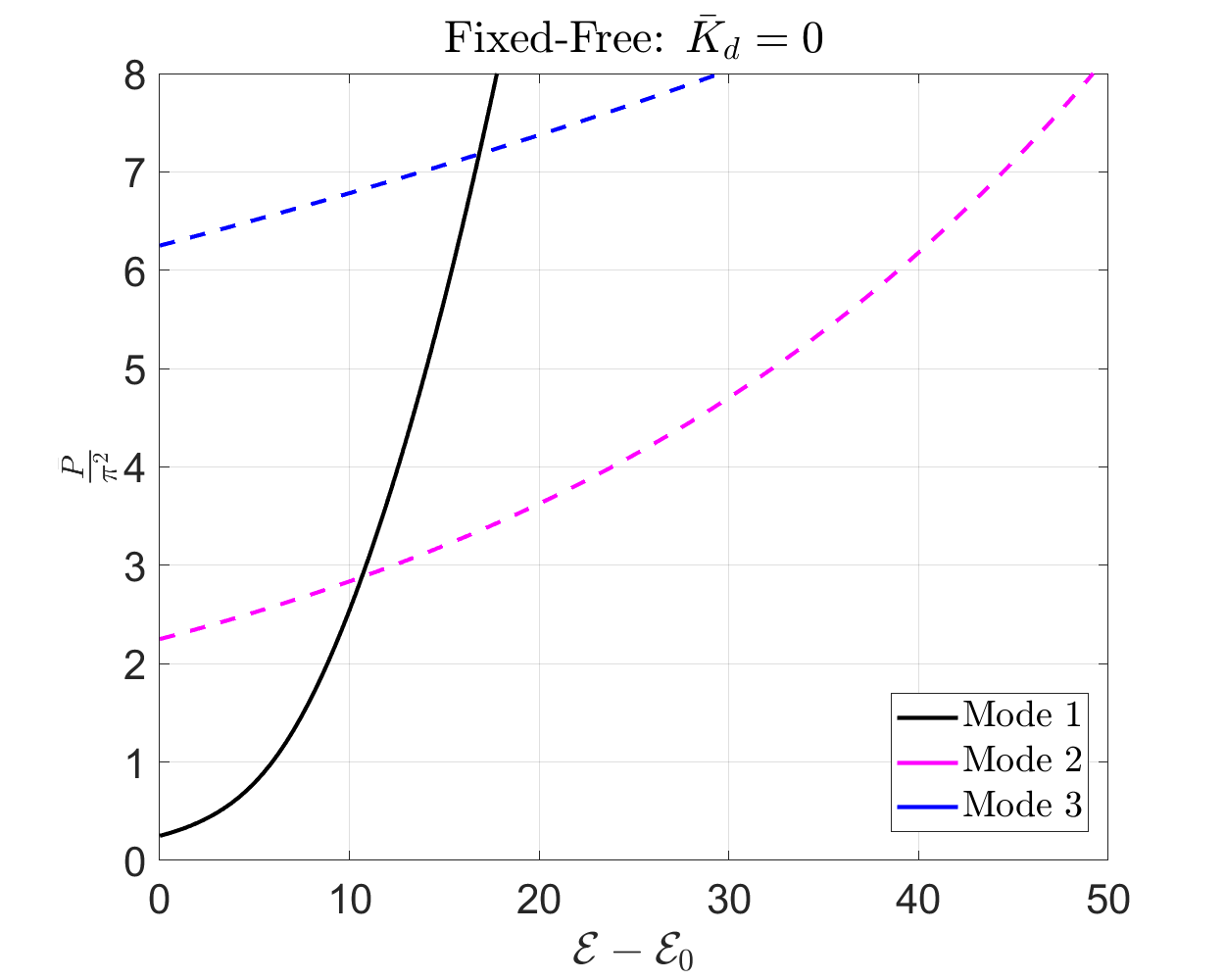}
		\caption{}
		\label{fig:purely-elastic-fixed-free-energy}
	\end{subfigure}
	\caption{Euler's elastica: (a) Stability diagram, note that the first critical load is compressive, that is, $\bar{P}_1 = +\frac{\pi^2}{4}$. (b) Total energy curve for fixed-free configuration; $\mathcal{E}_0 = 0$.}
	\label{fig:purely-elastic-fixed-free}
\end{figure}

Fig. \ref{fig:purely-elastic-fixed-free-stability} shows the stability diagram for the first three nonlinear modes when the elastica is fixed at one end while the other end remains free. We plot the horizontal displacement of the end-point, $u_3(\bar{s}=1) = 1 - \int_0^1 \cos\theta(\bar{s})d\bar{s}$, against loading parameter $\bar{P}$. Note that $\bar{P}>0$ corresponds to compressive loads and $\bar{P}<0$ corresponds to tensile loads. The solid curves denote stable deformations while the dotted lines represent unstable deformations. 

\begin{figure}[h!]
	\begin{subfigure}{.49\linewidth}
		\includegraphics[width=\linewidth]{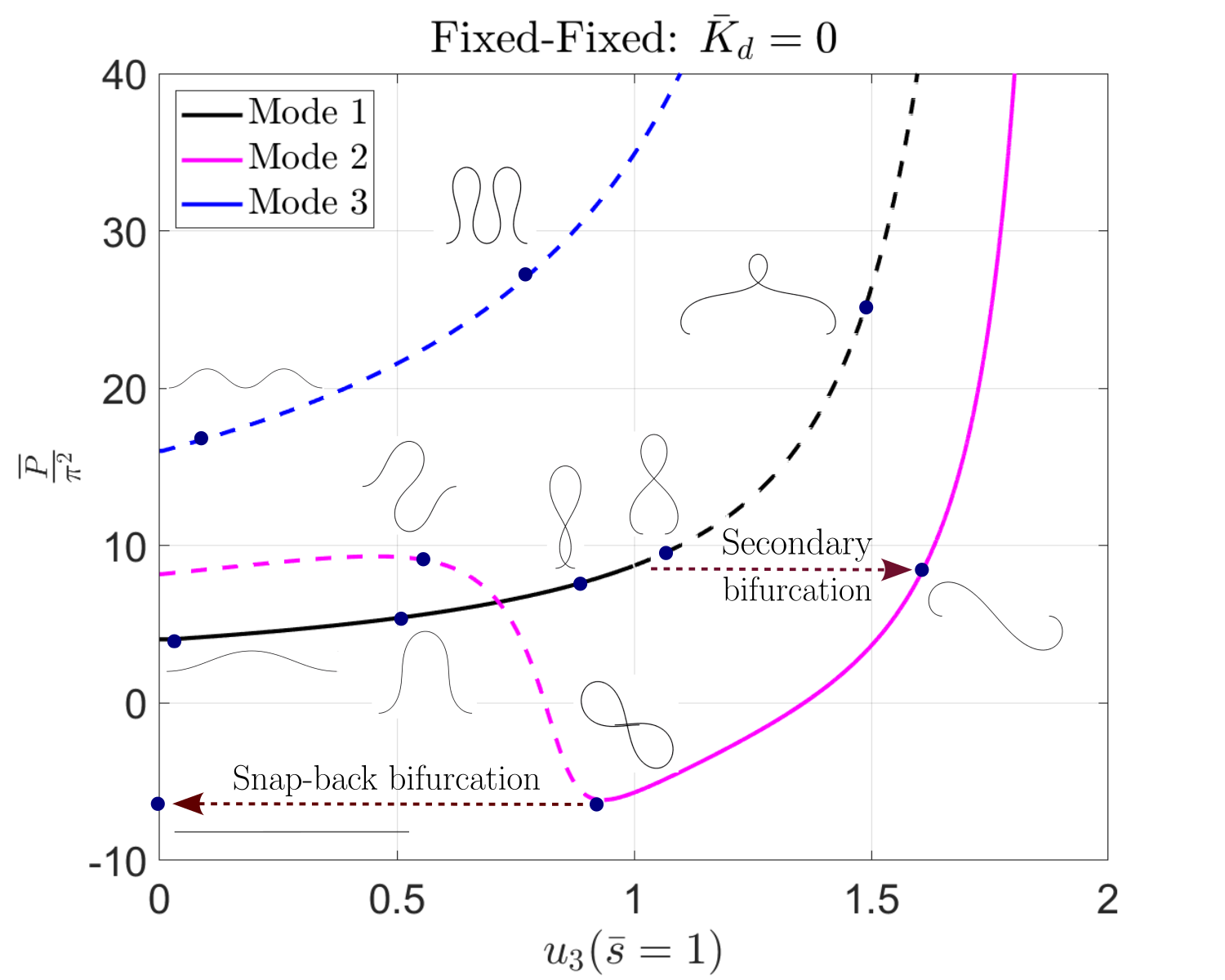}
		\caption{}
		\label{fig:purely-elastic-fixed-fixed-stability}
	\end{subfigure}\hfill 
	\begin{subfigure}{.49\linewidth}
		\includegraphics[width=\linewidth]{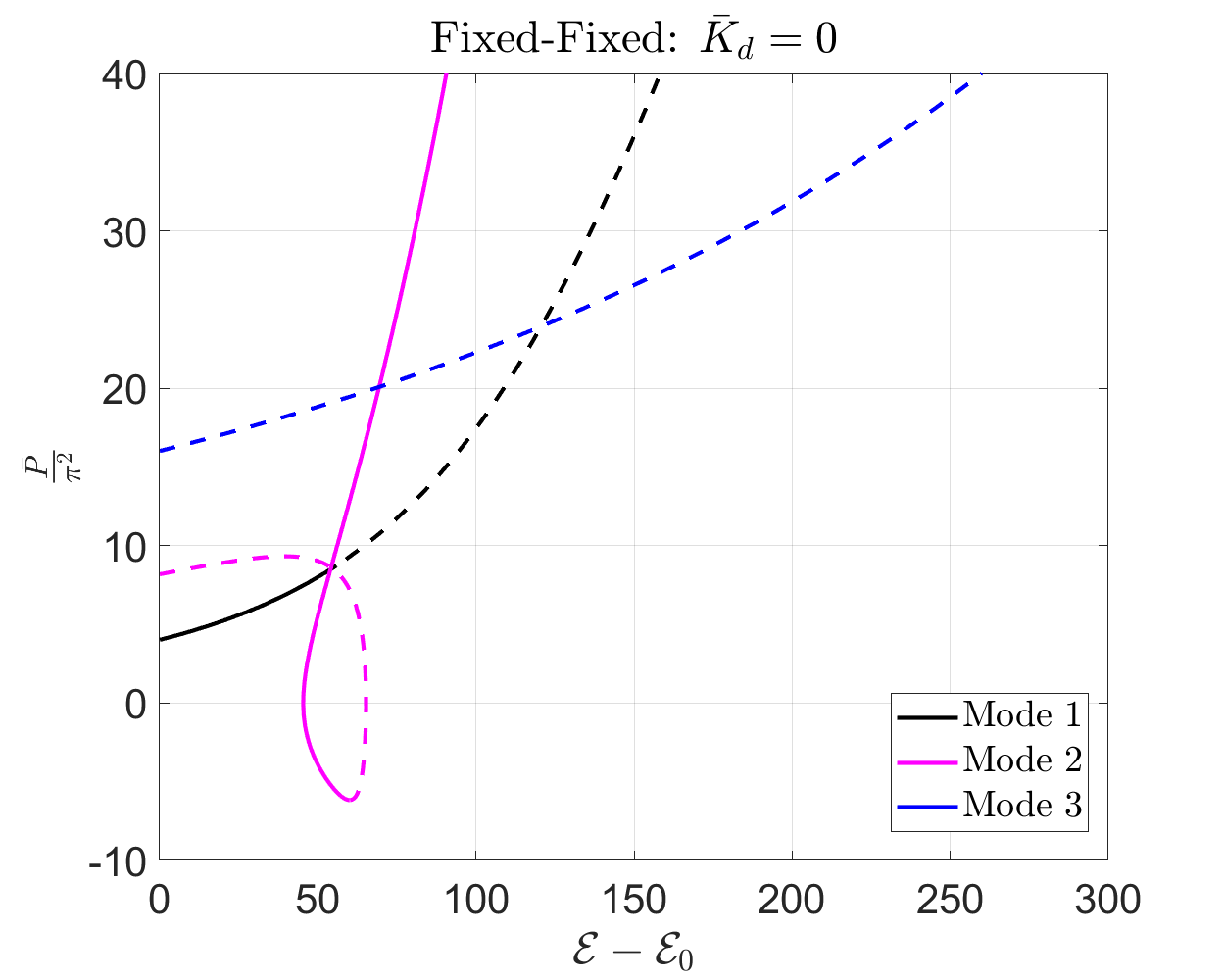}
		\caption{}
		\label{fig:purely-elastic-fixed-fixed-energy}
	\end{subfigure}
	\caption{Euler's elastica: (a) Stability diagram (b) Total energy curve for fixed-fixed configuration; $\mathcal{E}_0 = 0$.}
	\label{fig:purely-elastic-fixed-fixed}
\end{figure}

\begin{figure}[h!]
	\begin{subfigure}{.49\linewidth}
		\includegraphics[width=\linewidth]{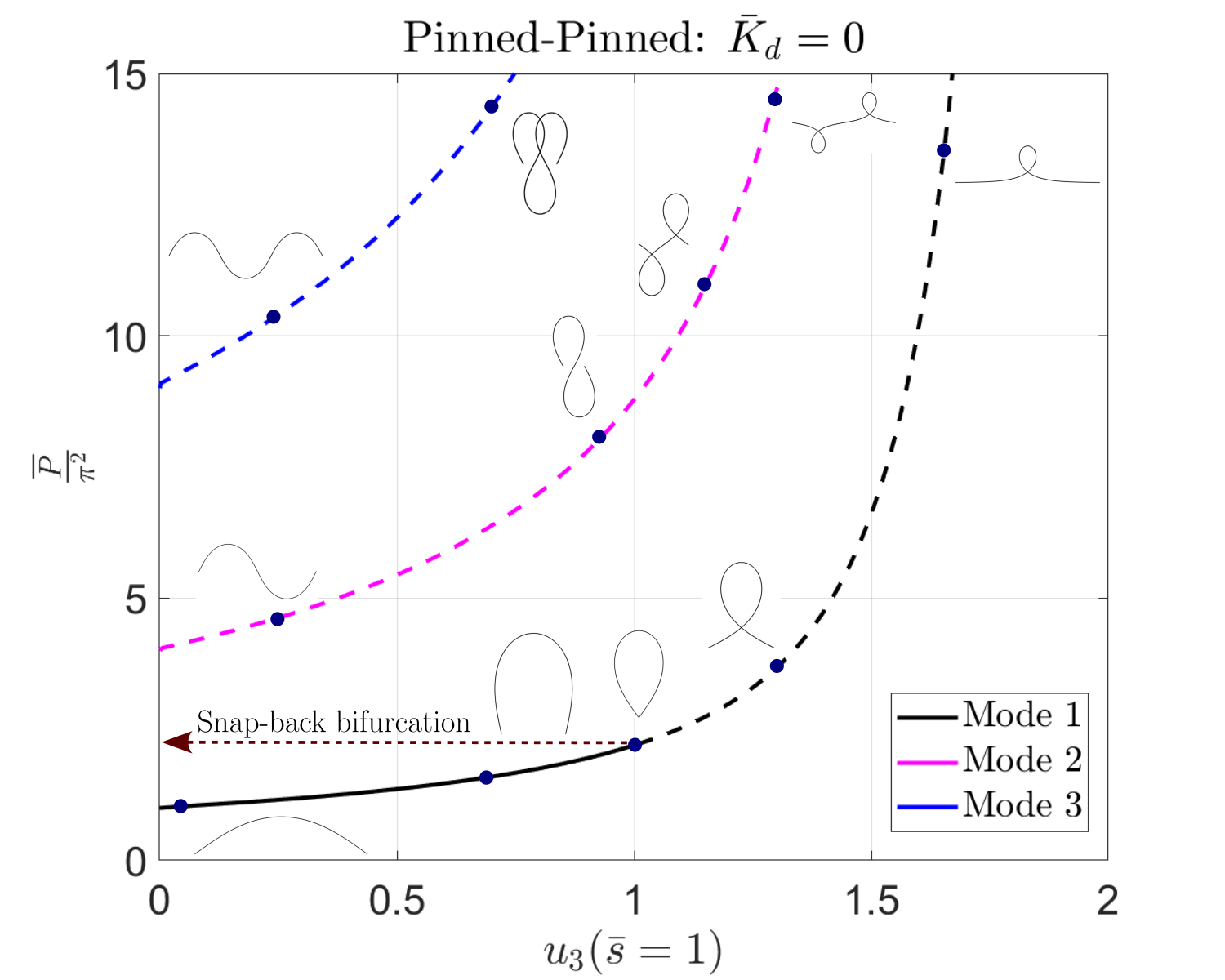}
		\caption{}
		\label{fig:purely-elastic-pinned-pinned-stability}
	\end{subfigure}\hfill 
	\begin{subfigure}{.49\linewidth}
		\includegraphics[width=\linewidth]{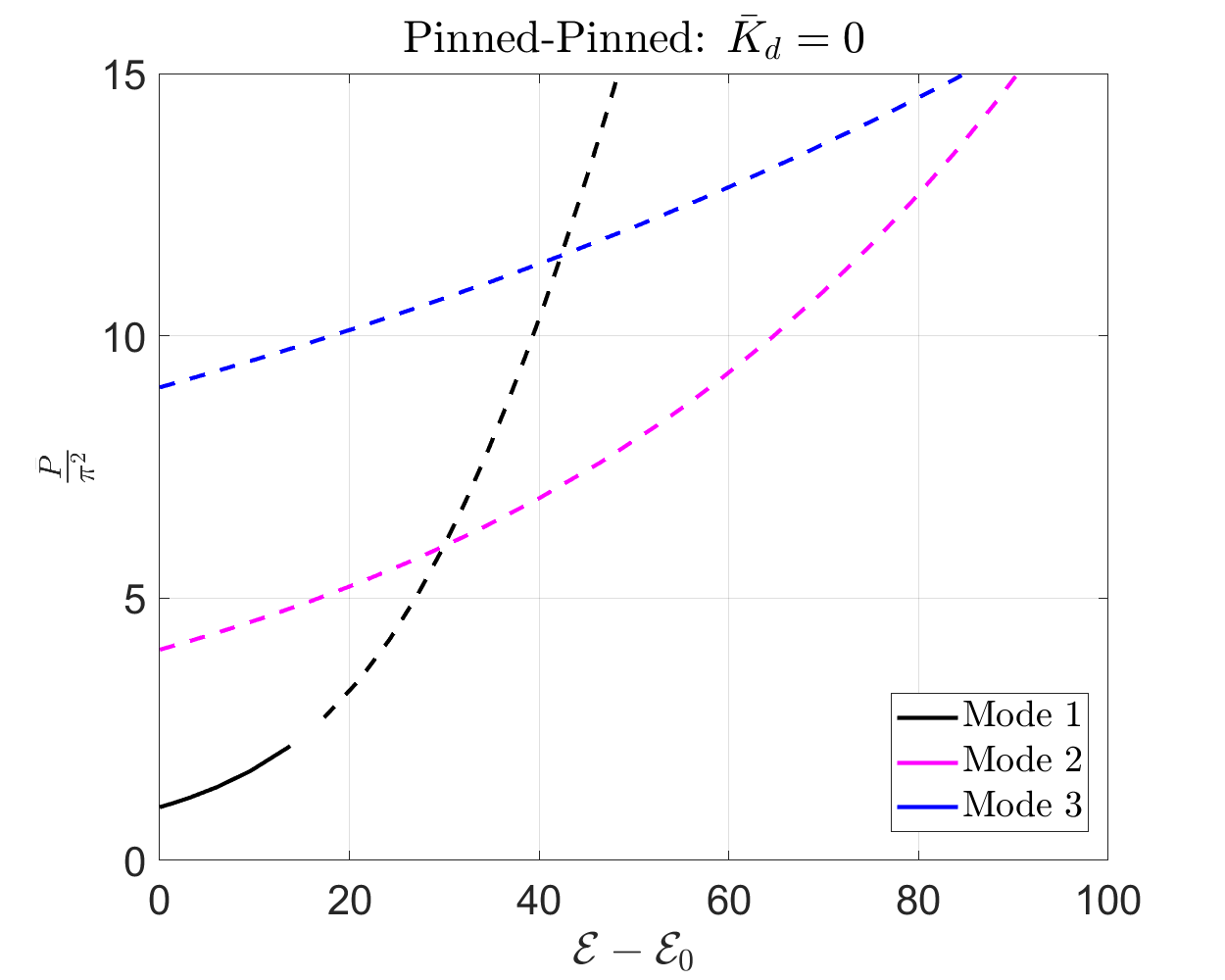}
		\caption{}
		\label{fig:purely-elastic-pinned-pinned-energy}
	\end{subfigure}
	\caption{Euler's elastica: (a) Stability diagram (b) Total energy curve for pinned-pinned configuration; $\mathcal{E}_0 = 0$.}
	\label{fig:purely-elastic-pinned-pinned}
\end{figure}

We observe that only the first nonlinear mode is stable. The stability diagram for the fixed-fixed scenario is shown in Fig. \ref{fig:purely-elastic-fixed-fixed-stability}. We observe that the elastica undergoes primary bifurcation along the first nonlinear mode initiating at the first critical $\bar{P}$ (obtained from linearized buckling analysis). The bifurcation at first critical $\bar{P}$ is continuous, as seen in Fig. \ref{fig:purely-elastic-fixed-fixed-energy}. As $\bar{P}$ is increased beyond the first critical load, the two ends of the ribbon meet. Thereafter, it undergoes snap-through (secondary) bifurcation at $\bar{P} = 10\pi^2$ to the mode-2 branch. At the secondary bifurcation, the total energy of mode-1 becomes larger than that of mode-2, see Fig. \ref{fig:purely-elastic-fixed-fixed}. The ribbon continues to deform on the second mode for  $\bar{P} > 10\pi^2$. If  $\bar{P}$ is quasi-statically reduced along the mode-2 branch, the nature of buckling load gradually switches from compressive to tensile until it snap backs to the pre-buckled (reference) tensile configuration at $\bar{P} = -6\pi^2$. The stability diagram corresponding to pinned-pinned case (Fig. \ref{fig:purely-elastic-pinned-pinned}) shows that only mode-1 is the stable mode beyond the first critical load and it stays so till the end-points meet, after which the ribbon snaps to the pre-buckled (reference) configuration \cite{Levyakov2009}.     

A close examination of Eqn. \ref{eqn:pure-elastica-equilibrium-equations} reveals that under the transformation of $\theta$, i.e., $\theta \rightarrow -\theta$, the equation possesses reflection symmetry if the condition $\bar{R} = 0$ is met. We observe that $\bar{R}$ always remains zero for all $\bar{P}_n~ (n=1,2,\dots)$ for pinned-pinned scenario. Nevertheless, for the fixed-fixed case, $\bar{R} = 0$ for odd $n$ ($n=1,3,\dots$) while it is non-zero for even $n$. The corresponding rich bifurcation behaviour in Fig. \ref{fig:purely-elastic-fixed-fixed-stability} could due attributed to this change of symmetry.



\subsection{Soft ferromagnetic ribbon: Transverse external magnetic field, $\bar{h}_e = \vb*{e}_2$}\label{subsec:soft-magnetic}
 We recall the demag. energy in soft magnetic case is $K_d at \int (\vb*{e}_2\cdot\vb*{n})^2 ds$.  To minimize the demag. energy, the curve's normal tends to align with the $\vb*{e}_3$ direction.
 This tends to align the ribbon in the $\vb*{e}_2$ direction. We consider $\bar{K}_d = 100$ for our analysis, since the magnetostatic energy becomes comparable to the mechanical energy in this regime. 
 
\begin{figure}[h!]
 	\begin{subfigure}{.49\linewidth}
 		\includegraphics[width=\linewidth]{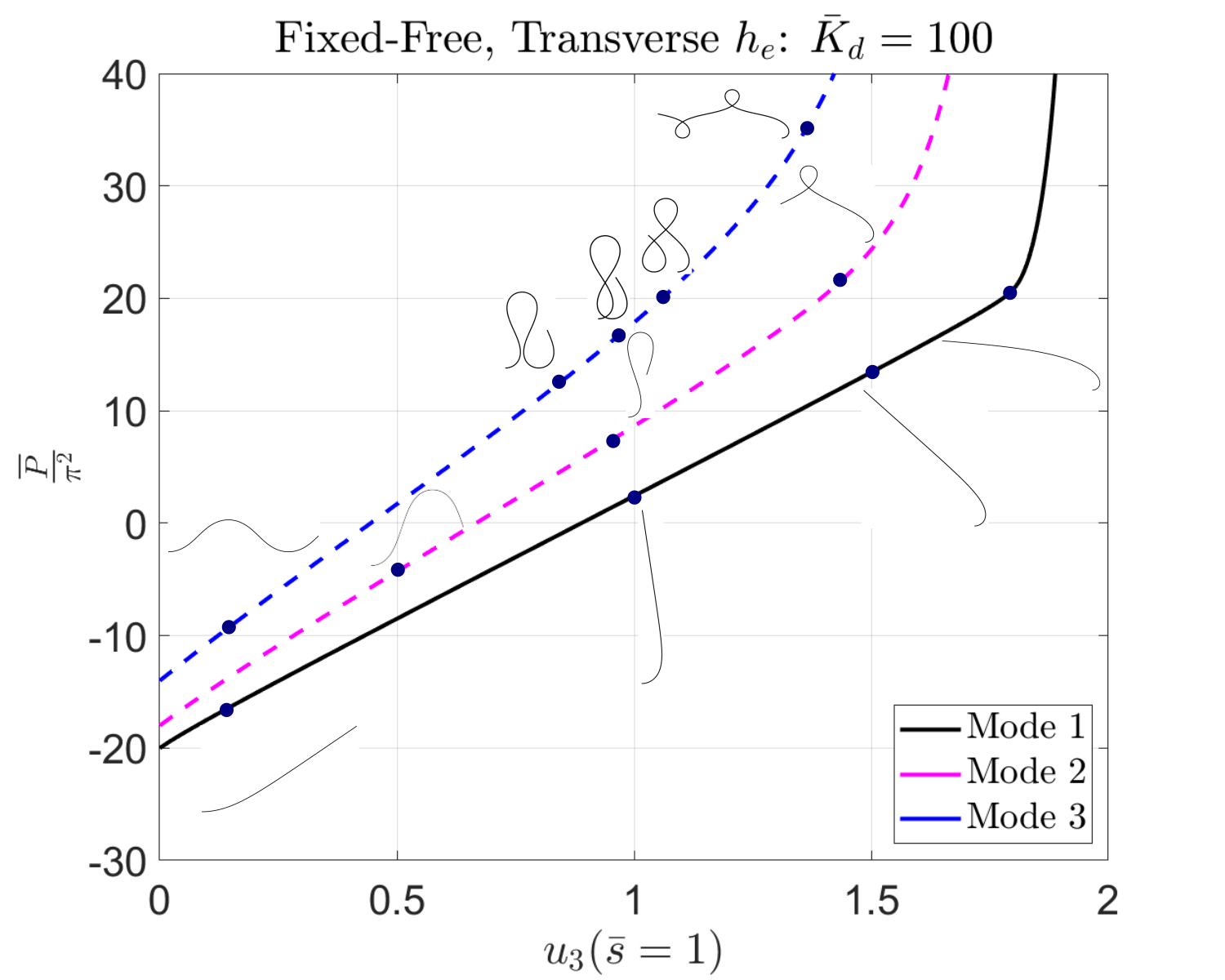}
 		\caption{}
 		\label{fig:vertical-field-fixed-free-kdbar-100-stability}
 	\end{subfigure}\hfill 
 	\begin{subfigure}{.49\linewidth}
 		\includegraphics[width=\linewidth]{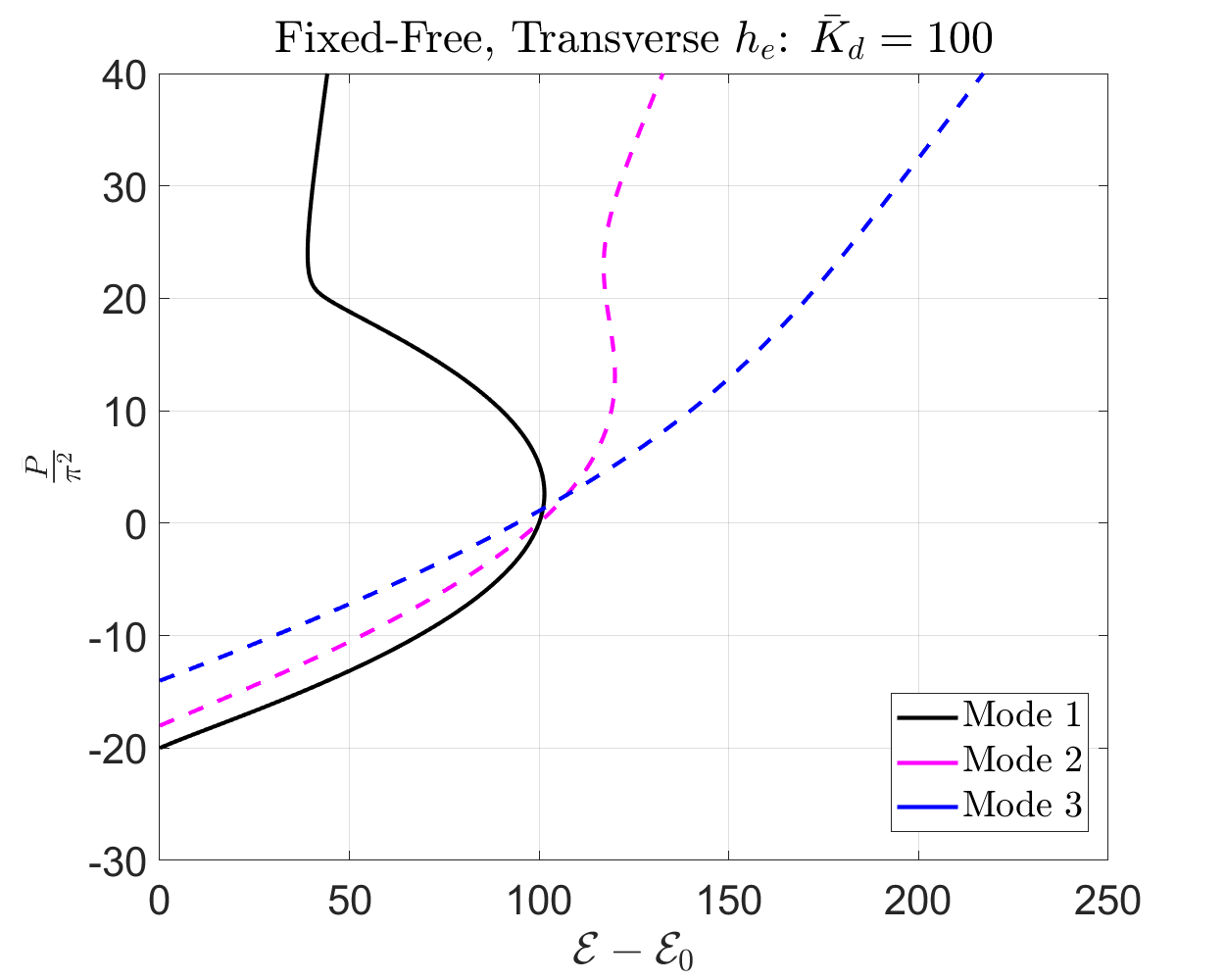}
 		\caption{}
 		\label{fig:vertical-field-fixed-free-kdbar-100-energy}
 	\end{subfigure}
 	\caption{Soft ferromagnetic ribbon,  $\bar{K}_d= 100$: (a) Stability diagram, note that the first critical load is tensile, that is, $\bar{P}_1 = -20.01 \pi^2$. (b) Total energy curve for fixed-free configuration under transverse magnetic field; $\mathcal{E}_0 = -100$.}
 	\label{fig:vertical-field-fixed-free-kdbar-100}
 \end{figure}

 Fig. \ref{fig:vertical-field-fixed-free-kdbar-100-stability} shows the stability diagram for fixed-free boundary condition ferromagnetic ribbon in the presence of a transverse $\vb*{h}_e$ external magnetic field, see Fig. \ref{fig:elastica-soft-magnetic-vertical-field}. Note that the critical load here is tensile. The first nonlinear mode remains stable for all $\bar{P}> -197.53$ and this bifurcation is continuous. 
 
\begin{figure}[h!]
	\begin{subfigure}{.49\linewidth}
		\includegraphics[width=\linewidth]{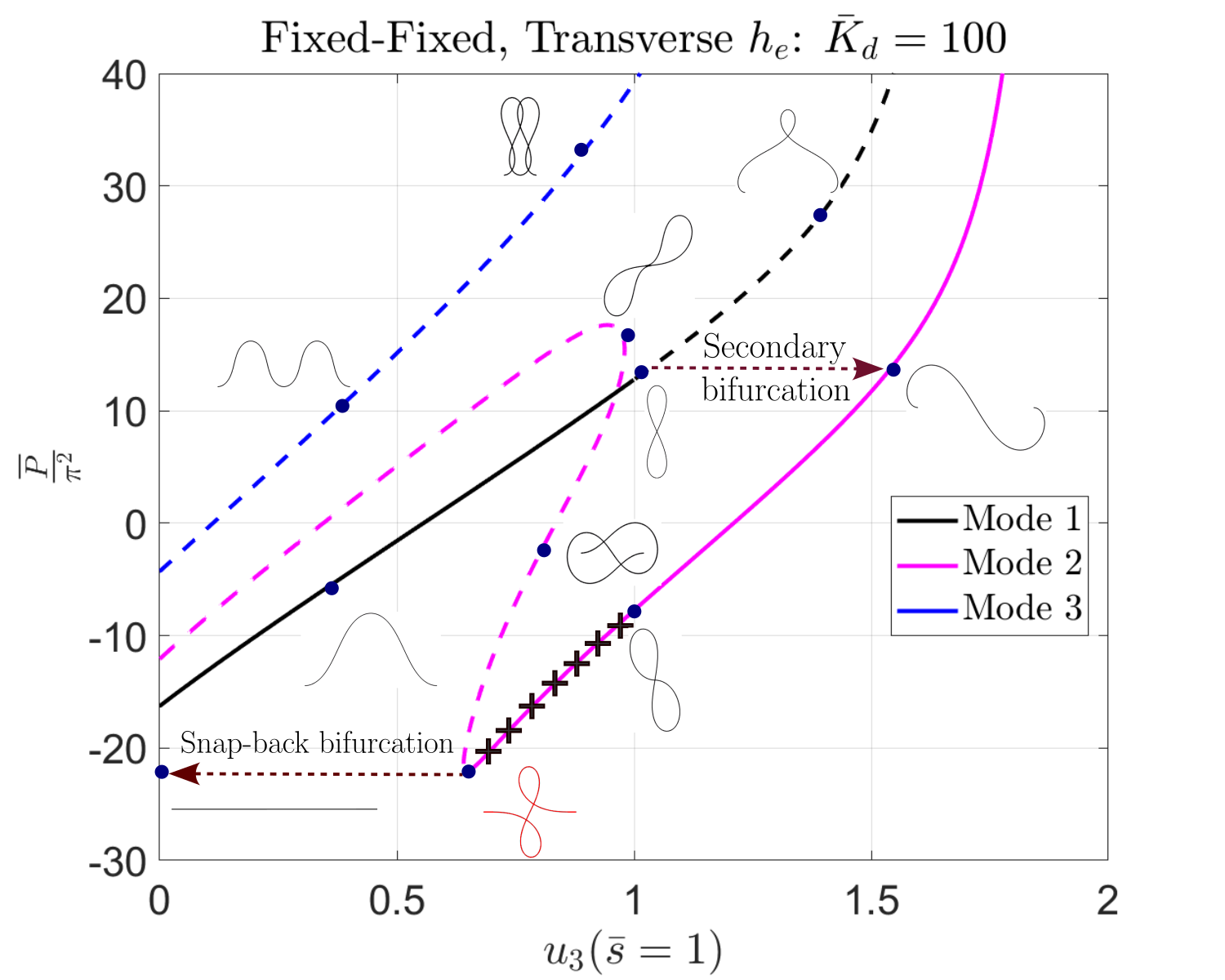}
		\caption{}
		\label{fig:vertical-field-fixed-fixed-kdbar-100-stability}
	\end{subfigure}\hfill 
	\begin{subfigure}{.49\linewidth}
		\centering
		\includegraphics[width=0.66\linewidth]{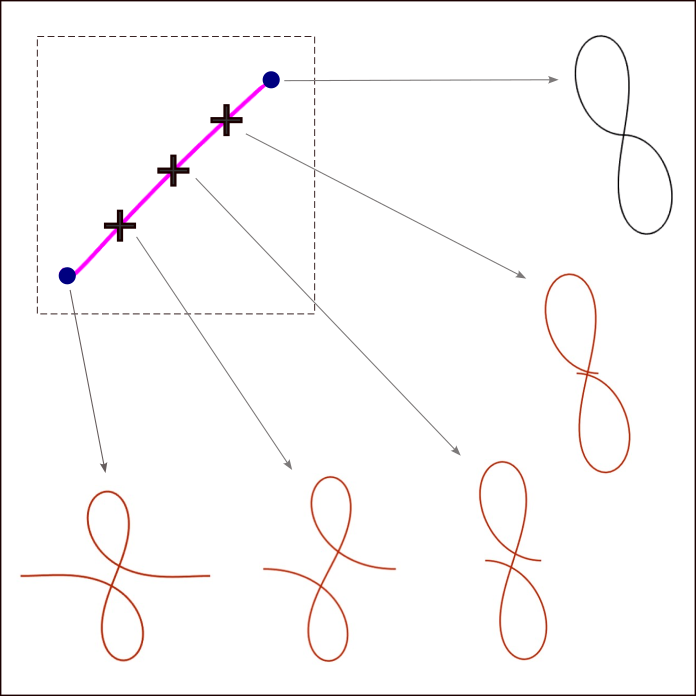}
		\caption{}
		\label{fig:fixed-fixed-kdbar-100-mode-2-stable-inset}
	\end{subfigure} \\
	\centering
	\begin{subfigure}{.49\linewidth}
		\includegraphics[width=\linewidth]{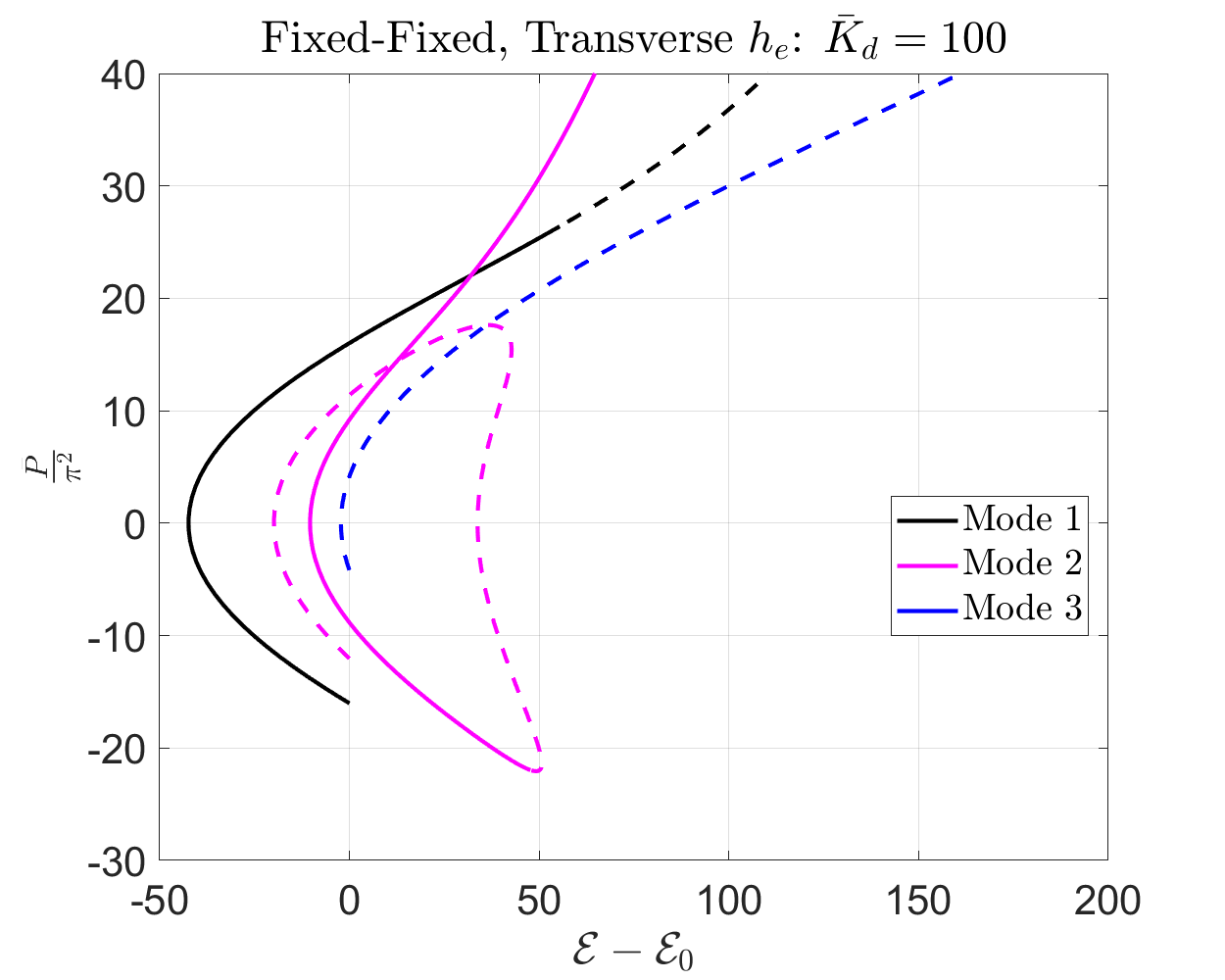}
		\caption{}
		\label{fig:vertical-field-fixed-fixed-kdbar-100-energy}
	\end{subfigure}
	
	\caption{Soft ferromagnetic ribbon,  $\bar{K}_d= 100$: (a) Stability diagram, segment of the mode-2 branch corresponding to novel stable curves is highlighted in `+' symbol. Deformed config. shown in red represents one such novel stable curve on this segment that cannot be observed in purely elastic ribbon.  (b) Segment of the mode-2 branch showing novel stable deformed configurations. (c) Total energy curve for fixed-fixed configuration under transverse magnetic field; $\mathcal{E}_0 = -100$. }
	\label{fig:vertical-field-fixed-fixed-kdbar-100}
\end{figure}

 For the fixed-fixed boundary condition, soft ferromagnetic ribbon buckles at a tensile load, and mode-1 deformation is observed for $\bar{P}> -160.52$, see Fig. \ref{fig:vertical-field-fixed-fixed-kdbar-100}. The ribbon snaps to the mode-2 branch as $\bar{P}$ is increased as seen in Fig. \ref{fig:vertical-field-fixed-fixed-kdbar-100}. The ribbon deforms further along the mode-2 branch with gradual decrease in $\bar{P}$. As we unload, that is, decrease $\bar{P}$ on the mode-2 branch, we observe the formation of new stable deformed configurations for tensile loads before it snaps back to the reference state. We observe a prolonged stretch of a stable segment in the mode-2 branch under the influence of $\vb*{h}_e$ for tensile loads ($\bar{P}<0$). This stable segment persists even after the two supports have crossed each other significantly. Interestingly, we observe novel and stable states in this segment of the mode-2 branch. This segment is highlighted in Fig.  \ref{fig:vertical-field-fixed-fixed-kdbar-100-stability}.  We have highlighted one such novel and stable deformed configuration in red in Fig. \ref{fig:vertical-field-fixed-fixed-kdbar-100-stability}, when the end displacement is $u_3= 0.65$. Note that in the deformed configuration depicted in red in Fig. \ref{fig:vertical-field-fixed-fixed-kdbar-100-stability}, there are two self-intersection points. {\color{black} A few intermediate novel deformed configurations are shown in red in Fig. \ref{fig:fixed-fixed-kdbar-100-mode-2-stable-inset}.} Curves with two self-intersection points are not observed in any stable configuration in the purely elastic case. As $\bar{P}$ is further reduced along this branch, that is, as we increase the tensile load, the ribbon undergoes a secondary bifurcation after which the ribbon snaps to the pre-buckled (reference) tensile configuration.

\begin{figure}[ht!]
	\begin{subfigure}{.49\linewidth}
	    \includegraphics[width=\linewidth]{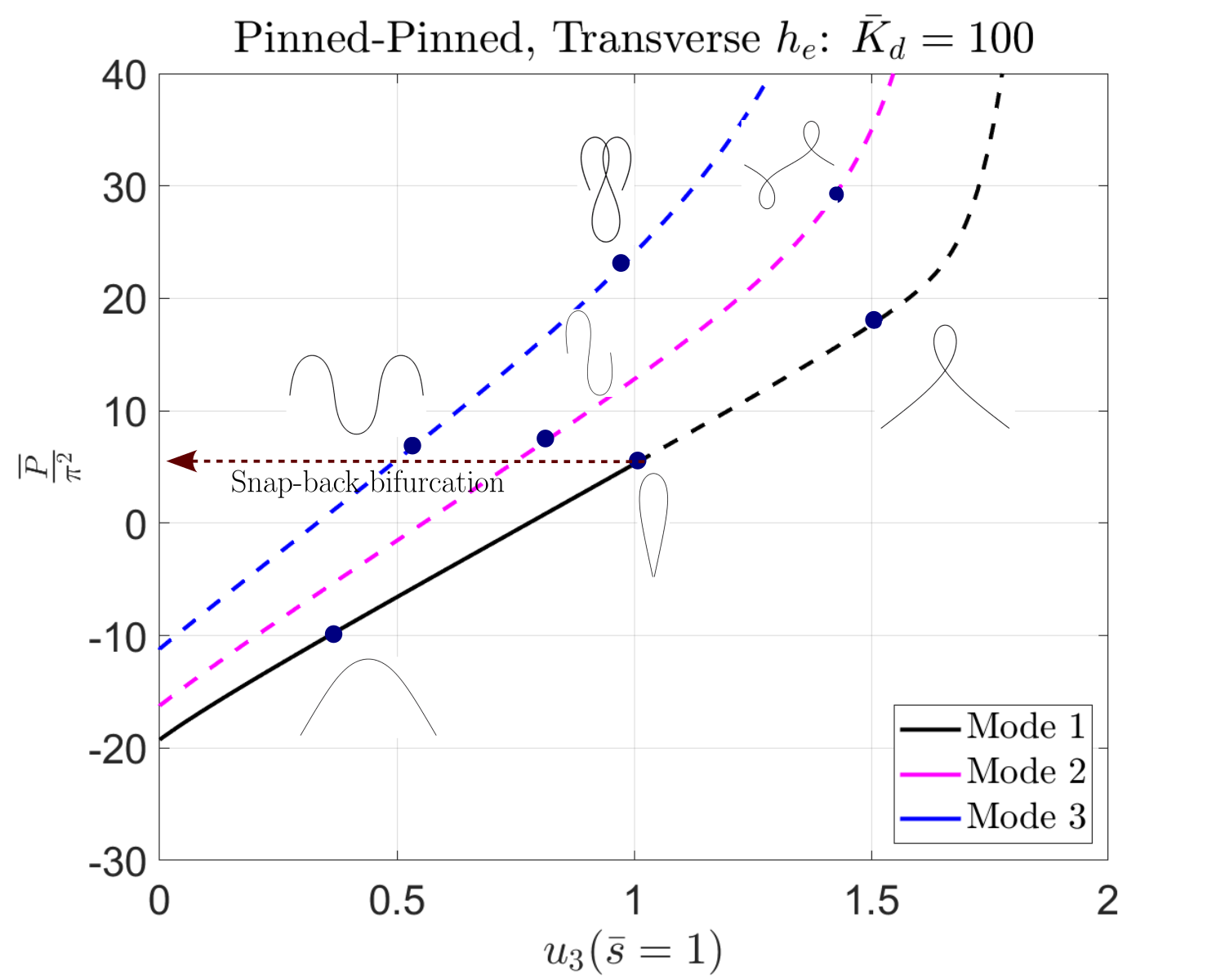}
		\caption{}
		\label{fig:vertical-field-pinned-pinned-kdbar-100-stability}
	\end{subfigure}\hfill 
	\begin{subfigure}{.49\linewidth}
	    \includegraphics[width=\linewidth]{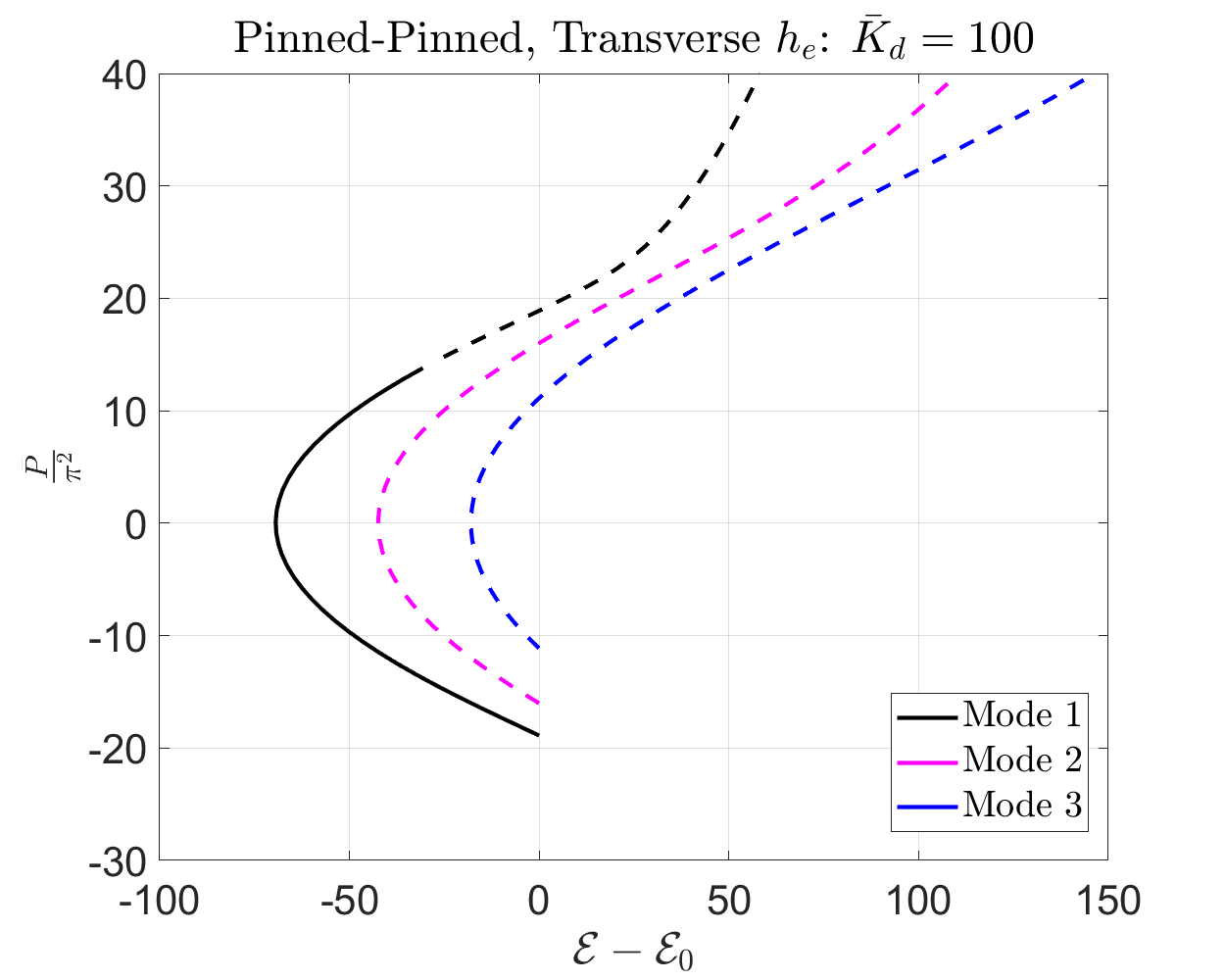}
		\caption{}
		\label{fig:vertical-field-pinned-pinned-kdbar-100-energy}
	\end{subfigure}
	\caption{Soft ferromagnetic ribbon,  $\bar{K}_d= 100$: (a) Stability diagram (b) Total energy curve for pinned-pinned configuration under transverse magnetic field; $\mathcal{E}_0 = -100$.}
	\label{fig:vertical-field-pinned-pinned-kdbar-100}
\end{figure}

For the pinned-pinned soft ferromagnetic ribbon, the stability diagram Fig. \ref{fig:vertical-field-pinned-pinned-kdbar-100-stability} is qualitatively similar to the purely elastic case, except that the deformed shape is aligned along the $\vb*{e}_2$ direction and the critical load is tensile. Thus, for a soft ferromagnetic ribbon for all the boundary conditions, the critical load is tensile and they can be determined from linearized equations (Eqn. \ref{eqn:soft-magnetic-elastica-equilibrium-equations}) about $\theta(\bar{s}) \approx 0$ such that $\sin\theta(\bar{s}) \approx \theta(\bar{s})$ and $\cos\theta(\bar{s}) \approx 1$ as follows:
\begin{equation}
		\theta''(\bar{s}) + (\bar{P} + 2\bar{K}_d)\theta(\bar{s}) + \bar{R} = 0,
\end{equation}
subject to the integral constraint (for fixed-fixed and pinned conditions): 
\begin{equation*}
   \bar{y}(1) = \int_{0}^{1} \theta(\bar{s})~ d\bar{s} = 0,
\end{equation*}
and gives the critical loads as
\begin{itemize}
	\item fixed-free: $\bar{P}_n = \left(n-\frac{1}{2}\right)^2\pi^2 - 2\bar{K}_d, \quad n=1,2,\dots$,
	\item fixed-fixed: $\left(2\tan\frac{\sqrt{\bar{P}_n + 2\bar{K}_d}}{2} = \sqrt{\bar{P}_n + 2\bar{K}_d}\right)$; $\bar{P}_1 = 4\pi^2 - 2\bar{K}_d, \bar{P}_2 = 8.183\pi^2 - 2\bar{K}_d, \bar{P}_3 = 16\pi^2 - 2\bar{K}_d, \dots,$
	\item pinned-pinned: $\bar{P}_n = n^2\pi^2 - 2\bar{K}_d$.
\end{itemize}

\subsection{Hard ferromagnetic ribbon: Transverse external magnetic field, $\bar{h}_e = \pm \vb*{e}_2$}\label{subsec:hard-magnetic}
\begin{figure}[h!]
	\begin{subfigure}{.49\linewidth}
		\includegraphics[width=\linewidth]{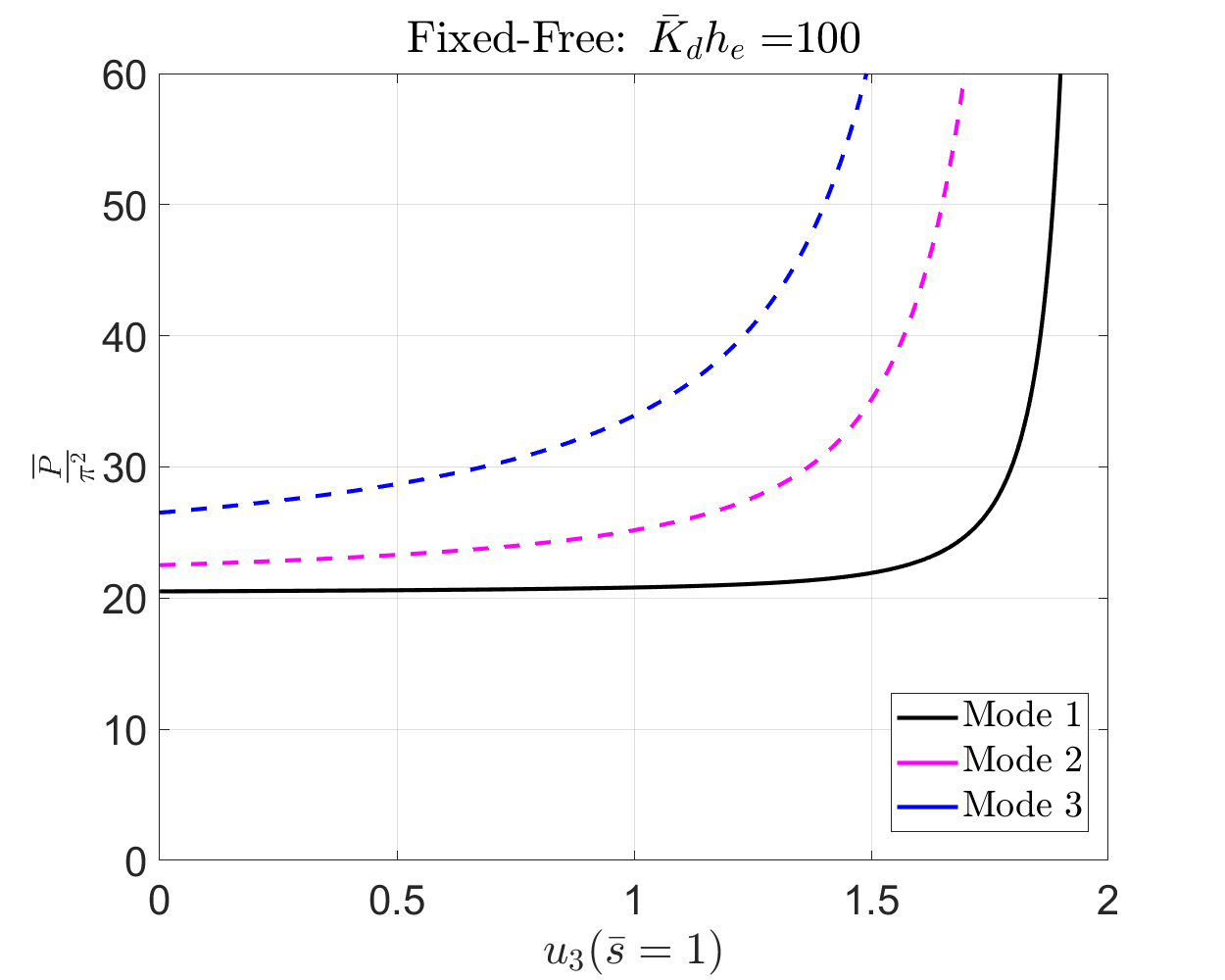}
		\caption{}
		\label{}
	\end{subfigure}\hfill 
	\begin{subfigure}{.49\linewidth}
		\includegraphics[width=\linewidth]{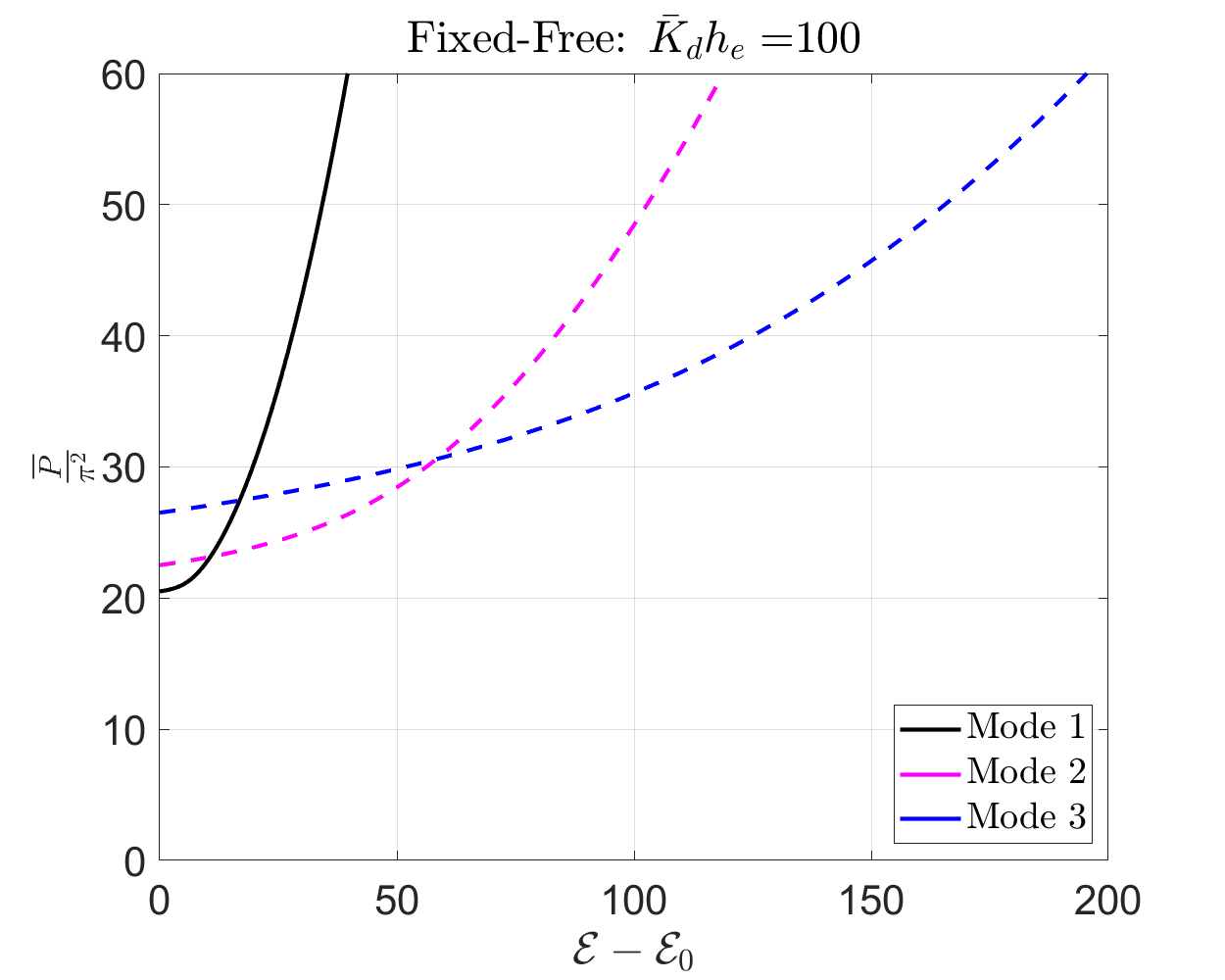}
		\caption{}
		\label{}
	\end{subfigure}
	\caption{Hard ferromagnetic ribbon: Axial $h_e$, Fixed-Free, $\vb*{m}=\vb*{t}(\bar{s})$, $\bar{K}_dh_e = 100$: (a) Stability diagram (b) Total energy curve; $\mathcal{E}_0 = 0$.}
	\label{fig:hard-tangent-fixed-free-kdbarhe-100}
\end{figure}

\begin{figure}[h!]
	\begin{subfigure}{.49\linewidth}
		\includegraphics[width=\linewidth]{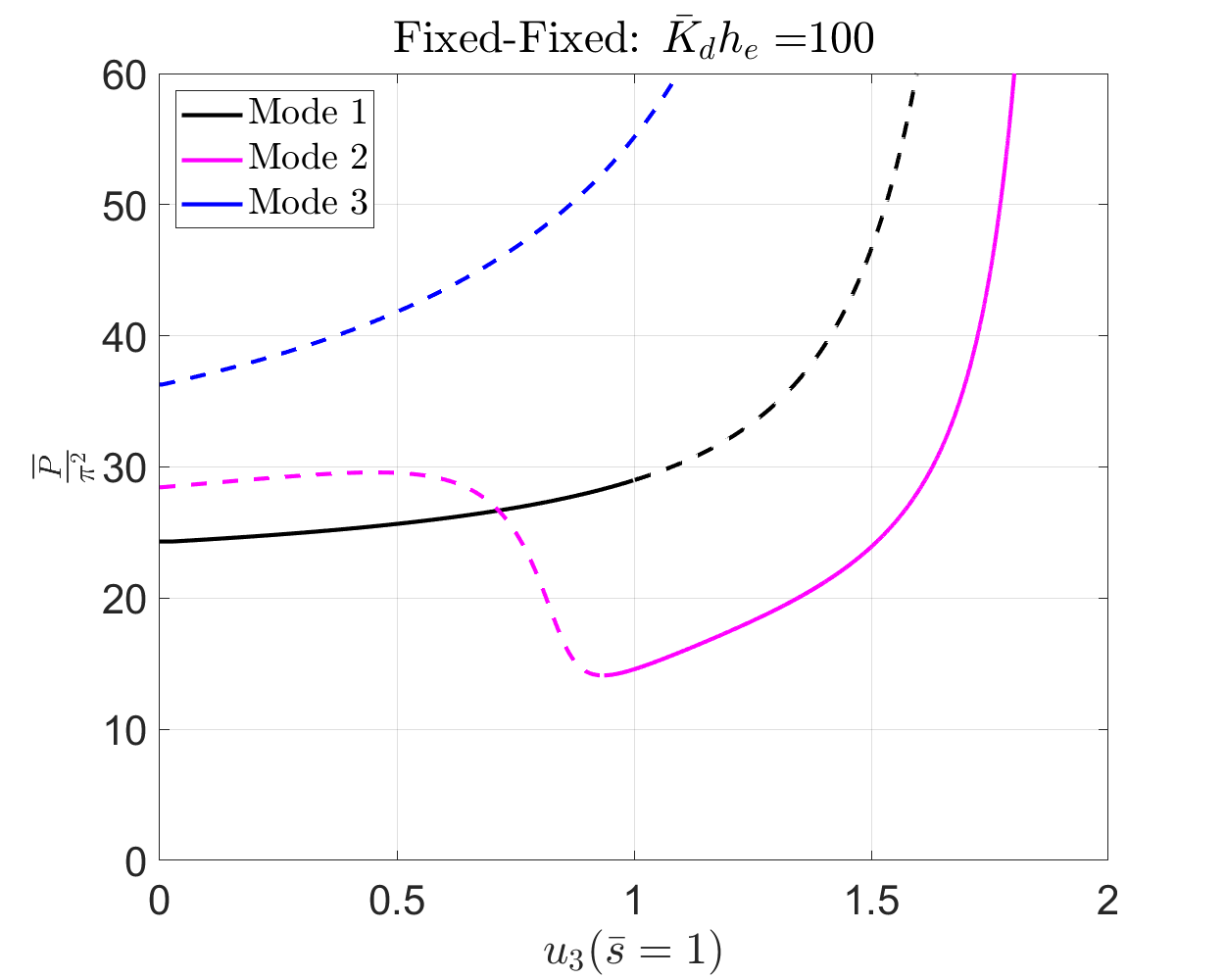}
		\caption{}
		\label{}
	\end{subfigure}\hfill 
	\begin{subfigure}{.49\linewidth}
		\includegraphics[width=\linewidth]{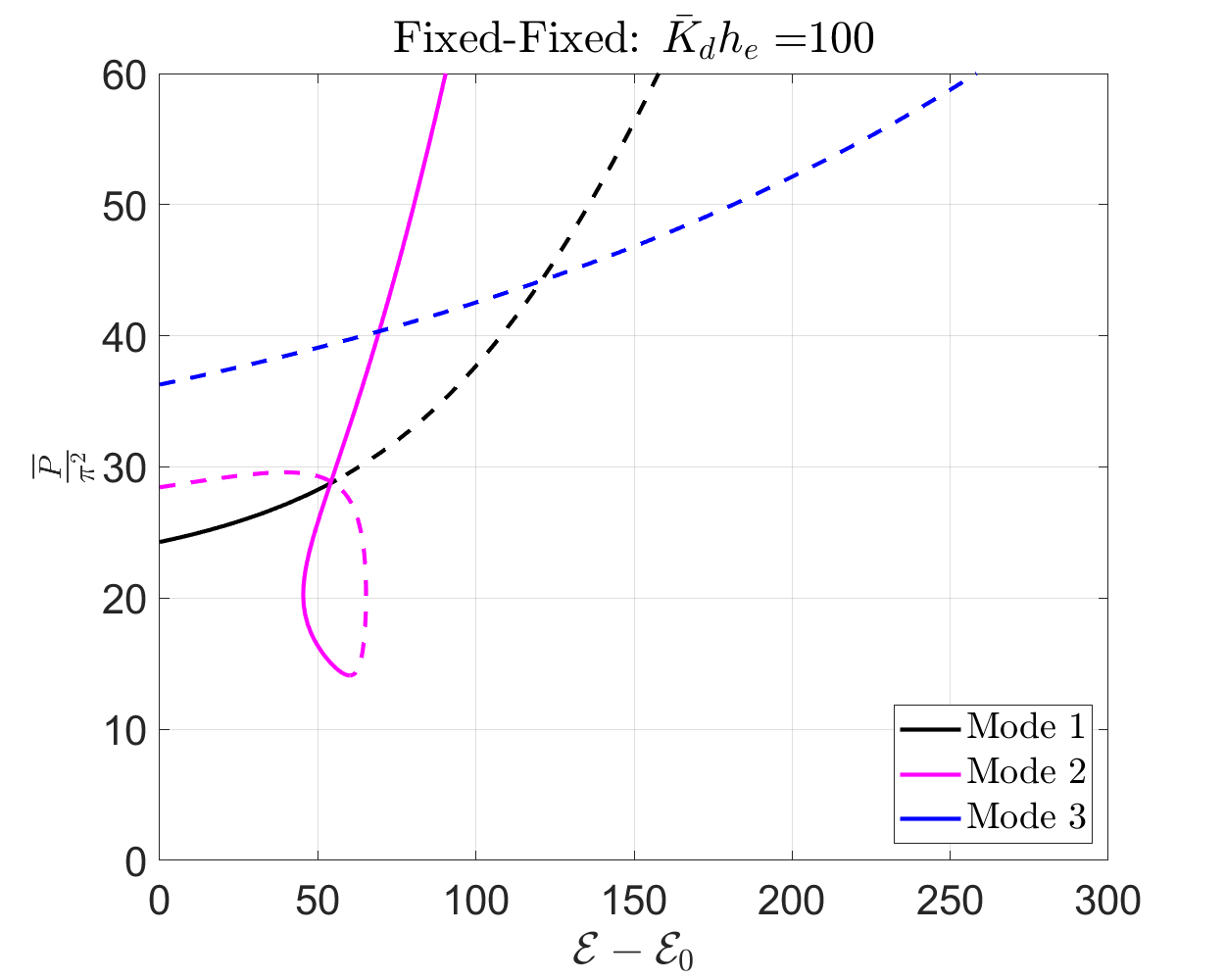}
		\caption{}
		\label{}
	\end{subfigure}
	\caption{Hard ferromagnetic ribbon: Axial $h_e$, Fixed-Fixed, $\vb*{m}=\vb*{t}(\bar{s})$, $\bar{K}_dh_e = 100$: (a) Stability diagram (b) Total energy curve; $\mathcal{E}_0 = 0$.}
	\label{fig:hard-tangent-fixed-fixed-kdbarhe-100}
\end{figure}

\begin{figure}[h!]
	\begin{subfigure}{.49\linewidth}
		\includegraphics[width=\linewidth]{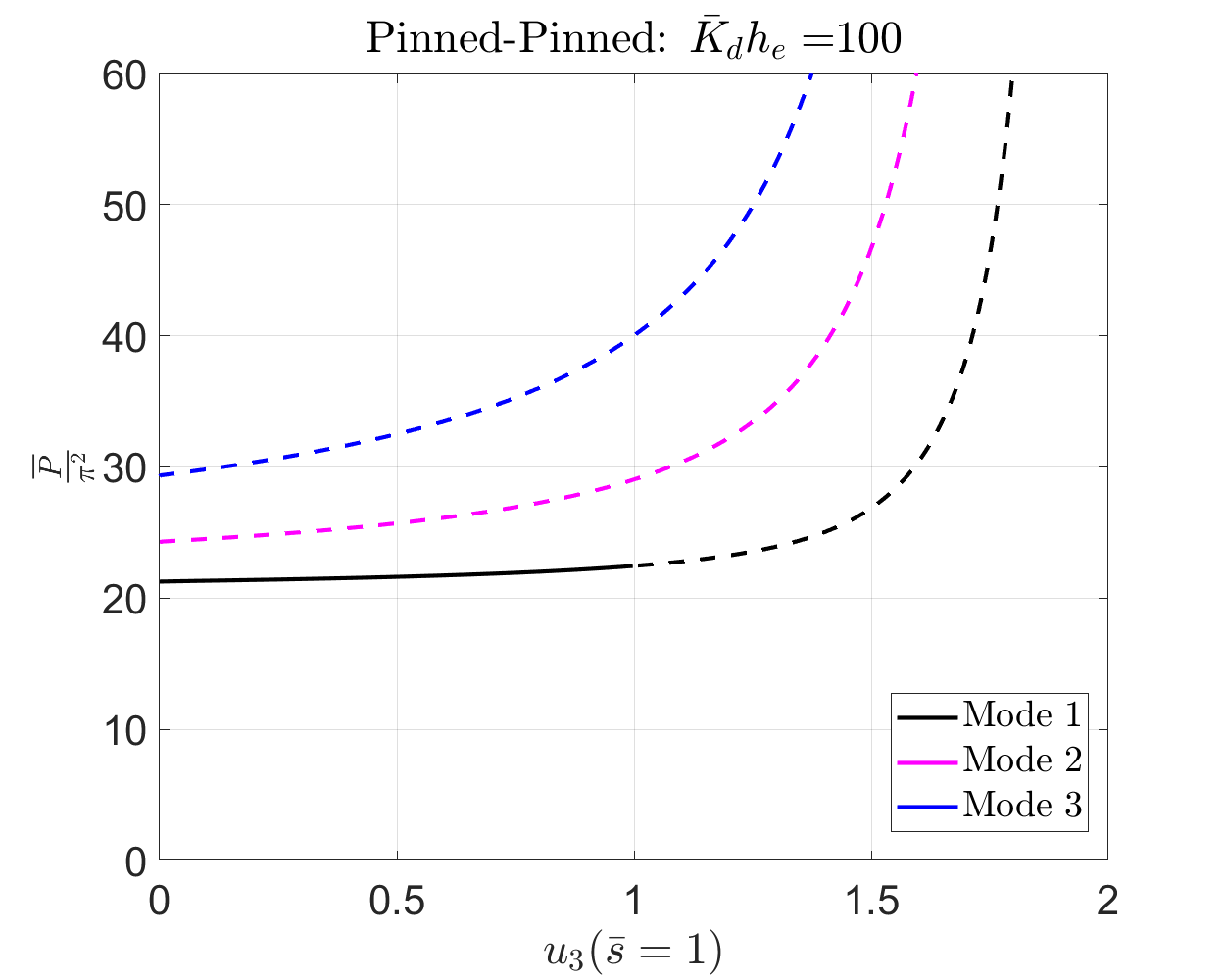}
		\caption{}
		\label{}
	\end{subfigure}\hfill 
	\begin{subfigure}{.49\linewidth}
		\includegraphics[width=\linewidth]{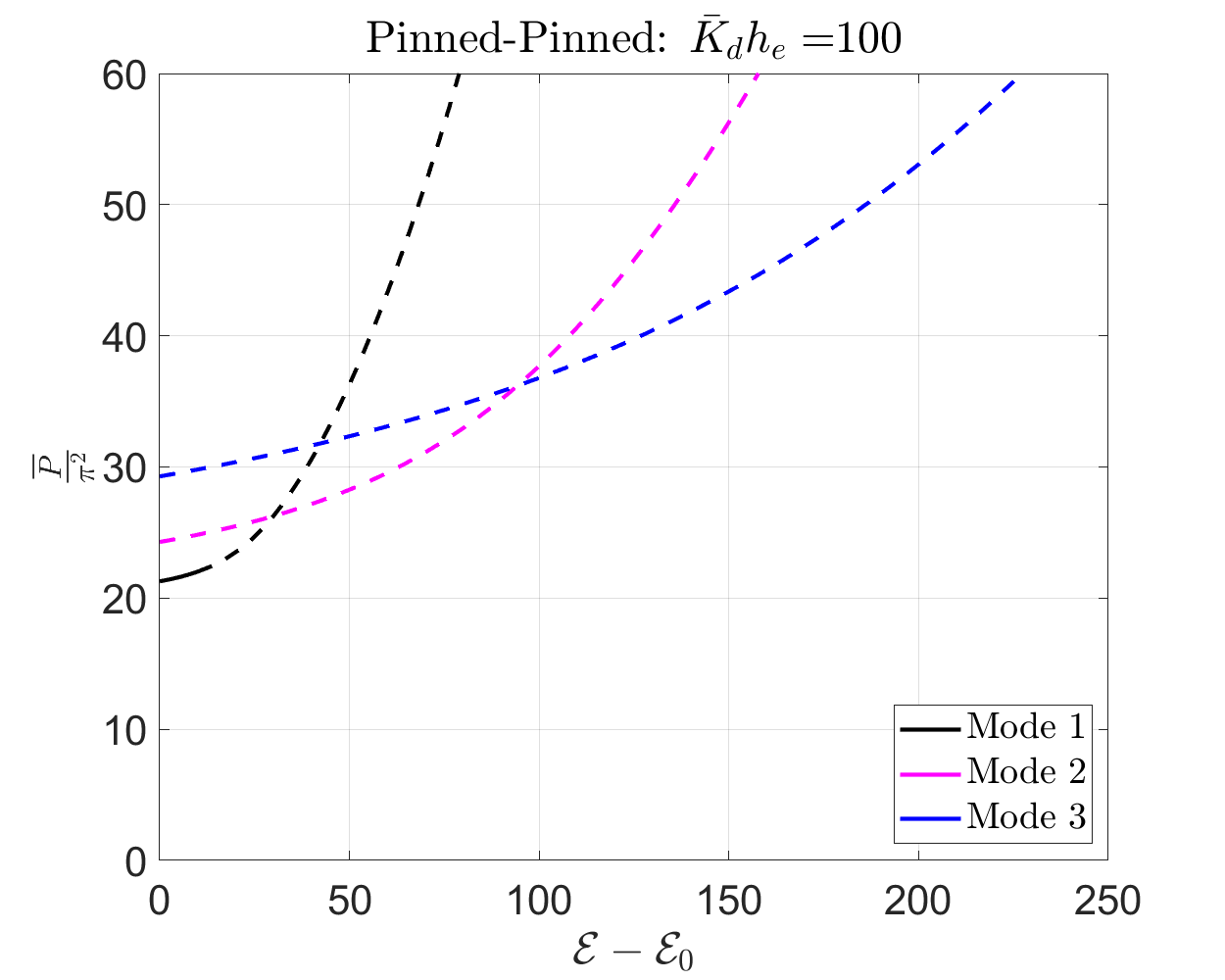}
		\caption{}
		\label{}
	\end{subfigure}
	\caption{Hard ferromagnetic ribbon: Axial $h_e$, Pinned-Pinned, $\vb*{m}=\vb*{t}(\bar{s})$, $\bar{K}_dh_e = 100$: (a) Stability diagram (b) Total energy curve; $\mathcal{E}_0 = 0$.}
	\label{fig:hard-tangent-pinned-pinned-kdbarhe-100}
\end{figure}

{ \color{black}
We now present our analysis for the hard ferromagnetic ribbon. Recall that, in this case that the magnetization vector makes a constant angle with the tangent at each point along the curve. Comparing Eqns. \ref{eqn:hard-magnetic-elastica-cases} and \ref{eqn:pure-elastica-equilibrium-equations} reveals that, in this case, magnetization does not change the structure of Euler's elastica equilibrium equations. For some combinations of $\left(\vb*{h}_e, \vb*{m}(\bar{s})\right)$, the deformed configurations corresponding to the hard magnetic case are obtained by a simple translation of the stability curves  of those of Euler's elastica  (Figs. \ref{fig:purely-elastic-fixed-free},\ref{fig:purely-elastic-fixed-fixed}, \ref{fig:purely-elastic-pinned-pinned}) along $\bar{P}$-axis, see Figs. \ref{fig:hard-tangent-fixed-free-kdbarhe-100}, \ref{fig:hard-tangent-fixed-fixed-kdbarhe-100} and \ref{fig:hard-tangent-pinned-pinned-kdbarhe-100}. For others, the stability curves of the hard magnetic case remain identical to those of Euler's elastica. In the case of the former $\left(\vb*{h}_e,\vb*{m}(\bar{s})\right)$-combinations, the critical loads are either significantly higher or lower than those corresponding to the classical elastica. This can easily be seen by linearizing Eqn. \ref{eqn:hard-magnetic-elastica-cases}$_2$.

Our proposed model and results matches with the equilibrium equation for planar deformation of  hard ferromagnetic ribbon under fixed-free configuration, as reported by Zhao et al \cite{Zhao2019} and Wang et al. \cite{Wang2020,Wang2022}, when subjected to an external magnetic field and no mechanical load. 

\paragraph{Case of $\vb*{m}(\bar{s}) = \vb*{t}(\bar{s})$}: We compare the equilibrium equation of the hard ferromagnetic ribbon with $\vb*{m}(\bar{s}) = \vb*{t}(\bar{s})$ (Eqn. \ref{eqn:hard-magnetic-elastica-cases}$_1$) and Eqn. \ref{eqn:pure-elastica-equilibrium-equations} and make the following observations:
\begin{itemize}
	\item The critical buckling load remains unaffected.
	\item The reactions $Q$ and $R_e$ are related via
	 \begin{equation}
	 	\bar{Q} = \bar{R}_e \mp 2\bar{K}_dh_e
	 	\label{eqn:reaction-transverse-he-hard-magnet-1}
	 \end{equation}
 when $\vb*{h}_e = \pm h_e\vb*{e}_2$.
\end{itemize}
The stability curves are identical to the classical elastica.

\paragraph{Case of $\vb*{m}(\bar{s}) = \vb*{n}(\bar{s})$}:
The stability curves of the hard ferromagnetic ribbon with normal magnetization distribution, $\vb*{m}(\bar{s}) = \vb*{n}(\bar{s})$ are obtained by translating those of Euler's elastica along $\bar{P}$ axis, with the measure of translation being proportional to $h_e$. Figs. \ref{fig:hard-tangent-fixed-free-kdbarhe-100}, \ref{fig:hard-tangent-fixed-fixed-kdbarhe-100} and \ref{fig:hard-tangent-pinned-pinned-kdbarhe-100} show the stability curves considering $\bar{K}_dh_e$ = 100. 

In the case of $\vb*{h}_e = h_e\vb*{e}_2$, the critical load significantly exceeds that of the classical elastica, as is evident from Eqns. \ref{eqn:hard-magnetic-elastica-cases}$_2$ and \ref{eqn:pure-elastica-equilibrium-equations}. On the other hand, the critical load becomes much lower when $\vb*{h}_e = -h_e\vb*{e}_2$. The reactions are related as 
\begin{equation}
	\bar{Q} = \bar{R}_e. \label{eqn:reaction-transverse-he-hard-magnet-2}
\end{equation}

}

\subsection{Limit as  ($\bar{K}_a\rightarrow 0$ \text{ and} $\bar{K}_d \rightarrow \infty$) or $\left(K_a at \ll  \frac{EI}{l} \ll K_d at \right)$}
\label{subsec:kd-infinity}

In this section, we explore the deformation of our planar ferromagnetic ribbons as the limit of $\bar{K}_a \rightarrow 0$ and $ \bar{K}_d \rightarrow \infty$ approaches infinity. {\color{black} This limit is also attained for a soft ferromagnetic ribbon ($K_a \approx 0$) with $E$, $a$, $l$, and $K_d$ are held constant, and as $t\rightarrow 0$. This represents a physically relevant limit for soft ferromagnetic  nano rods/ribbons used in MEMS devices \cite{Samy2023}}.

\begin{figure}[h!]
	\centering
	\includegraphics[width=0.4\linewidth]{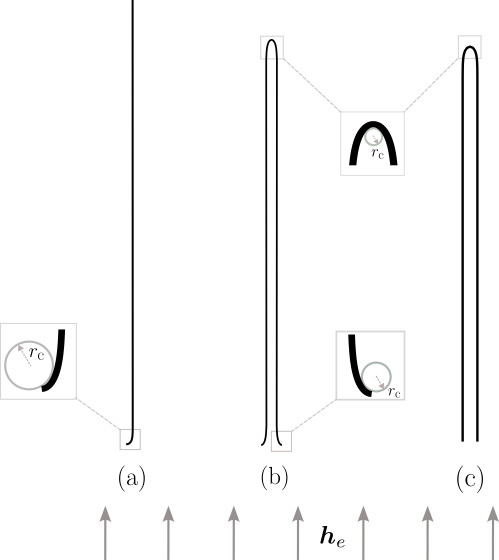}
    \caption{Deformed shapes of soft ferromagnetic ribbon for $\bar{K}_d = 10^4$ showing osculating circles at the curved bends: (a) fixed-free (b) fixed-fixed, and (c) pinned-pinned cases. The respective values of ($\bar{P}$,$u_3(\bar{s}=1)$) are (0.4$\pi^2$,0.99), ($6\pi^2$,0.96) and ($2\pi^2$,0.98).}
    \label{fig:asymptotic}
\end{figure}

As $\bar{K}_d \rightarrow \infty$, the magnetostatic energy is much larger than the elastic energy.
 In this regime, the magnetisation $\vb*{m}$ and the normal $\vb*{n}$ are  perpendicular almost all along the length of the ribbon, except in short intervals where this condition cannot be met due to either the imposed mechanical boundary condition or due to the isoperimetric constraint imposed due to  inextensibility. 
 These short intervals exhibit a curvature with a radius denoted as $r_{c}$, as shown in  Fig. \ref{fig:asymptotic}. An estimate of $r_{c} = (\kappa_{c}^{-1})$ can be obtained by balancing the elastic energy and demagnetisation energy, as outlined below:
 
\begin{equation}
	\begin{split}
		\text{Total energy}: &\quad \frac{EI}{2}\int_{0}^{l} \kappa^2(s) ds + K_d at \int_{0}^{l} (\vb*{m} \cdot \vb*{n})^2 ds \approx EI \int_{0}^{2\pi r} \frac{1}{r^2} ds + K_d at \int_{0}^{2\pi r} ds \\
		 &\qquad = \frac{EI \cdot 2\pi}{r} + K_d at \cdot 2 \pi r.
	\end{split}
\end{equation}
Optimising the total energy with respect to $r$, we obtain $r_{c} \sim \mathcal{O}\bigg( t\sqrt{\frac{E}{K_d}}\bigg)$. The deformed configurations in this regime of fixed-free, fixed-fixed, and pinned-pinned soft ferromagnetic ribbons are shown in Figure \ref{fig:asymptotic}.

A similar analysis for a hard ferromagnetic ribbon indicates that the radius of curvature, $r_{c}$, is approximately on the order of $\mathcal{O}\bigg( t\sqrt{\frac{E}{K_d h_{e}}}\bigg)$. This implies that the radius of curvature decreases as the external magnetic field is increased in the case of a hard ferromagnetic ribbon.

\section{Conclusions}\label{sec:conclusion}

In this paper, we have presented a novel model that combines ideas from Euler's elastica and continuum theory of micromagnetics  to predict the deformation of ferromagnetic ribbons. We have analysed the buckling of a planar inextensible ribbon subjected to a large but constant external magnetic field and a gradually applied quasi-static load. While maintaining fixed magnetisation, we investigate the influence of magnetisation on the deformation of the planar ferromagnetic ribbon. Exploring how deformation influences magnetisation is an  area for future research, although it presents a challenging problem. Our analysis commences by deriving and incorporating the magnetostatic energy of a curved structure into the total energy of the system. We derive the equilibrium equations and solve them numerically to obtain the equilibrium path as the load is increased. Further, we also determine the stability of the equilibrium solutions by casting the second variation of the total energy as a Sturm-Liouville eigenvalue problem, which is solved numerically. 

We examine ribbons composed of both hard and soft ferromagnetic properties, under a transversely  applied external magnetic field and various mechanical boundary conditions. We observe that the critical buckling load is tensile mainly for a soft ferromagnetic ribbon under various canonical boundary conditions. Interestingly, our findings reveal the presence of novel stable configurations in the case of a fixed-fixed setup with a transversally applied external magnetic field. We anticipate that these stable configurations can be observed through meticulously conducted experiments.

Our model is applicable within specific parameter ranges. For a typical ferromagnetic material such as iron (Fe) or nickel (Ni), the Young's modulus ($E$) and the magnetostatic energy density ($K_{d}$) are approximately $E\sim 10^{11}$ and $K_{d}\sim 10^5$, respectively. The aspect ratio, defined as the ratio of the width ($a$) to the thickness ($t$), can be determined by ensuring that $\bar\kappa<<1$, yielding $\nicefrac{a}{t}\sim$. The ratio of the thickness to the length of the ribbon is determined by the values of $\bar K_{d}$ derived from the conducted simulations. Our simulations reveal that in soft ferromagnets, the magnetic effect becomes significant when $\bar K_{d}\sim 100$. Hence, $\nicefrac{l}{t}\sim 10\sqrt{\nicefrac{E}{K_{d}}}$. Similarly, for hard ferromagnets, magnetic effects are observed when $\bar K_{d} h_{e}\sim 100$, resulting in: $\nicefrac{l}{t}\sim 10\sqrt{\nicefrac{E}{K_{d}h_{e}}}$. Furthermore, for a hard ferromagnet, the maximum external magnetic field is determined by ensuring the exchange energy remains non-negative, expressed as $\kappa_{c}<\kappa_{max}=\nicefrac{2}{t}$, consequently leading to a maximum external magnetic field $h_{e} < \sqrt{\nicefrac{E}{K_{d}}}$.

Our model is valid for the following range of parameters. The Young's modulus ($E$), the magnetostatic energy density ($K_{d}$) for a typical ferromagnetic material (Fe, Ni) are $E\sim 10^{11}$ and $K_{d}\sim 10^5$. The ratio of the width to the thickness can be evaluated by ensuring $\bar\kappa<<1$. That is we obtain $\nicefrac{a}{t}\sim$. The ratio of the thickness to the length of the ribbon is determined by the values of $\bar K_{d}$ obtained from the reported simulations. Finally, for a hard ferromagnet the maximum external magnetic field is evaluated by ensuring that the exchange energy is non negative. That is, $\kappa_{c}<\kappa_{max}=\nicefrac{2}{t}$, and hence the maximum external magnetic field $h_{e} < \sqrt{\nicefrac{E}{K_{d}}}$.   


Our analysis can be readily extended to understanding the deformation of planar ferromagnetic rods. The magnetostatic energy of planar ferromagnetic rod scales as $\mathcal{O}(K_{d}\pi r^2 l)$ \cite{Slastikov2011}. A balance of the elastic energy $\mathcal{O}(E \pi \big(\frac{r^4}{8l})\big)$, and the magnetostatic energy for ferromagnetic rods would suggest that our analysis remains valid for rods with an aspect ratio of $\mathcal{O}\bigg(\sqrt{\frac{E}{8Kd}}\bigg)$.

The stability analysis conducted in our study solely examines planar perturbations. It is widely recognized that, in the case of Euler's elastica, stability under planar perturbations does not guarantee stability of the equilibrium configurations under fully three-dimensional perturbations \cite{Maddocks1984}. Motivated by this study, we outline the following future research directions:
\begin{description}[font=$\bullet$~\normalfont\scshape\color{red!50!black}]
\item Explore stability under three-dimensional perturbations, and ascertain the potential role of the vertical reaction, $\bar{R}$, in this analysis.
\item Investigating the influence of twist and out-of-plane deformation on  ferromagnetic ribbons.
\item Design laboratory experiments to observe novel stable deformed configurations of ferromagnetic ribbon structures, as identified in our analysis.
\end{description}

\begin{acknowledgements}
We sincerely thank Professor R.D. James (University of Minnesota) for his valuable assistance in helping us understand the work of Moon and Pao and comparing it with our own. We would also like to thank Professor Kaushik Bhattacharya  (Caltech) for suggesting the calculations carried out in  section \ref{subsec:kd-infinity}. We thank Dr. Raghavendra Venkatraman (Courant Institute) for his valuable suggestions and insights on the draft of this paper. We also acknowledge Professor Ajeet Kumar (IIT, Delhi) for pointing out the validity of the Kirchhoff model. Finally, we would like to acknowledge the Indian Institute of Science Startup Grant and the Prime Minister's Research Fellowship (PMRF ID: 0201857) for providing financial support for this research.

\end{acknowledgements}

%
\section*{Conflict of interest}
The authors declare that they have no conflict of interest.

\bibliographystyle{spmpsci}      


\appendix

\section{Derivation of magnetostatic energy} \label{app:magnetostatic}
In this section, we present the calculation of the leading order demag. energy of a deformed planar ferromagnetic ribbon. The magnetostatic energy is evaluated by solving Maxwell's equations of magnetostatics
\begin{equation}
	\begin{split}
		\curl{\vb*{h}_m(\vb*{x})} &= 0, \\
		\div(\vb*{h}_m + \vb*{m}(\vb*{x})) &= 0.
	\end{split}
\end{equation}
Applying Fourier transform to the above equations imply
\begin{align}
	\vb*{\xi} \cross \hat{\vb*{h}}_m &= 0, \label{eqn:maxwell-equation-fourier-1} \\
	\vb*{\xi} \cdot (\hat{\vb*{h}}_m + \hat{\vb*{m}}) &= 0, \label{eqn:maxwell-equation-fourier-2}
\end{align}
where $\hat{\vb*{f}}(\vb*{\xi}) := \int_{\mathbb{R}^3} \vb*{f}(\vb*{x}) \exp(-i 2\pi \vb*{x}\cdot\vb*{\xi}) d\vb*{x}$.
Eqn. \ref{eqn:maxwell-equation-fourier-1} implies $\hat{\vb*{h}}_m \parallel \vb*{\xi}$ and hence,
\begin{equation}
	\hat{\vb*{h}}_m  = (\hat{\vb*{h}}_m  \cdot \hat{\vb*{\xi}})\hat{\vb*{\xi}} \text{ where } \hat{\vb*{\xi}} = \frac{\vb*{\xi}}{\abs{\vb*{\xi}}}.
\end{equation}
Eqn. \ref{eqn:maxwell-equation-fourier-2} implies that $\vb*{\xi}\cdot\hat{\vb*{h}}_m = - \vb*{\xi} \cdot \hat{\vb*{m}}$ and hence $\hat{\vb*{h}}_m = - (\hat{\vb*{\xi}}\cdot\hat{\vb*{m}})\hat{\vb*{\xi}}$.
Using Planchevel's identity, we can then write the magnetostatic energy as follows
\begin{equation}
	\int_{\mathbb{R}^3} \abs{\vb*{h}_m}^2 d\vb*{x} = 	\int_{\mathbb{R}^3} \abs{\hat{\vb*{h}}_m}^2 d\vb*{\xi} = \int_{\mathbb{R}^3} (\hat{\vb*{\xi}}\cdot\hat{\vb*{m}})^2 d\vb*{\xi}  = \int_{\mathbb{R}^3} \frac{(\vb*{\xi}\cdot\hat{\vb*{m}})^2}{\abs{\vb*{\xi}}^2} d\vb*{\xi} = \int_{\mathbb{R}^3} \frac{\abs{\widehat{\div\vb*{m}}}^2}{\abs{\vb*{\xi}}^2} d\vb*{\xi}.
\end{equation}
We carry out the above integration in the material frame $(\vb*{d}_1(s),\vb*{d}_2(s),\vb*{d}_3(s))$ and $\vb*{x} \mapsto \vb*{r}(s) + \tilde{a} \vb*{d}_1(s) + \tilde{t} \vb*{d}_2(s)$.
\noindent The Fourier transform in the material frame is given as follows:
\begin{equation}
	\mathcal{F}[\div\vb*{m}](\vb*{\xi}) = \widehat{\div\vb*{m}}(\vb*{\xi}) = \int_{\Omega} (\div\vb*{m}) \exp(-i2\pi\vb*{x}\cdot\vb*{\xi}) d\vb*{x} = \int_{\Omega} (\div\vb*{m}) \exp(-i2\pi(\vb*{r}(s) + \tilde{a} \vb*{d}_1(s) + \tilde{t} \vb*{d}_2(s))\cdot \vb*{\xi}) J ds d\tilde{t} d\tilde{a},
	\label{eqn:div-m-fourier}
\end{equation}
here, $\mathcal{F}(v)$ denotes Fourier transform of $v$ and $J = \abs{\frac{\partial(x_1,x_2,x_3)}{\partial(s,\tilde{a},\tilde{t})}}$ is the Jacobian associated with the change of variables.
\noindent The divergence of $\vb*{m}$ is invariant upon coordinate transformation and is expressed as follows in the material frame:
\begin{equation}
	\div\vb*{m} = \frac{1}{h_s h_{\tilde{a}} h_{\tilde{t}}}  \left\{\pdv{(h_{\tilde{a}} h_{\tilde{t}} m_{d_3})}{s} + \pdv{(h_{\tilde{t}}h_s m_{d_1})}{\tilde{a}} +  \pdv{(h_s h_{\tilde{a}} m_{d_2})}{\tilde{t}}  \right\},
\end{equation}
here, $h_s = \abs{\pdv{\vb*{x}}{s}}, h_{\tilde{a}} = \abs{\pdv{\vb*{x}}{\tilde{a}}}$ and $h_{\tilde{t}} = \abs{\pdv{\vb*{x}}{\tilde{t}}}$.\\
Note:
\begin{itemize}
	\item $\pdv{\vb*{x}}{s} = \vb*{r}'(s) + \tilde{a} \vb*{d}_1'(s) + \tilde{t} \vb*{d}_2'(s) = \vb*{d}_3(s) + \tilde{t} \kappa \vb*{d}_3(s) = (1+\tilde{t}\kappa) \vb*{d}_3(s) \implies h_s = (1 + \tilde{t}\kappa)$,
	
	\item $\pdv{\vb*{x}}{\tilde{a}} = \vb*{d}_1(s) \implies h_{\tilde{a}} = 1$,
	
	\item $\pdv{\vb*{x}}{\tilde{t}} = \vb*{d}_2(s) \implies h_{\tilde{t}} = 1$.   	
\end{itemize} 
Furthermore, $J = h_s h_{\tilde{a}} h_{\tilde{t}}$, and hence
\begin{align}
	\div\vb*{m} &= \frac{1}{J}\left\{\pdv{m_{d_3}}{s} + (1 + \tilde{t}\kappa)\pdv{m_{d_1}}{\tilde{a}} + (1 + \tilde{t}\kappa)\pdv{m_{d_2}}{\tilde{t}} +  \kappa m_{d_2} \right\} \notag \\
	&= \frac{1}{J} \left\{(\partial_{\tilde{a}} m_{d_1} + \partial_{\tilde{t}} m_{d_2} + \partial_s m_{d_3} ) + \kappa (\tilde{t} \partial_{\tilde{a}} m_{d_1} + \tilde{t}\partial_{\tilde{t}} m_{d_2} + m_{d_2})\right\}.
\end{align}
Substituting the above expression for $\div\vb*{m}$ in Eqn. \ref{eqn:div-m-fourier} can now be written as follows:
\begin{equation}
	\mathcal{F}[\div\vb*{m}](\vb*{\xi}) = \int_{\Omega} \frac{1}{J}\left\{(\partial_{\tilde{a}} m_{d_1} + \partial_{\tilde{t}} m_{d_2} + \partial_s m_{d_3} ) + \kappa (\tilde{t} \partial_{\tilde{a}} m_{d_1} + \tilde{t}\partial_{\tilde{t}} m_{d_2} + m_{d_2})\right\} \exp(-i2\pi(\vb*{r}(s) + \tilde{a} \vb*{d}_1(s) + \tilde{t} \vb*{d}_2(s))\cdot \vb*{\xi}) J ds d\tilde{t} d\tilde{a}.
\end{equation}
Since $\vb*{r}(0) = \vb*{0}$ and $\vb*{r}(s)\cdot \vb*{\xi} = \int_{0}^{s} \vb*{d}_3(s')ds' \cdot \vb*{\xi} = s \xi_{d_3}$ and therefore, $(\vb*{r}(s) + \tilde{a} \vb*{d}_1(s) + \tilde{t} \vb*{d}_2(s))\cdot\vb*{\xi} = \tilde{a}\xi_{d_1} + \tilde{t}\xi_{d_2} + s\xi_{d_3}$.
\noindent The divergence of $\vb*{m}$ in the material frame is written as a sum of these six integrals:
\begin{equation}
	\mathcal{F}[\div\vb*{m}](\vb*{\xi}) = I_1 + I_2 + I_3 + I_4 + I_5 + I_6,
\end{equation}
where
\begin{equation}
	\begin{split}
		I_1 &= \mathcal{F}[\partial_{\tilde{a}} m_{d_1}](\vb*{\xi}_d) = \xi_{d_1} F_0(a,\xi_{d_1}) F_0(t,\xi_{d_2}) \mathcal{F}[m_{d_1} \chi_{(0,l)} (s)](\xi_{d_3})  \quad \text{(even in $\xi_{d_2}$)}, \\
		I_2 &= 	\mathcal{F}[\partial_{\tilde{t}} m_{d_2}](\vb*{\xi}_d)= \xi_{d_2} F_0(a,\xi_{d_1}) F_0(t,\xi_{d_2}) \mathcal{F}[m_{d_2} \chi_{(0,l)} (s)](\xi_{d_3})\quad \text{(odd in $\xi_{d_2}$)}, \\
		I_3 &= 	\mathcal{F}[\partial_{s} m_{d_3}](\vb*{\xi}_d)=  F_0(a,\xi_{d_1}) F_0(t,\xi_{d_2}) \mathcal{F}[\partial_{s} m_{d_3} \chi_{(0,l)} (s)](\xi_{d_3}) \quad \text{(even in $\xi_{d_2}$)}, \\
		I_4 &= 	\mathcal{F}[\kappa\tilde{t}\partial_{\tilde{a}} m_{d_1}](\vb*{\xi}_d)=  \xi_{d_1} F_0(a,\xi_{d_1}) F_1(t,\xi_{d_2}) \mathcal{F}[\kappa m_{d_1} \chi_{(0,l)} (s)](\xi_{d_3}) \quad \text{(odd in $\xi_{d_2}$)},  \\
		I_5 &= 	\mathcal{F}[\kappa\tilde{t}\partial_{\tilde{t}} m_{d_2}](\vb*{\xi}_d)=  F_0(a,\xi_{d_1})\xi_{d_2} F_1(t,\xi_{d_2}) \mathcal{F}[\kappa m_{d_2} \chi_{(0,l)} (s)](\xi_{d_3})  \quad \text{(even in $\xi_{d_2}$)}, \\
		I_6 &= 	\mathcal{F}[\kappa m_{d_2}](\vb*{\xi}_d)=  F_0(a,\xi_{d_1}) F_1(t,\xi_{d_2}) \mathcal{F}[\kappa m_{d_2} \chi_{(0,l)} (s)](\xi_{d_3})  \quad \text{(even in $\xi_{d_2}$)}, 
	\end{split}
\end{equation}
where $F_0(v,\eta)$ and $F_1(v,\eta)$ are defined as follows:
\begin{equation*}
	F_0(v,\eta) = \int_{-v/2}^{v/2} \exp(-i2\pi\eta v') dv' \quad \text{ and } \quad F_1(v,\eta) = \int_{-v/2}^{v/2} v' \exp(-i2\pi\eta v') dv'.
\end{equation*}
The magnetostatic energy can now be expressed in terms of $I_1,I_2,I_3,I_4,I_5,I_6$ as follows:
\begin{multline}
	\int_{\mathbb{R}^3} \frac{\abs{\widehat{\div\vb*{m}}(\vb*{\xi}_d)}^2}{\abs{\vb*{\xi}_d}^2} d\vb*{\xi}_d = \int_{\mathbb{R}^3} \frac{(I_1 + I_2 + I_3 + I_4 + I_5 + I_6)^2}{\abs{\vb*{\xi}_d}^2} \\ = \underbrace{\int \frac{I_1^2}{\abs{\vb*{\xi}_d}^2}}_{\mathcal{O}(t^2)} + \underbrace{\int  \frac{I_2^2}{\abs{\vb*{\xi}_d}^2}}_{\mathcal{O}(t)} + \underbrace{\int  \frac{I_3^2}{\abs{\vb*{\xi}_d}^2}}_{\mathcal{O}(t^2)} + \underbrace{\int \frac{I_4^2}{\abs{\vb*{\xi}_d}^2}}_{\mathcal{O}(t^5)} + \underbrace{\int \frac{I_5^2}{\abs{\vb*{\xi}_d}^2} }_{\mathcal{O}(t^3)} + \underbrace{\int \frac{I_6^2}{\abs{\vb*{\xi}_d}^2}}_{\mathcal{O}(t^3)} \\
	+ \underbrace{2\int \frac{I_1 I_3}{\abs{\vb*{\xi}_d}^2}}_{\mathcal{O}(t^2)} +  \underbrace{2\int \frac{I_1 I_5}{\abs{\vb*{\xi}_d}^2}}_{\mathcal{O}(t^3)} +  \underbrace{2\int \frac{I_1 I_6}{\abs{\vb*{\xi}_d}^2}}_{\mathcal{O}(t^2)} +  \underbrace{2\int \frac{I_2 I_4}{\abs{\vb*{\xi}_d}^2}}_{\mathcal{O}(t^3)} +  \underbrace{2\int \frac{I_3 I_5}{\abs{\vb*{\xi}_d}^2}}_{\mathcal{O}(t^3)} +  \underbrace{2\int \frac{I_3 I_6}{\abs{\vb*{\xi}_d}^2}  }_{\mathcal{O}(t^2)} +  \underbrace{2\int \frac{I_5 I_6}{\abs{\vb*{\xi}_d}^2}}_{\mathcal{O}(t^3)}.
\end{multline}
The remaining product terms are odd functions in $\xi_{d_2}$ and hence they integrate out to zero. Therefore we have the leading order term for the magnetostatic energy as follows:
\begin{equation}
	\int_{\mathbb{R}^3} \frac{\abs{\widehat{\div\vb*{m}}(\vb*{\xi}_d)}^2}{\abs{\vb*{\xi}_d}^2} d\vb*{\xi}_d =  	\int_{\mathbb{R}^3} \frac{I_2^2}{\abs{\vb*{\xi}_d}^2} d\vb*{\xi}_d + \mathcal{O}(t^2).
\end{equation}
Now,
\begin{equation}
	\lim_{t\to 0} \int_{\mathbb{R}^3} \frac{I_2^2}{\abs{\vb*{\xi}_d}^2} d\vb*{\xi}_d = at\int_{\mathbb{R}} \abs{\widehat{m_{d_2}}(\xi_{d_3}) }^{2}d\xi_{d_3} + \mathcal{O}(t^2) = at \int_{s=0}^l (m_{d_2})^2 ds + \mathcal{O}(t^2).
\end{equation}
Hence, we have 
\begin{equation}
	\int_{\mathbb{R}^3} \abs{\vb*{h}_m}^2 d\vb*{x} = at \int_{s=0}^l (m_{d_2})^2 ds + \mathcal{O}(t^2).
\end{equation}

\section{Derivation of exchange energy} \label{app:exchange}
The centerline representation is given by
\begin{equation}
	\vb*{x}(s,\tilde{a},\tilde{t}) = \vb*{r}(s) + \tilde{a}\vb*{d}_1(s) + \tilde{t}\vb*{d}_2(s). \label{eqn:centerline-exchange}
\end{equation}
We define $\{\vb*{g}_a, \vb*{g}_t, \vb*{g}_s\}$ is the natural basis for the mapping defined in Eqn. \ref{eqn:centerline-exchange}, are expressed as
\begin{equation}
		\vb*{g}_a(s) = \pdv{\vb*{x}}{\tilde{a}}, \qquad
		\vb*{g}_t(s) = \pdv{\vb*{x}}{\tilde{t}}, \qquad
		\vb*{g}_s(s) = \pdv{\vb*{x}}{s}.
\end{equation}
Evidently, the basis $\{\vb*{g}_a, \vb*{g}_t, \vb*{g}_s\}$ is orthogonal but not orthonormal.

Let $\{\vb*{g}^a, \vb*{g}^t, \vb*{g}^s\}$ be the reciprocal basis, that is, $\vb*{g}^i\cdot\vb*{g}_j = \delta_j^i$, $i,j\in \{a,t,s\}$. We express the magnetisation vector as 
\begin{equation}
	\vb*{m} = m^a\vb*{g}_a + m^t\vb*{g}_t + m^s\vb*{g}_s, \text{ where } m^i = \vb*{m}\cdot\vb*{g}^i~ (i=1,2,3),
\end{equation}
and its gradient as $\grad\vb*{m} = \grad_i m^k \vb*{g}_k\otimes \vb*{g}^i$, here, $\grad_i m^k$ are the covariant derivatives of $\vb*{m}$. Now, $\grad_i m^k = \left(m_{,i}^k + \Gamma_{ij}^{k}m^j\right)$, where $\Gamma_{ij}^k$ are the Christoffel symbols, and is given by $\Gamma_{ij}^k := \vb*{g}^k\cdot\vb*{g}_{i,j}$.
Out of the 27 Christoffel symbols, only the following three are non-zero:
\begin{equation}
 \begin{split}
 	\Gamma_{ts}^s &= \vb*{g}^s\cdot\vb*{g}_{t,s} = \frac{\vb*{d}_3(s)}{(1+\kappa t)}\cdot\vb*{g}_{t,s} = \frac{\vb*{d}_3\cdot\vb*{d}_2'(s)}{1+\kappa t} = \kappa \frac{\vb*{d}_3(s)\cdot\vb*{d}_3(s)}{(1 + \kappa t)} = \frac{\kappa}{(1 + \kappa \tilde{t})} = \Gamma_{st}^s, \\
 	\Gamma_{ss}^t &= \vb*{g}^t\cdot\vb*{g}_{s,s} = \vb*{d}_2(s)\cdot (1+\kappa \tilde{t})\vb*{d}_3'(s) = -\kappa(1+\kappa \tilde{t})\vb*{d}_2(s)\cdot\vb*{d}_2(s) = -\kappa(1+\kappa \tilde{t}).
 \end{split}
\end{equation}
Hence, 
\begin{equation}
	\begin{split}
		\grad\vb*{m} &= \left(\pdv{m^a}{\tilde{a}}\right)\vb*{g}_a\vb*{g}^a + \left(\pdv{m^a}{\tilde{t}}\right)\vb*{g}_a\vb*{g}^t + \left(\pdv{m^a}{s}\right)\vb*{g}_a\vb*{g}^s \\
		&+ \left(\pdv{m^t}{\tilde{a}}\right)\vb*{g}_t\vb*{g}^a + \left(\pdv{m^t}{\tilde{t}}\right)\vb*{g}_t\vb*{g}^t + \left(\pdv{m^t}{s} - \kappa(1 + \kappa t)m^s\right) \vb*{g}_t\vb*{g}^s \\
		&+ \left(\pdv{m^s}{\tilde{a}}\right)\vb*{g}_s\vb*{g}^a + \left(\pdv{m^s}{\tilde{t}} + \frac{\kappa}{(1 + \kappa \tilde{t})} m^s\right)\vb*{g}_s\vb*{g}^t + \left(\pdv{m^s}{s} + \frac{\kappa m^t}{(1+\kappa \tilde{t})}\right)\vb*{g}_s\vb*{g}^s.
	\end{split}
\end{equation}
We can write the above expression in terms of orthonormal basis of $(\vb*{d}_1,\vb*{d}_2,\vb*{d}_3)$ using the physical components of $\vb*{m} (m_{d_1}, m_{d_2}, m_{d_3})$ as follows:
\begin{equation}
  \begin{split}
  	\grad\vb*{m} &= \left(\pdv{m_{d_1}}{\tilde{a}}\right)\vb*{d}_1\otimes\vb*{d}_1 + \left(\pdv{m_{d_1}}{\tilde{t}}\right)\vb*{d}_1\otimes\vb*{d}_2 + \frac{1}{(1 + \kappa \tilde{t})}\pdv{m_{d_1}}{s} \vb*{d}_1\cdot\\vb*{d}_3 \\
  	&+ \left(\pdv{m_{d_2}}{\tilde{a}}\right)\vb*{d}_1\otimes\vb*{d}_2 + \left(\pdv{m_{d_2}}{\tilde{t}}\right)\vb*{d}_2\otimes\vb*{d}_2 + \frac{1}{(1+\kappa\tilde{t})}\left(\pdv{m_{d_2}}{s}- \kappa m_{d_3}\right)\vb*{d}_2\otimes\vb*{d}_3 \\
  	&+ \left(\pdv{m_{d_3}}{\tilde{a}}\right)\vb*{d}_3\otimes\vb*{d}_1 + \left(\pdv{m_{d_3}}{\tilde{t}}\right)\vb*{d}_3\otimes\vb*{d}_2 + \frac{1}{(1+\kappa \tilde{t})}\left(\pdv{m_{d_3}}{s} + \kappa m_{d_2}\right)\vb*{d}_3\otimes\vb*{d}_3,
  \end{split}
\end{equation}
where $\vb*{g}^s = \frac{1}{(1+\kappa \tilde{t})}\vb*{d}_3(s),~\vb*{g}_s = \left(1 + \kappa \tilde{t}\right)\vb*{d}_3(s)$. We note that $m_{d_1} = \vb*{m}\cdot\vb*{d}_1(s) = \vb*{m}\cdot\vb*{g}^a,~m_{d_2} =\vb*{m}\cdot\vb*{d}_2(s) = \vb*{m}\cdot\vb*{g}^t = m^t,~m_{d_3} = \vb*{m}\cdot\vb*{d}_3(s) = \vb*{m}\cdot (1+\kappa\tilde{t})\vb*{g}^s = (1+\kappa\tilde{t})m^s$. Also, 
\begin{equation}
 \begin{split}
 	\left(\pdv{m^t}{\tilde{t}}\right)\vb*{g}_t\vb*{g}^t + \left(\pdv{m^t}{s} - \kappa(1 + \kappa t)m^s\right) \vb*{g}_t\vb*{g}^s &= \left(\pdv{m_{d_2}}{s} - \kappa m_{d_3}\right)\frac{1}{(1 + \kappa\tilde{t})}\vb*{d}_2\otimes\vb*{d}_3    \\
 	\left(\pdv{m^s}{\tilde{t}} + \frac{\kappa}{(1 + \kappa \tilde{t})} m^s\right)\vb*{g}_s\vb*{g}^t  &= \left(\pdv{}{\tilde{t}}\left(\frac{m_{d_3}}{(1+\kappa \tilde{t})}\right) + \frac{\kappa}{(1+\kappa \tilde{t})^2}m_{d_3}\right)(1 +\kappa \tilde{t})\vb*{d}_3\otimes\vb*{d}_2.
 \end{split}
\end{equation}
Hence,
\begin{equation}
	[\grad \vb*{m}]_{\vb*{d}_i} = 
	\begin{bmatrix}
	\pdv{m_{d_1}}{\tilde{a}} & \pdv{m_{d_1}}{\tilde{t}} & \frac{1}{(1+\kappa \tilde{t})}\pdv{m_{d_1}}{s} \\
	\pdv{m_{d_2}}{\tilde{a}} & \pdv{m_{d_2}}{\tilde{t}} & \frac{1}{(1+\kappa\tilde{t})}\left(\pdv{m_{d_2}}{s}-\kappa m_{d_3}\right) \\
	\frac{1}{(1+\kappa\tilde{t})}\pdv{m_{d_3}}{\tilde{a}} & \pdv{m_{d_3}}{\tilde{t}} & \frac{1}{(1+\kappa \tilde{t})}\left(\pdv{m_{d_3}}{s} + \kappa m_{d_2}\right)	
	\end{bmatrix}.
\end{equation}
The exchange energy is thus expressed as
\begin{equation}
	\mathcal{E}_{ex} = A\int_{\Omega} \abs{\grad\vb*{m}}^2 d\tilde{t}d\tilde{a}ds.
\end{equation}

\section{Numerical discretization of equilibrium equations} \label{app:equilibrium}
We demonstrate the discretization process for Eqn. \ref{eqn:soft-magnetic-elastica-equilibrium-equations} invoking fixed-fixed boundary conditions. We discretize the domain into $N+1$ nodal points as $\bar{s}_i = ih; i=0,1,\dots,N+1$ where $h = \frac{1}{N+1}$. We utilize second-order accurate central difference scheme to approximate the second-order derivative in  Eqn. \ref{eqn:soft-magnetic-elastica-equilibrium-equations}. The dicrete system of equations alongwith the boundary conditions is 
\begin{equation}
	\begin{aligned}
		\frac{\theta_{i-1} - 2\theta_i + \theta_{i+1}}{h^2} + \bar{P} \sin\theta_i + \bar{K}_d\sin 2\theta_i + \bar{R} \cos\theta_i &= 0; \qquad i=1,2,\dots, N,\\
		\theta_0 = 0, \theta_{N+1} = 0, &
	\end{aligned}
\end{equation}
where $\theta_i$ denotes the numerical counterpart of $\theta(\bar{s} = \bar{s}_i)$. \\
Upon incorporating the boundary conditions, we assemble the discrete equations to form the following non-linear system
\begin{align}
	\underbrace{\frac{1}{h^2} 
	\begin{bmatrix}
		-2 & 1 & 0 &  &  &  &  \\
		1 & -2 & 1 &  &  &  &  \\
		0 & \ddots & \ddots & \ddots &  &  &  \\
		& \ddots & 1 & -2 & 1 & \ddots &  \\
		&  &  & \ddots & \ddots & \ddots &  \\
		&  &  &  & 1 & -2 & 1 \\
		&  &  &  & 0 & 1 & -2 
	\end{bmatrix}}_{\vb*{K}}~ \underbrace{\begin{pmatrix}
		\theta_1 \\
		\theta_2 \\
		\vdots \\
		\theta_i \\
		\vdots \\
		\theta_{N-1} \\
		\theta_N 
	\end{pmatrix}}_{\vb*{\theta}} + 
	\bar{P}\begin{pmatrix}
		\sin\theta_1 \\
		\sin\theta_2 \\
		\vdots \\
		\sin\theta_i \\
		\vdots \\
		\sin\theta_{N-1} \\
		\sin\theta_N 
	\end{pmatrix} + 
	\bar{K}_d \begin{pmatrix}
		\sin 2\theta_1 \\
		\sin 2\theta_2 \\
		\vdots \\
		\sin 2\theta_i \\
		\vdots \\
		\sin 2\theta_{N-1} \\
		\sin 2\theta_N 
	\end{pmatrix} + 
	\bar{R} \begin{pmatrix}
		\cos\theta_1 \\
		\cos\theta_2 \\
		\vdots \\
		\cos\theta_i \\
		\vdots \\
		\cos\theta_{N-1} \\
		\cos\theta_N 
	\end{pmatrix}
	= \begin{pmatrix}
		0 \\
		0 \\
		\vdots \\
		0 \\
		\vdots \\
		0 \\
		0 
	\end{pmatrix}.
\end{align}
In compact form, the discretized system can be written for all three cases (for fixed-free case where $\bar{R}=0$) as 
\begin{align}
	\vb*{K\theta} + \bar{P}\sin\vb*{\theta} + \bar{K}_d \sin 2\vb*{\theta} + \bar{R} \cos\vb*{\theta} = \vb*{0},
\end{align}
where $\vb*{K}$ is the stiffness matrix, and $\vb*{\theta}$ is the vector containing the inclination angles at discrete points. \\
Upon discretizing the integral (or isoperimetric) constraint using trapezoidal rule, we have
\begin{align}
	\int_{0}^{1} \sin \theta(\bar{s}) d\bar{s} = 0 &\implies \frac{h}{2} [\underbrace{\sin\theta_0}_{0} + 2(\sin\theta_1 + \dots + \sin\theta_N) + \underbrace{\sin\theta_{N+1}}_{0}] = 0, \\
	&\implies \sin\theta_1 + \dots + \sin\theta_N = 0.
\end{align}
Forming a combined nonlinear set of equations
\begin{align}
	\begin{pmatrix}
		\vb*{K\theta} + \bar{P}\sin\vb*{\theta} + \bar{K}_d \sin 2\vb*{\theta} + \bar{R} \cos\vb*{\theta} \\
		\sin\theta_1 + \dots +  \sin\theta_N
	\end{pmatrix} &=
	\begin{pmatrix}
		\vb*{0} \\
		0
	\end{pmatrix},  \\
	\implies \vb*{f}(\vb*{\theta}, \bar{R}, \bar{P}) = \vb*{0}.
\end{align}

\section{Bifurcation analysis} \label{app:bifurcation}
We briefly describe the bifurcation analysis for planar deformation of ribbons. We begin with taking the second variation of the total magnetoelastic energy functional, followed by Fourier expansion for kinematically-compatible variations, we establish stability criterion for deformed configuration under various loading scenarios and end conditions. The numerical procedure to implement this criterion has been explained in Section \ref{sec:bifurcation-analysis}.
\subsection{Derivation of stability condition}
Let us non-dimensionalize the energy functional for planar-deformed magnetoelastic rods and ribbons. The external magnetic field is applied along the Cartesian $\vb*{e}_2$-axis, that is, $\vb*{h}_e = h_e \vb*{e}_2$  such that $\vb*{m}(s) = \vb*{e}_2$. Upon substituting the values in Eqn. \ref{eqn:elastica-energy-functional-augmented-1} and non-dimensionalizing the energy functionals appropriately, we have
\begin{equation}
	\mathcal{\overline{E}}  (\theta) =  \frac{1}{2} \int_{0}^{1} (\theta'(\bar{s}))^2 d\bar{s} + \bar{K}_d \int_{0}^{1} \cos^2\theta(\bar{s}) at d\bar{s} - \underbrace{2 \bar{K}_d h_e at \int_{0}^{1} 1 \cdot d\bar{s}}_{\text{constant}} - \bar{P} \bigg(1 - \int_{0}^{1} \cos\theta(\bar{s}) d\bar{s}\bigg) - \bar{R} \int_{0}^{1} \sin\theta(\bar{s}) d\bar{s}.
\end{equation}
Introducing first order perturbation to the assumed extremum $\theta$ as $\hat{\theta}(\bar{s}) = \theta (\bar{s}) + \epsilon \eta (\bar{s})$ where $\eta (\bar{s})$ is a kinematically admissible planar variation and $\epsilon$ is a small parameter. Now,
\begin{equation}
	\mathcal{\overline{E}}  (\theta + \epsilon \eta) =  \frac{1}{2} \int_{0}^{1} (\theta' + \epsilon \eta')^2 d\bar{s} + \bar{K}_d \int_{0}^{1} \cos^2 (\theta + \epsilon \eta) at d\bar{s} + \bar{P} \int_{0}^{1} \cos (\theta + \epsilon \eta) d\bar{s} - \bar{R} \int_{0}^{1} \sin (\theta + \epsilon \eta ) d\bar{s} + \text{constant}.
\end{equation}
Differentiating the above with respect to $\epsilon$ twice and putting $\epsilon = 0$ results in the second variational of $\mathcal{\overline{E}} (\theta)$ 
\begin{equation}
	\delta^2\mathcal{\overline{E}}  (\theta) = \dv[2]{\mathcal{\overline{E}}  (\theta + \epsilon \eta)}{\epsilon}\bigg\rvert_{\epsilon=0} = \int_{0}^{1} \eta'^2 d\bar{s} - 2\bar{K}_d\int_{0}^{1} \cos 2\theta \eta^2 d\bar{s} - \bar{P} \int_{0}^{1} \cos \theta \eta^2 d\bar{s} + \bar{R}\int_{0}^{1} \sin \theta \eta^2 d\bar{s}.
\end{equation}
Integrating by parts the first term,
\begin{align}
	\delta^2\mathcal{\overline{E}} (\theta) &= \underbrace{\eta'\eta\bigg\rvert_{0}^{1}}_{=0} - \int_{0}^{1} \eta''\eta d\bar{s} - 2\bar{K}_d\int_{0}^{1} \cos 2\theta \eta^2 d\bar{s} - \bar{P} \int_{0}^{1} \cos \theta \eta^2 d\bar{s} + \bar{R}\int_{0}^{1} \sin \theta \eta^2 d\bar{s}, \\
	\implies \delta^2\mathcal{\overline{E}} (\theta) &= - \int_{0}^{1} [\eta'' + 2\bar{K}_d\cos 2\theta \eta + \bar{P} \cos \theta \eta - \bar{R}\sin \theta \eta] \eta d\bar{s}, \label{eqn:second-variational-derivative-appendix}
\end{align}
for all kinematically admissible functions $\eta(\bar{s})$. The stability criterion requires that
\begin{align}
	\delta^2\mathcal{\overline{E}}  (\theta) \begin{cases}
		> 0 & \text{ stable} \\
		< 0 & \text{ unstable}.
	\end{cases}
\end{align}
Introducing first variation to the integral constraint $\int_{0}^{1} \sin\theta(\bar{s})d\bar{s} = 0$ results in 
\begin{align}
	\int_{0}^{1} \cos \theta(\bar{s}) \eta(\bar{s}) d\bar{s} = 0. \label{eqn:constraints-first-variation-appendix}
\end{align}

\subsection{Construction of Sturm-Liouville problem} A meticulous study of the second variation $\delta^2\mathcal{\overline{E}}  (\theta)$ necessitates the construction of the following Sturm-Liouville problem whose non-trivial solutions are $\phi_n (\bar{s})$:
\begin{align}
	\phi_n''(\bar{s}) + \lambda_n (2\bar{K}_d\cos 2\theta + \bar{P} \cos \theta - \bar{R} \sin \theta) \phi_n (\bar{s}) = C_{R_n} \cos \theta (\bar{s}), \label{eqn:sturm-liouville-problem-appendix}
\end{align}
where $\lambda_n$ are the eigenvalues, $\phi_n$ the corresponding eigenmodes of Eqn. \ref{eqn:sturm-liouville-problem-appendix} and $L(\bar{s}) = (2\bar{K}_d\cos 2\theta(\bar{s}) + \bar{P} \cos \theta(\bar{s}) - \bar{R} \sin \theta(\bar{s}))$ denotes the weight function. $C_{R_n}$ is meant to enforce the isoperimetric constraint (Eqn. \ref{eqn:constraints-first-variation-appendix}). The conditions on $\phi_n(\bar{s})$ are
\begin{itemize}
	\item fixed-fixed case: $\phi_n(0) = \phi_n(1) = 0$ and $\int_{0}^{1} \cos \theta(\bar{s}) \phi_n(\bar{s}) d\bar{s} = 0$,
	\item pinned-pinned case: $\phi_n'(0) = \phi_n'(1) = 0$ and $\int_{0}^{1} \cos \theta(\bar{s}) \phi_n(\bar{s}) d\bar{s} = 0$,
	\item fixed-free case:  $\phi_n(0) = \phi_n'(1) = 0$.
\end{itemize}
Multiplying both sides of Eqn. \ref{eqn:sturm-liouville-problem-appendix} by $\phi_n$, integrating over the domain and using the conditions stipulated on $\phi_n(\bar{s})$, we get
\begin{align}
	\int_{0}^{1} [\phi_n''(\bar{s}) + \lambda_n L(\bar{s})\phi_n(\bar{s})]\phi_n(\bar{s}) d\bar{s}  &= C_{R_n} \underbrace{\int_{0}^{1}\cos \theta (\bar{s})\phi_n(\bar{s}) d\bar{s}}_{0}, \notag \\
	\implies \lambda_n  \int_{0}^{1} L(\bar{s}) \phi_n^2(\bar{s})d\bar{s} &= \int_{0}^{1} \phi_n'^2(\bar{s})d\bar{s}.
\end{align}
We multiply Eqn. \ref{eqn:sturm-liouville-problem-appendix} by $\phi_m$ and integrate over the domain to get
\begin{align}
	\int_{0}^{1} [\phi_n''(\bar{s}) + \lambda_n L(\bar{s}) \phi_n (\bar{s})]\phi_m d\bar{s}  &= C_{R_n} \underbrace{\int_{0}^{1}\cos \theta (\bar{s})\phi_m(\bar{s}) d\bar{s}}_{0}, \notag \\
	\implies -  \int_{0}^{1} \phi_n'(\bar{s})\phi_m'(\bar{s}) d\bar{s} + \lambda_n  \int_{0}^{1} L(\bar{s}) \phi_n(\bar{s})\phi_m(\bar{s})d\bar{s} &= 0.
\end{align}
Similarly, consider Eqn. \ref{eqn:sturm-liouville-problem} for $\phi_m$, multiply it by $\phi_n$ and integrate by parts to get
\begin{align}
	-  \int_{0}^{1} \phi_m'(\bar{s})\phi_n'(\bar{s}) d\bar{s} + \lambda_m  \int_{0}^{1} L(\bar{s}) \phi_n(\bar{s})\phi_m(\bar{s})d\bar{s} &= 0.
\end{align}
Subtracting the second equation from the first results in
\begin{align}
	(\lambda_n - \lambda_m) \int_{0}^{1} L(\bar{s}) \phi_n(\bar{s})\phi_m(\bar{s})d\bar{s} &= 0.
\end{align}
For $n\neq m$, $\lambda_n \neq \lambda_m$, we obtain the orthogonality condition as
\begin{align}
	\int_{0}^{1} L(\bar{s}) \phi_n(\bar{s})\phi_m(\bar{s})d\bar{s} &= 0. 
\end{align}
\paragraph{Spectral decomposition} Let us use $\phi_n(\bar{s}$ alongwith the weight function $L(\bar{s})$ to construct a Fourier series representation (converging in the mean) to the square-integrable function $\eta(\bar{s})$,
\begin{align}
	\eta (\bar{s}) &= \sum_{n=1}^\infty c_n \phi_n (\bar{s}), \qquad c_n \text{ are Fourier coefficients}.
\end{align}
Substitute the above representation in Eqn. \ref{eqn:second-variational-derivative-appendix},
\begin{align*}
	\delta^2\mathcal{\overline{E}}  (\theta) &= - \int_{0}^{1} [c_n \phi_n''(\bar{s}) + L(\bar{s}) c_n\phi_n(\bar{s})] c_m\phi_m(\bar{s}) d\bar{s} \\
	&=  - c_m c_n \int_{0}^{1} [C_{R_n} \cos \theta(\bar{s}) - \lambda_n L(\bar{s}) \phi_n(\bar{s})]\phi_m(\bar{s}) d\bar{s} - c_n c_m \int_{0}^{1}  L(\bar{s}) \phi_m(\bar{s})\phi_n(\bar{s}) d\bar{s} \\
	&= -c_m c_n C_{R_n} \underbrace{\int_{0}^{1} \cos \theta(\bar{s}) \phi_m(\bar{s}) d\bar{s}}_{=0} + c_n c_m  (\lambda_n - 1) \underbrace{\int_{0}^{1} L(\bar{s}) \phi_m(\bar{s}) \phi_n(\bar{s}) d\bar{s}}_{\text{(use orthogonality condition)}} \\
	&= c_n^2 (\lambda_n - 1) \int_{0}^{1} L(\bar{s}) \phi_n^2(\bar{s}) d\bar{s} \\
	&= c_n^2 (\lambda_n - 1) \frac{1}{\lambda_n} \int_{0}^{1} (\phi_n'(\bar{s}))^2 d\bar{s}.
\end{align*}
The stability criterion is
\begin{align}
	\implies    \delta^2\mathcal{\overline{E}}  (\theta) = \sum_{n=1}^{\infty} c_n^2 \bigg(1 - \frac{1}{\lambda_n}\bigg)\int_{0}^{1} (\phi_n'(\bar{s}))^2 d\bar{s} \quad  \begin{cases}
		> 0 \text{ if } \lambda_n \notin [0,1]  & \text{ stable} \\
		< 0 \text{ if } \lambda_n \in [0,1] & \text{ unstable}.
	\end{cases}
\end{align}



\bibliography{JoE_Avatar_Dabade}

\end{document}